\documentclass{article}

\usepackage{microtype}
\usepackage{graphicx}
\usepackage{subcaption}
\usepackage{booktabs} 

\usepackage{xurl}
\usepackage{hyperref}

\usepackage[accepted]{icml2026}

\usepackage{amsmath}
\usepackage{amssymb}
\usepackage{mathtools}
\usepackage{amsthm}

\usepackage[capitalize,noabbrev]{cleveref}

\theoremstyle{plain}

\theoremstyle{definition}

\theoremstyle{remark}

\usepackage[textsize=tiny]{todonotes}

\usepackage{graphicx}
\usepackage{subcaption}  
\usepackage{booktabs,multirow,pifont}
\usepackage{booktabs,array,xcolor,pifont}
\usepackage{wrapfig}
\usepackage{tcolorbox}
\tcbuselibrary{breakable,skins,hooks} 
\usepackage{placeins}                 
\usepackage{listings}
\usepackage[shortlabels]{enumitem}

\newcommand{\modelname}[1]{\textsc{#1}}
\newcommand{\cmark}{\textcolor{green}{\ding{51}}} 
\newcommand{\xmark}{\textcolor{red}{\ding{55}}}   

\newcommand{\camera}[1]{\textcolor{black}{#1}}
\newcommand{\icmlrevision}[1]{\textcolor{black}{#1}}
\newcommand{\ours}{\textsc{QuArch}}

\usepackage{xcolor}
\definecolor{customfirst}{HTML}{48A057}
\definecolor{customsecond}{HTML}{2E6CB1}
\definecolor{customthird}{HTML}{8E4C97}

\definecolor{myblue}{RGB}{20,60,120} 
\definecolor{mylightblue}{RGB}{220,230,250} 

\newcommand{\first}[1]{\textcolor{customfirst}{\textbf{#1}}}
\newcommand{\second}[1]{\textcolor{customsecond}{\textbf{#1}}}
\newcommand{\third}[1]{\textcolor{customthird}{\textbf{#1}}}
\newcommand{\ballnumber}[1]{\tikz[baseline=(myanchor.base)] \node[circle,fill=.,inner sep=1pt] (myanchor) {\color{-.}\bfseries\footnotesize #1};}

\icmltitlerunning{QuArch: A Benchmark for Evaluating LLM Reasoning in Computer Architecture}

\begin{document}
\twocolumn[
  \icmltitle{\textsc{QuArch}: A Benchmark for Evaluating\\LLM Reasoning in Computer Architecture}




  \begin{icmlauthorlist}
    \icmlauthor{Shvetank Prakash}{harvard}
    \icmlauthor{Andrew Cheng}{harvard}
    \icmlauthor{Mark Mazumder}{harvard}
    \icmlauthor{Arya Tschand}{harvard}
    \icmlauthor{Varun Gohil}{mit}\\
    \icmlauthor{Jeffrey Ma}{harvard}
    \icmlauthor{Jason Yik}{harvard}
    \icmlauthor{Zishen Wan}{harvard,gatech}
    \icmlauthor{Jessica Quaye}{harvard}
    \icmlauthor{Elisavet Lydia Alvanaki}{columbia}
    \icmlauthor{Avinash Kumar}{utexas}
    \icmlauthor{Chandrashis Mazumdar}{ucsc}
    \icmlauthor{Tuhin Khare}{gatech}
    \icmlauthor{Alexander Ingare}{harvard}
    \icmlauthor{Ikechukwu Uchendu}{harvard}
    \icmlauthor{Radhika Ghosal}{harvard}
    \icmlauthor{Abhishek Tyagi}{rochester}
    \icmlauthor{Chenyu Wang}{harvard}
    \icmlauthor{Andrea Mattia Garavagno}{harvard,genoa,santanna}
    \icmlauthor{Sarah Gu}{harvard}
    \icmlauthor{Alice Guo}{harvard}
    \icmlauthor{Grace Hur}{harvard}
    \icmlauthor{Luca P. Carloni}{columbia}
    \icmlauthor{Tushar Krishna}{gatech}
    \icmlauthor{Ankita Nayak}{qualcomm}
    \icmlauthor{Amir Yazdanbakhsh}{deepmind}
    \icmlauthor{Vijay Janapa Reddi}{harvard}
  \end{icmlauthorlist}

  \icmlaffiliation{harvard}{Harvard University}
  \icmlaffiliation{mit}{Massachusetts Institute of Technology}
  \icmlaffiliation{gatech}{Georgia Institute of Technology}
  \icmlaffiliation{columbia}{Columbia University}   
  \icmlaffiliation{utexas}{University of Texas at Austin}
  \icmlaffiliation{ucsc}{UC Santa Cruz}
  \icmlaffiliation{rochester}{University of Rochester}
  \icmlaffiliation{genoa}{University of Genoa}
  \icmlaffiliation{santanna}{Scuola Superiore Sant'Anna}
  \icmlaffiliation{qualcomm}{Qualcomm AI Research}
  \icmlaffiliation{deepmind}{Google DeepMind}

  \icmlcorrespondingauthor{Shvetank Prakash}{sprakash@g.harvard.edu}

  \icmlkeywords{Machine Learning, ICML, Computer Architecture, Systems, Benchmark, Dataset, Question-Answering, AI Agents, Large Language Models}
  \vskip 0.3in
]

\printAffiliationsAndNotice{}

\begin{abstract}
The field of computer architecture, which bridges high-level software abstractions and low-level hardware implementations, remains absent from current large language model (LLM) evaluations. 
To this end, we present \ours{} (pronounced `quark'), the first benchmark designed to facilitate the development and evaluation of LLM knowledge and reasoning capabilities specifically in computer architecture.
\ours{} \camera{v1.0} provides a comprehensive collection of 2,671 expert-validated question-answer (QA) pairs covering various aspects of computer architecture, including processor design, memory systems, and interconnection networks. 
Our evaluation reveals that while frontier models possess domain-specific knowledge, they struggle with skills that require higher-order thinking in computer architecture.
Frontier model accuracies vary widely (from 34\% to 73\%) on these advanced questions, highlighting persistent gaps in architectural reasoning across analysis, design, and implementation QAs.
\icmlrevision{
Furthermore, via fine-tuning we find that \ours{} can translate to improved performance on a realistic memory hierarchy design task, resulting in up to $1.99\times$ more area-efficient solutions and up to 40\% more viable solutions overall.
}
By holistically assessing fundamental skills, \ours{} provides a foundation for building and measuring LLM capabilities that can accelerate innovation in computing systems.
\camera{The \ours{} benchmark and leaderboard are publicly available at: \textcolor{blue}{\url{https://quarch.ai/}}.}
\end{abstract}

\vspace{-0.25in}

\section{Introduction}
\begin{figure}[t!]
  \centering
  \includegraphics[width=0.49\textwidth]{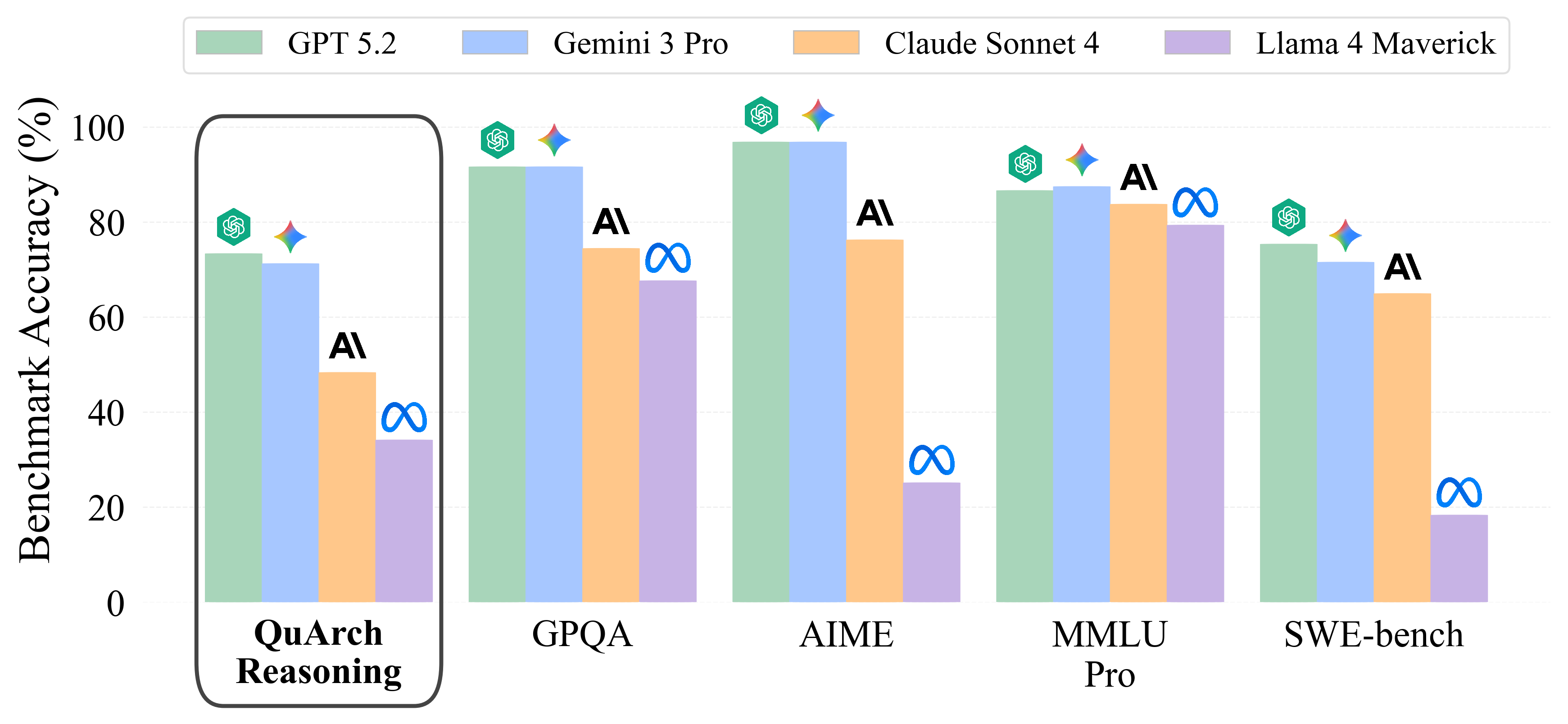}
  \caption{Reported results ~\citep{valsai_benchmarks} for flagship models across \ours{} and other \textsc{SoTA} benchmarks to date.}
  \vspace{-1.5em}
  \label{fig:benchmarks_vs_quarch}
\end{figure}

\begin{figure*}[t]
  \centering
  \includegraphics[width=0.9\textwidth]{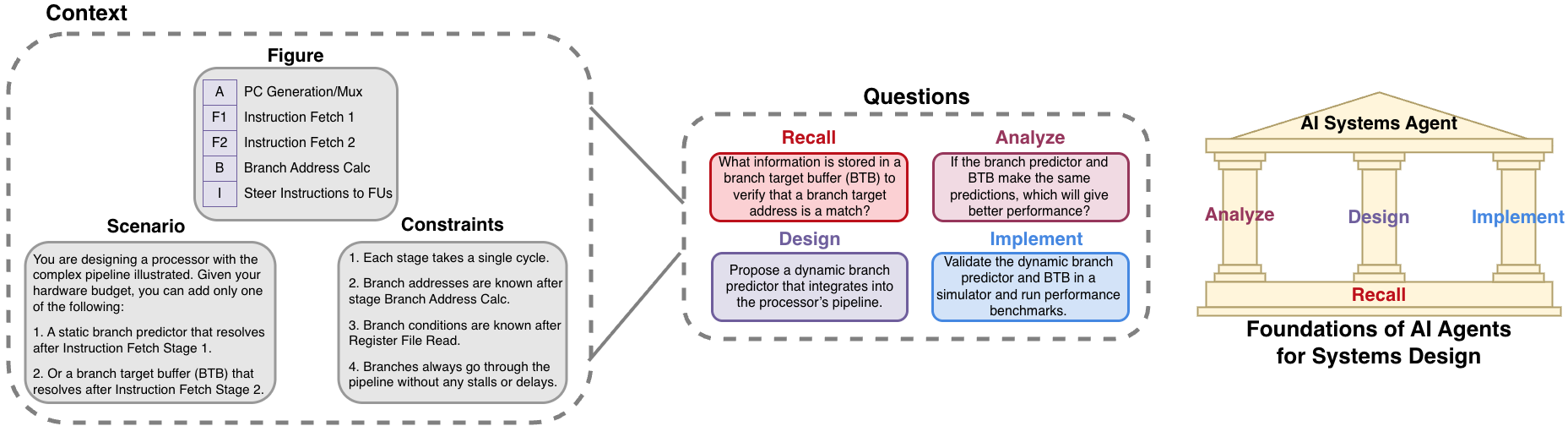}
  \caption{\ours{} QA Skills Framework. The benchmark evaluates four core competencies in systems design and computer architecture: Recall, Analyze, Design, and Implement. QAs in \ours{} contain relevant context describing the scenario, constraints, and figures when appropriate. The illustrative example shows how distinct question styles derived from the same context can probe different skills.
  } 
  \vspace{-0.1in}
  \label{fig:quarch_pillars}
\end{figure*}

Benchmarks that require reasoning from large language models (LLMs) are highly sought after, as they evaluate critical thinking beyond surface-level
knowledge and pattern matching. 
State-of-the-art (\textsc{SoTA}) progress is now measured by benchmarks which elicit multi-step reasoning, and models with explicit test-time deliberation (i.e., ``thinking'') consistently climb leaderboards.
Widely adopted datasets such as GSM8K~\citep{gsm8k}, AIME~\citep{aime2025}, SWE-bench~\citep{jimenez2024swebench}, GPQA~\citep{gpqa}, and MMLU-Pro~\citep{wang2024mmlu} 
serve as proxies for measuring math, software engineering, and natural and physical science expertise.

Reasoning is equally central to \emph{computer architecture}, which emphasizes evaluating trade-offs within a multi-objective optimization design space. 
For example, computer architects decide how to organize and balance components of systems (e.g., compute, memory, interconnects) and their power, performance, and area trade-offs.
However, computer architecture remains an area without dedicated LLM benchmarks. 

Existing benchmarks in computing systems target engineering tasks 
for software or chip implementation such as code generation~\citep{jimenez2024swebench, yang2024sweagent, openai_swebench_verified_2024, he2025swe}, register-transfer level (RTL) generation~\citep{verilogeval, pinckney2025comprehensive}, system-on-chip (SoC) integration~\citep{alvanaki2025sldb}, and chip verification~\citep{wan2025fixme}.
While these are important, they primarily evaluate whether a model can produce or manipulate programmatic artifacts, not whether it can reason about the principles that guide design decisions. 
Computer architecture plays a different role in the computing stack: it serves as the vital interface between software and hardware to define how these complex pieces interact, where careful orchestration of system components and their trade-offs \textit{informs and influences implementation}.
These decisions rely on conceptual understanding and analytical reasoning that is guided by application workloads and technology trends, rather than just code synthesis.
Importantly, the skills required by architects to navigate these multi-objective design space problems can be systematically evaluated through a question–answering (QA) paradigm.

To this end, we introduce \ours{}: a question-answering benchmark to assess the architectural knowledge and reasoning capabilities of LLMs required in computing systems design. 
Figure~\ref{fig:benchmarks_vs_quarch} presents reported performance of frontier models across other reasoning domains in comparison to \ours{}, demonstrating that reasoning models are not yet able to solve advanced architecture questions.
This gap underscores the need for focused evaluation on architectural reasoning to translate LLM progress into agentic methodologies that can accelerate innovation
in computing systems.

\ours{} aims to capture the expert domain knowledge and skills that architects possess by constructing a benchmark around four foundational competencies: Recall, Analyze, Design, and Implement. 
Existing systems benchmarks primarily target technical implementation skills (Table~\ref{tab:related-work-landscape}), but all four competencies---recalling foundational principles, analyzing workloads and objectives, designing systems that balance constraints, and implementing solutions via code---are complementary and critical for effective architecture design. 
While we focus on computer architecture, these skills are broadly applicable to many systems tasks. 
\ours{}'s evaluation framework enables LLM progress to be measured with fine-grained skills and compared over time. 

In summary, our work makes the following contributions:
\protect\ballnumber{1} \textbf{\ours{} \camera{v1.0} is the first benchmark designed to evaluate advanced computer architecture knowledge and reasoning in LLMs and is comprised of 2,671 expert-validated QAs.} 1,124 questions were curated through academic crowdsourcing and community competitions, and 1,547 questions were synthetically generated and human-verified.
\protect\ballnumber{2} \textbf{To promote holistic evaluation of LLMs for systems tasks, we formalize a skills framework to systematically assess 10 frontier models on \ours{}.} Our evaluation reveals that even flagship LLMs today struggle with skills requiring higher-order thinking. Notably, \ours{} uncovers a significant performance gap between LLMs’ architectural knowledge and reasoning abilities.
\protect\ballnumber{3} \textbf{We conduct an in-depth analysis to offer key insights and observations on model trends and failure points}. This includes incorrect architectural assumptions made, difficulties with modeling system state, absence of architecture-semantics in code execution, and heterogeneity in LLM topic expertise.
\protect\ballnumber{4} \textbf{We establish a trustworthy and scalable methodology for evaluating the correctness of free-form responses in \ours{} by comparing LLM judgments with human domain-expert verdicts across 100 QAs and 10 frontier models.} We show that LLM judgments agree with human experts at a rate of 85.48\%, which is comparable to human-human grading agreement rates of 90.75\% on the \ours{} benchmark.
\icmlrevision{
\protect\ballnumber{5} \textbf{We demonstrate knowledge transfer from \ours{} to a concrete and realistic architecture design task.} 
We show that fine-tuning open-source language models on \ours{} QAs can lead to improved outcomes for memory system design: Compared to base models, fine-tuned models proposed solutions that satisfied strict chip area and energy budgets in up to 40\% more trials and discovered designs that were up to $1.99\times$ more area-efficient.
}

\camera{
\textbf{Conflict of Interest Disclosure.}
The author A. Y. is employed by Google DeepMind, which leads the development of the Gemini and Gemma families of models, which were among the models evaluated in this paper.
}

\begin{table*}[t]
\caption{ML benchmarks \& datasets across the computing stack. \ours{} broadens the scope of current benchmarks by focusing on conceptual and analytical reasoning skills required for computer architecture and systems design. Benchmarks above \ours{} target more software-oriented tasks, while those below focus on more hardware-centric, chip design tasks.}
\label{tab:related-work-landscape}
\centering
\setlength{\tabcolsep}{4pt}
\renewcommand{\arraystretch}{1.1}
\resizebox{\textwidth}{!}{
    \begin{tabular}{l p{2.6cm} c c c c c}
    \toprule
      \textbf{\shortstack{Benchmark \& Dataset\\for Computing Systems}} 
    & \textbf{\shortstack{Focus in\\Computing Stack}} 
    & \textbf{\shortstack{Conceptual \&\\Analytical QA}} 
    & \textbf{\shortstack{Design QA \&\\Program Impl.}} 
    & \textbf{\shortstack{Multimodal\\Assessment}} 
    & \textbf{\shortstack{Expert\\Verified}} 
    & \textbf{\shortstack{Benchmark\\Size}} \\
    \midrule
    SWE-bench~\citep{jimenez2024swebench}            & Software Eng.              & \xmark & \cmark & \xmark & \xmark & 2294 \\
    SWE-bench Verified~\citep{openai_swebench_verified_2024}   & Software Eng.              & \xmark & \cmark & \xmark & \cmark & 500 \\
    SWE-Perf~\citep{he2025swe}             & Performance Eng.       & \xmark & \cmark & \xmark & \xmark & 140 \\
    KernelBench~\citep{ouyang2025kernelbench}         & Performance Eng.       & \xmark & \cmark & \xmark & \xmark & 250 \\
    CodeMMLU~\citep{nguyen2025codemmlu}             & Code Reasoning        & \cmark & \xmark & \xmark & \xmark & 19912 \\
    CRUXEval~\citep{gu2024cruxeval} & Code Reasoning & \cmark & \cmark & \xmark & \xmark & 800 \\
    
    \midrule
    \textbf{\ours{} (This Work)} & \textbf{Architecture} & \textbf{\cmark} & \textbf{\cmark} & \textbf{\cmark} & \textbf{\cmark} & \textbf{2671} \\
    \midrule
    SLDB~\citep{alvanaki2025sldb}                 & System Design   & \xmark & \cmark & \xmark & \cmark & 10 \\
    CreativEval~\citep{delorenzoCreativEvalEvaluatingCreativity2024}          & HW Design        & \xmark & \cmark & \xmark & \xmark & 120 \\
    VerilogEval~\citep{verilogeval}          & RTL Generation              & \xmark & \cmark & \xmark & \xmark & 156 \\
    CVDP~\citep{pinckney2025comprehensive}                 & RTL Generation              & \cmark & \cmark & \xmark & \cmark & 783 \\
    MG-Verilog~\citep{mgverilog}           & RTL Generation              & \xmark & \cmark & \xmark & \xmark & 11000 \\
    EDA Corpus~\citep{eda1}           & EDA Tooling              & \cmark & \cmark & \xmark & \cmark & 1533 \\
    FIXME~\citep{wan2025fixme}                & Verification     & \xmark & \cmark & \xmark & \xmark & 180 \\
    ChiPBench~\citep{chipbench}            & Layout           & \xmark & \cmark & \xmark & \xmark & 20 \\
    \bottomrule
    \end{tabular}
}
\end{table*}

\vspace{-0.1in}
\section{\ours{}}

\subsection{Towards AI Agents for Computer Architecture}
\label{sec:building_an_architect}
\textbf{Skill Requirements.} To systematically assess progress towards agentic design of computing systems, we first introduce a conceptual framework to decompose the fundamental skills that computer architects and systems engineers require. 
Figure~\ref{fig:quarch_pillars} illustrates these skills: within a single problem scenario, we exemplify how different styles of QAs exercise different skills, from fundamental domain knowledge recall to advanced analysis, design, and implementation.
\camera{Our framework is intentionally designed to align with how architects reason in practice through these coarse-grained skills, while also being informed by prior educational frameworks that employ similar competency verbs~\cite{abet2025criteria, acm_blooms_computing}.}

\textit{Recall}: Retrieving domain knowledge, definitions, and facts.
\textit{``What information is stored in a branch target buffer (BTB) to verify that a branch target address is a match?''}
This includes the ability to identify components and roles in a diagram or specification such as standard digital logic elements.
Critically, domain knowledge underpins advanced reasoning~\citep{krieger2004domain, duncan2007role}.
\looseness=-1

\textit{Analyze}: Deducing, inferring, calculating, or interpreting data and information from a scenario to reason about workload implications and system behavior. 
Identifying bottlenecks and being able to explain ``why'' is critical for choosing trade-offs.
 \textit{``If the branch predictor and BTB make the same predictions, which will give better performance?''}
\textit{Design}: Proposing, inventing, or improving an architectural feature (method, component, or policy) while satisfying system requirements and constraints. 
It requires balancing nuanced performance, power, area, and cost trade-offs. 
Synthesizing a design requires iterating over architectural block diagrams and system specifications. 
\textit{``Suggest a dynamic branch prediction system for this processor's pipeline.''}

\textit{Implement}: Translating a design into executable artifacts (e.g., code/RTL/simulation scripts). 
Typically, this skill is used to validate a solution via modeling or measurement.
\textit{``Implement the dynamic branch predictor and BTB in a simulator and run performance benchmarks.''}

Crucially, all of these skills are significant pillars exercised in different scenarios at different times by architects and systems engineers, with domain knowledge being the foundation upon which other higher-order skills can be built. For example, without first knowing the basics of how processor execution,
memory hierarchy, concurrency, parallelism, and communication work, it is difficult (if not impossible) to reason about design and performance trade-offs within a complex multi-core system.

\camera{We use this framework in Section~\ref{sec:quarch_characterization} for benchmark characterization and Section~\ref{sec:skill_trends} to analyze model capabilities across different skills.}
\looseness=-1

\textbf{Knowledge Breadth Requirements.} 
Computer architecture has a multitude of specialized areas. 
Historical focus on microprocessor design has expanded towards many-core systems and domain-specific accelerators~\citep{5230801, dally2020domain} due to memory and power walls~\citep{wulf1995hitting, esmaeilzadeh2011dark}, elevating the importance of memory systems, interconnects, and system-level methodology~\citep{4167779, 7299248} to first-class concerns. 
\emph{Effective architectural reasoning requires understanding relationships and interactions across these areas.} 
For example, a processor aggressively optimized without considering the connected memory subsystem will exhibit more performance bottlenecks than if the two were co-designed. Thus, a benchmark should capture topic breadth to properly assess architectural knowledge.

\subsection{Benchmark Construction}

\label{sec:quarch_construction}
\begin{figure*}[t]
    \centering
      \includegraphics[width=\textwidth]{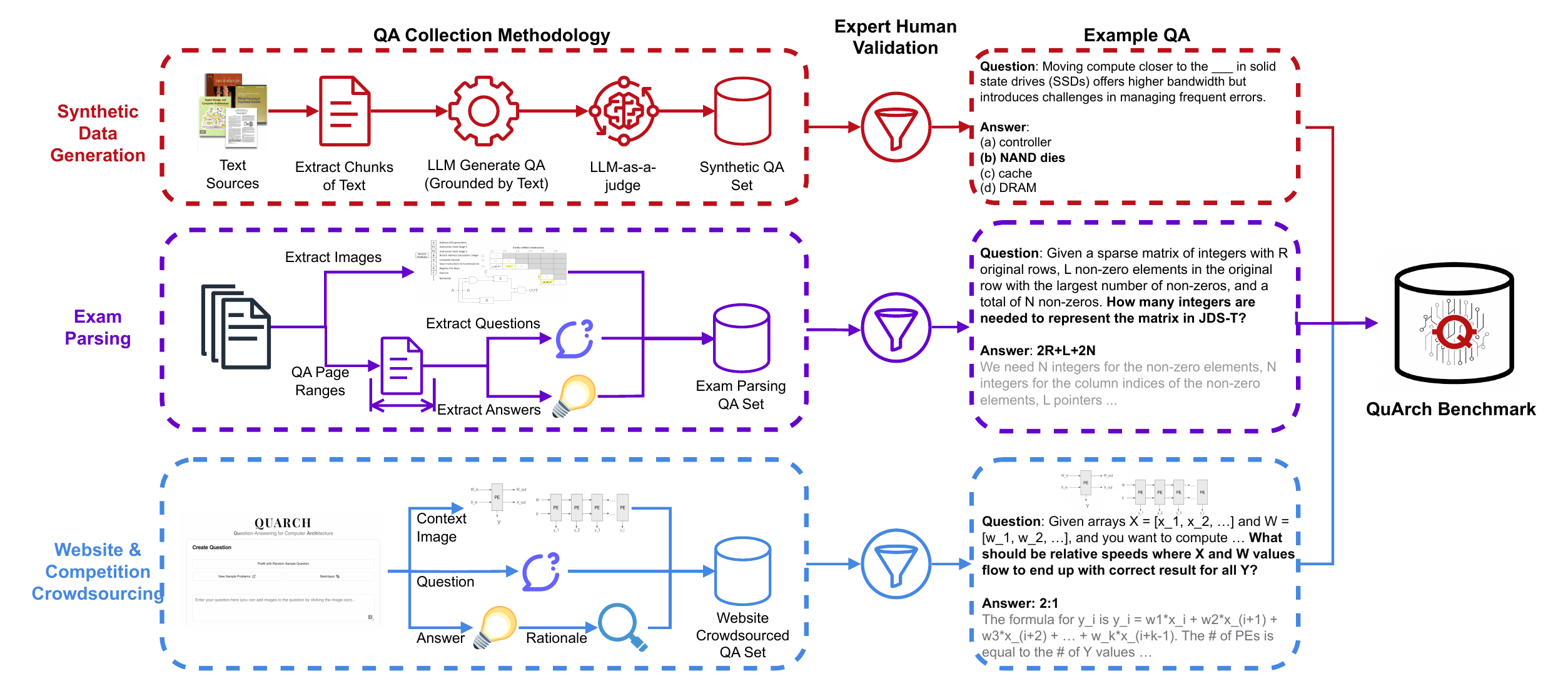}
    \caption{We construct \ours{} with a three-pronged approach including a blend of synthetic data generation, community crowdsourcing, and academic exams. All QAs are validated by a human expert to curate \ours{}'s final benchmark set of 2,671 question-answer pairs.}
    \label{fig:quarch-construction}
\end{figure*}

\textbf{Construction Approach.}
Curating a computer-architecture benchmark is particularly difficult because high-quality, openly usable sources are scarce relative to other domains~\citep{10857820} and authoring or validating benchmark entries requires substantial domain expertise to ensure technical correctness. 
\icmlrevision{We adopt a non-agentic QA task formulation for this domain's first benchmark because (1) it is efficient to evaluate and (2) it can assess fundamental architecture knowledge required for effective design (demonstrated in Section~\ref{sec:finetuning-case-study}).}
\camera{
To ensure realistic and challenging QAs while abstracting away iterative toolchain use, our benchmark construction process grounds questions in high-quality technical sources and incorporates both authoring and review of QAs from experts throughout \ours{}'s curation.
}
Specifically, we adopt a three-pronged strategy that combines synthetic data generation, academic exams, and expert crowdsourcing and competitions (Fig.~\ref{fig:quarch-construction}).
\icmlrevision{\textit{All three strategies importantly draw from artifacts that are used to train human architects to ensure benchmark relevance to computer architecture skills and knowledge.}}
\looseness=-1

\textbf{Synthetic Data Generation.} 
We collected open-source materials to curate a large corpus of computer architecture knowledge spanning technical manuals, academic publications, and comprehensive online resources.
This corpus reflects a diverse and thorough survey of publicly available knowledge in the field and serves as a foundation for \ours{}.
Using this corpus, LLMs generated cloze-style multiple-choice QAs~\citep{rogers2023qa} to balance educational value with practical assessment.
QAs then underwent two-stage validation: LLM-as-a-judge~\citep{zheng2023judging} for initial filtering (as these cloze-style QAs naturally involve little reasoning) followed by independent review of each QA by three experts. 
This approach enabled the identification and removal of questions lacking definitive answers or those too narrowly scoped for meaningful assessment. 
Prompt details are in Appendix~\ref{appendix:mcq_generation_prompt} and ~\ref{appendix:mcq_filtering_prompt}.
\textbf{Expert Crowdsourcing \& Competitions.}  
We developed a web-based portal specifically for crowdsourcing architectural reasoning questions to target more advanced analysis, design, and implementation skills that are difficult to synthetically generate.
QAs were collected via an open submission platform for individuals with technical backgrounds and time-boxed competitions.
Similar to other recent benchmark curation methodologies such as Humanity's Last Exam~\citep{phan2025humanity}, the interactive portal provided exemplary reasoning examples and real-time feedback on submitted questions to encourage participants to submit challenging questions and a solution rationale (Appendix~\ref{appendix:website}).
The individual submissions and competition submissions underwent expert review to check for ambiguity and correctness before final acceptance.

\textbf{Academic Exams.} 
We additionally curated QAs from university computer-architecture exams obtained via our community crowdsourcing process and manual web scraping.
A custom pipeline was developed to convert PDFs into standalone QAs.
Llamaparse was first used to extract diagrams~\citep{llamaparse2025}.
An LLM then segmented the exam into per-question PDFs to decompose the large exam PDF and parse each QA into context, question, and solution fields. 
To verify parsing, QAs underwent similar two-stage validation as our synthetic data generation process that employed LLM-as-a-judge for initial filtering followed by expert review. 
This pipeline yielded exam-level, multimodal QAs suitable for benchmarking.
Prompt details for this pipeline are in Appendix~\ref{appendix:text_extraction}, ~\ref{appendix:image_extraction}, and ~\ref{appendix:exam_verification}.

\subsection{Benchmark Characterization}
\label{sec:quarch_characterization}
We characterize the 2,671 QA pairs of \ours{} \camera{v1.0} along architecture topics, skill focus, question format, and modality, establishing a framework for fine-grained tracking of benchmark growth over time.
\looseness=-1

\begin{figure}%
  \centering
  \includegraphics[width=0.8\linewidth]{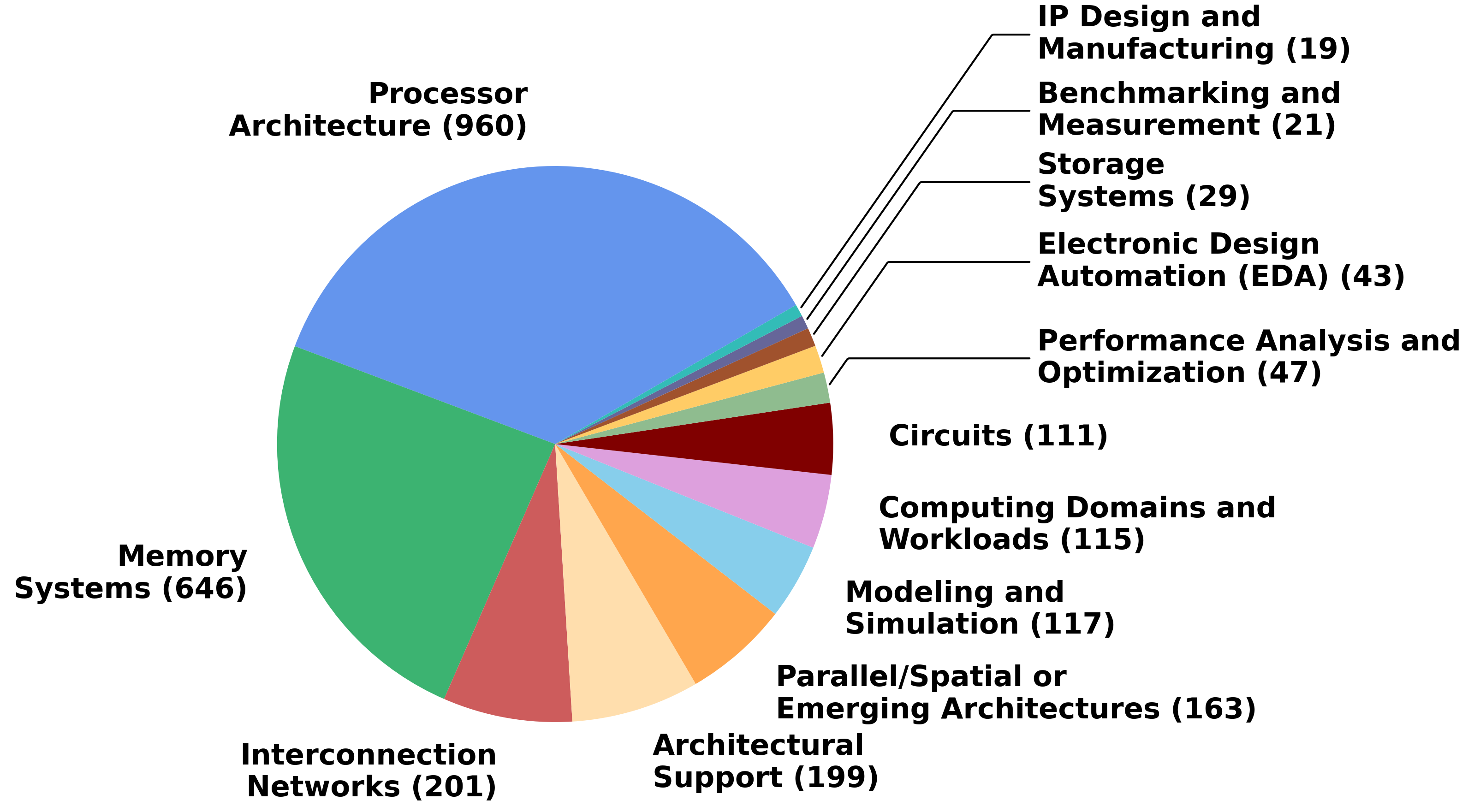}
  \caption{Distribution of topics in \ours{}. 
  }
  \label{fig:topic-dist}
\vspace{0.1in}
\end{figure}

\begin{figure}[!t]
    \centering
    \includegraphics[width=\linewidth]{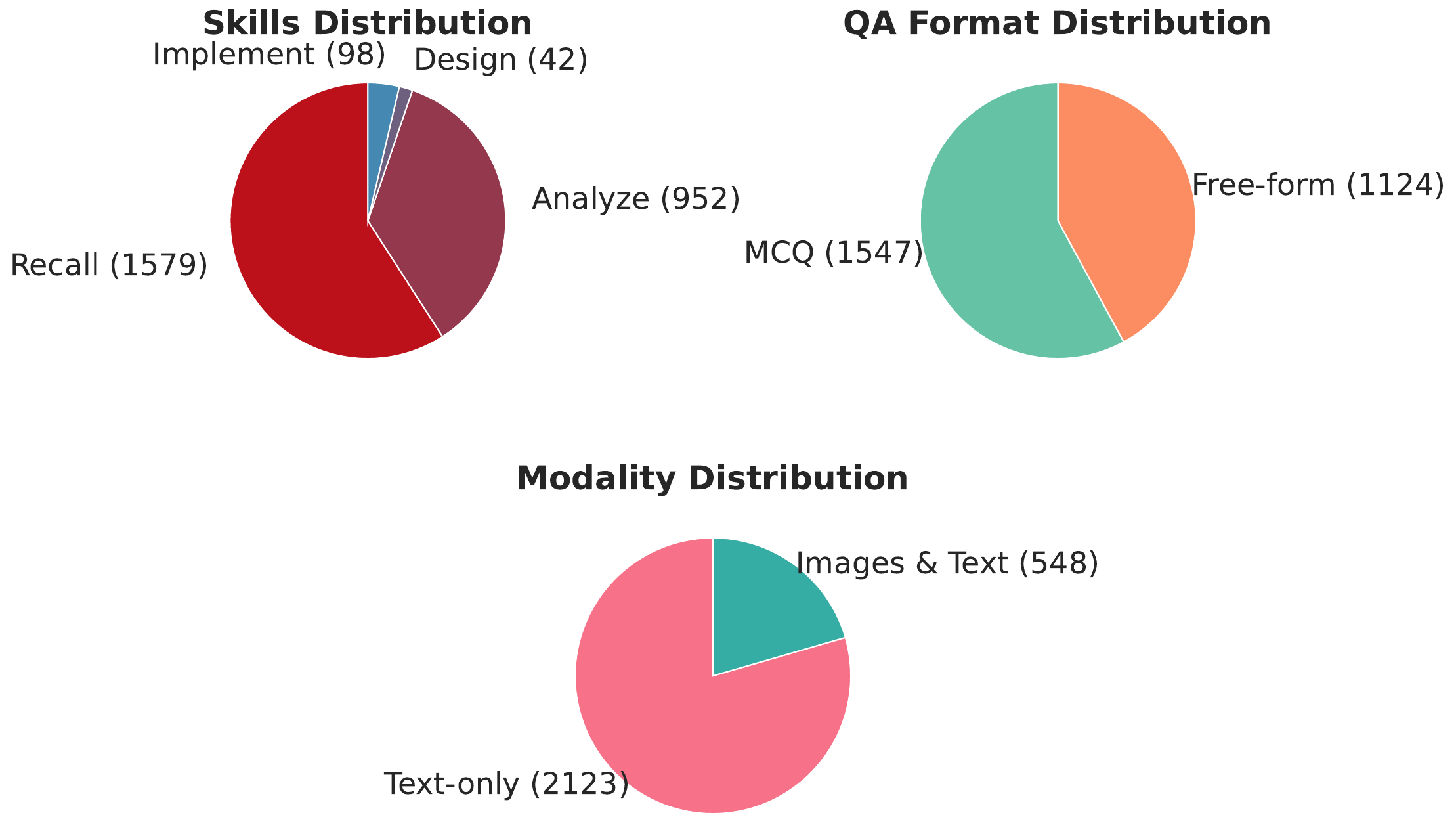}
    \caption{Breakdown of skill, format, and modality distributions in \ours{}. 
    }
    \vspace{-.25in}
    \label{fig:qa_distributions}
\end{figure}

\textbf{Architecture Topic Diversity.} 
\ours{} captures diverse topics in 13 core areas derived from key themes in modern computer architecture research (Figure~\ref{fig:topic-dist}).
Processor architecture accounts for the largest proportion of QAs (37\%), followed by memory systems (25\%) and interconnection networks (8\%). 
This distribution mirrors the field’s current and historical emphasis, with niche areas containing fewer QAs.
Appendix \ref{examples-appendix} provides example QAs that show the breadth and depth of topics covered in \ours{}.
The topic distribution was estimated via two-stage classification using a text embedding model and LLM labeling (Appendix~\ref{appendix:architecture_classification_prompt}).
\textbf{Skills Coverage.} 
Figure~\ref{fig:qa_distributions} characterizes QAs by the skills in Sec.~\ref{sec:building_an_architect} with examples for each given in Appendix~\ref{sec:example_taxonomy}.
\camera{We observed that many QAs naturally require multiple skills. To better capture this nuance, all QAs were manually annotated by domain experts using a primary and optional secondary skill label when appropriate. For consistency, our analysis and characterizations use the primary skill label.}
We term \textit{recall-focused} QAs \textsc{QuArch-Recall} and \textit{higher-order skill} QAs (e.g., analyze, design, implement) \textsc{QuArch-Reasoning}.
In particular, \textsc{QuArch-Reasoning} targets advanced reasoning by providing nearly 1000 analysis QAs and 140 design \& implementation QAs.
This delta reflects the intrinsic difficulty of authoring and validating design \& implementation QAs, mirroring their natural frequency in our sources (e.g., exams typically have few design QAs relative to analysis).
\camera{
We importantly note that despite the limited quantity of design and implementation QAs, the graduate-level academic provenance of these questions ensures their rigor and complexity.
Moreover, as shown in Section~\ref{sec:key_insights} and reflected by prior art in Table~\ref{tab:related-work-landscape}, small but high-quality benchmarks can meaningfully discriminate LLM reasoning capabilities and drive substantial progress.
}

\textbf{Question Format and Input Modalities.} 
In line with prior QA and code-reasoning benchmarks~\citep{gpqa,mmlu,nguyen2025codemmlu}, \ours{} includes 1,547 multiple-choice questions (MCQs), which are amenable to synthetic generation (Section~\ref{sec:quarch_construction}) and have clear evaluation criteria.
However, academic evaluation of domain-expert architecture knowledge is highly open-ended in structure, requiring deeper critical thinking, and thus cannot be formulated as MCQs. \ours{} therefore includes 1,124 free-response questions (FRQ), with examples in Appendix~\ref{examples-appendix}.
Furthermore, \ours{} contains both 2,123 text-only and 548 multimodal (images \& text) questions.  
These multimodal examples assess image interpretation and reasoning capabilities on structured and spatial information, such as architecture datapath diagrams, circuit schematics, pipeline timing charts, roofline performance plots, and specification tables.
\looseness=-1
\subsection{Example Questions}

\begin{figure*}
    \centering
    \includegraphics[width=0.95\textwidth]{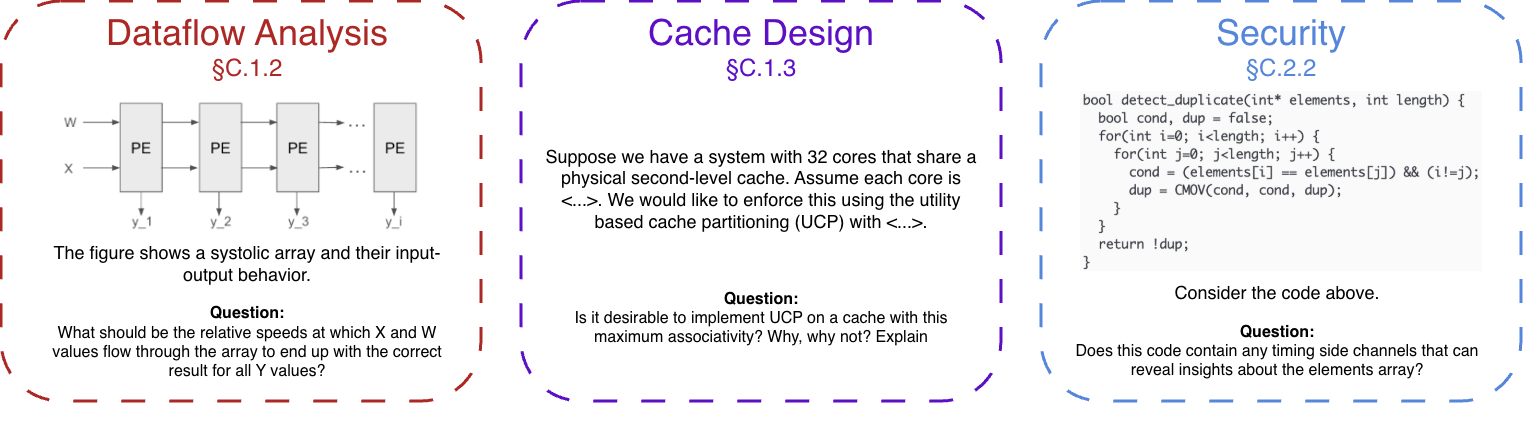}
    \vspace{-0.1in}
    \caption{
    Example Questions from \ours{}. Full question details in Appendix~\ref{examples-appendix}.
    }
    \vspace{-0.1in}
    \label{fig:example-questions}
\end{figure*}

Figure~\ref{fig:example-questions} presents sample questions from \ours{}.
As shown, questions in \ours{} exercise a variety of practical skills core to computer architecture. The examples include designing dataflow accelerators, configuring memory systems, and identifying hardware vulnerabilities of low-level code.
Fully expanded versions of these questions, along with others, can be found in Appendix~\ref{examples-appendix}.

\subsection{\textcolor{black}{Scope \& Extensibility}}
\camera{
\ours{} v1.0 is designed to evaluate the conceptual reasoning that underlies real-world architectural decision-making and design. 
Recent work~\citep{mhapsekar2026cachemind,llm_for_codesign} has similarly highlighted that architectural reasoning in a QA format is a critical component of the design process and architect's workflows.
Accordingly, \ours{} probes core cognitive tasks such as analyzing system performance, evaluating strategy trade-offs under constraints, and identifying bottlenecks. 
More broadly, \ours{} complements code and agentic benchmarks in the systems domain by isolating and measuring reasoning capabilities that influence downstream agentic performance, consistent with trends in LLM evaluation that distinguish reasoning-focused tasks from toolchain-heavy agentic evaluations~\citep{nguyen2025codemmlu,dinella2024crqbench,gu2024cruxeval}.
Critically, \ours{} is extensible by design:
our public submission portal (Appendix~\ref{appendix:website}) continues to receive new submissions from academic and industry contributors, and our skills framework (Fig.~\ref{fig:quarch_pillars}, Sec.~\ref{sec:building_an_architect}) provides a foundation for extending \ours{} toward future evaluations that incorporate executable artifacts for design and implementation tasks, enabling the benchmark to grow in scope and difficulty alongside advancing frontier model capabilities.}

\vspace{-0.10in}
\section{Experimental Setup} 
\label{sec:evaluation}

\label{subsec:setup}

\textbf{Models.} 
We evaluate 10 frontier models from
Anthropic, DeepSeek, Google, Meta, Mistral, and OpenAI on \ours{} v1.0 across the four skills (Recall, Analyze, Design, Implement) presented in Section~\ref{sec:building_an_architect}.
Evaluation results for an additional 20 models are reported in Appendix~\ref{sec:non-sota-models}.

\textbf{Evaluation grading.} All models are 
evaluated in a zero-shot setting. Full evaluation prompts are provided in Appendix~\ref{appendix:prompts}.
For MCQ-style responses, models must conclude with the correct choice of A, B, C, or D.
For FRQ-style questions, we employ LLM-as-a-judge~\citep{zheng2023judging}, tasking an external model to assess the correctness of an answer with respect to the ground truth.
We further motivate and rigorously validate our use of LLM-as-a-judge in Section~\ref{sec:llm-judge-analysis}.

\textbf{Metrics.} For both MCQ- and FRQ-style questions, we report model performance using "per-generation accuracy", the percentage of correct answers received out of $n$ total responses, to provide a better estimate of the \textit{pass@k}=1 metric~\citep{pinckney2025revisiting} under stochastic generation. In all evaluations, we generate $n=3$ samples per question.

\section{Evaluation \& Analysis}\label{sec:analysis}
\subsection{Model Performance}
\label{sec:model-performance}

Table~\ref{tab:headline-results} reports headline accuracy of 11 frontier models on \ours{}. 
As defined in Section~\ref{sec:quarch_characterization}, \textsc{QuArch-Reasoning} covers higher\mbox{-}order skills in our framework, while \textsc{QuArch-Recall} captures domain\mbox{-}knowledge retrieval (rather than reasoning).
\textsc{QuArch-Recall} performance is consistently strong across all frontier models. 
Including a recall split is useful to establish a baseline in a field that lacks a dedicated benchmark: the split distinguishes ``don't know'' from ``can't reason,'' and informs whether fundamental domain knowledge is present in models. Unlike frontier models, current small language models (SLMs) exhibit gaps on recall performance (Appendix~\ref{sec:full-model-results}).
\icmlrevision{This is significant as SLMs are becoming increasingly important for agentic AI~\citep{belcak2025small}.}
Overall, the reasoning variant of GPT\mbox{-}5 leads on \ours{} today with Gemini models forming the next tier. 
We note that \modelname{GPT-OSS-120B} and \modelname{DeepSeek-R1} are evaluated only on text (no images), so their scores reflect text-only capability.
We focus the rest of our analysis on \textsc{QuArch-Reasoning} because it offers the most headroom for today's frontier models to improve.
\begin{table}[t]
    \small
    \centering
    \caption{Frontier model performance on \ours{}. Reported values are the per-generation accuracy across 3 generations. All models struggle much more on \ours{}-\textsc{Reasoning} compared to \ours{}-\textsc{Recall}. We highlight the \first{first}, \second{second}, and \third{third} best performing models.}
    \label{tab:headline-results}
    \resizebox{\columnwidth}{!}{%

    \begin{tabular}{
        l
        c
        c
        c
    }
    \toprule
    Model & \multicolumn{3}{c}{\ours{}} \\
    \cmidrule(lr){2-4}
     & Recall & Reasoning & $\Delta$ \\
    \midrule
    \textit{Multimodal Models} \\
    \modelname{GPT-5.2} & \third{88.4} & \first{73.2} & \first{-15.2} \\
    \modelname{GPT-5} (Non-Reasoning) & 86.4 & 48.4 & -38.1 \\
    \modelname{Gemini-3-Pro} & \first{90.9} & \third{71.0} & \third{-19.9} \\
    \modelname{Gemini-3-Flash} & \second{90.1} & \second{71.9} & \second{-18.2} \\
    \modelname{Claude-Sonnet-4} & 85.3 & 48.1 & -37.2 \\
    \modelname{Claude-3.7-Sonnet-Thinking} & 85.7 & 51.8 & -33.9 \\
    \modelname{Llama-4-Maverick} & 85.2 & 33.7 & -51.5 \\
    \modelname{Mistral-Medium-3.1} & 84.2 & 33.8 & -50.4 \\
    \midrule
    \textit{Text-Only Models} \\
    \modelname{GPT-OSS-120B} & 84.5 & 63.6 & -20.8 \\
    \modelname{DeepSeek-R1} & 86.9 & 55.4 & -31.5 \\
    \bottomrule
    \end{tabular}}

\end{table}

\subsection{Skill Performance Trends}
\label{sec:skill_trends}
Table~\ref{tab:skill-accuracy} 
provides fine-grained skill-wise performance across models, with key trends shared below:

\textbf{(1) Recall is Mastered, Higher-Order Skills are Not.}
Frontier models have largely mastered recall, but fall short on advanced skills. 
Recall accuracy ranges between 84\%-91\%, suggesting architectural knowledge is present.
However, analyze, design, and implement skills are lower than recall by 31\%, 31\%, and 38\% on average respectively. 
Notably, multiple models with strong recall accuracy fall below 35\% on other skills (e.g., \modelname{Llama-4-Maverick}, \modelname{Mistral-Medium-3.1}).
In particular, implement skills exhibit the widest performance gaps, ranging between 24-72\%.
This suggests the design QAs, despite comprising a small proportion of the dataset, provide a strong discriminative signal that
correlates with performance across the full \ours{} benchmark.

\textbf{(2) Reasoning Matters for Advanced Skills.}
Results from (1) indicate that translating domain knowledge into advanced skills will require targeted training and test-time deliberation mechanisms.
Comparison of \modelname{GPT-5.2} with a non-reasoning
variant supports this.  
\modelname{GPT-5.2} outperforms \modelname{GPT-5} (Non-Reasoning) by 24\%, 22\%, and 33\% on analyze, design, and implement QAs respectively, compared to a much smaller 2\% lift on recall QAs.

\textbf{(3) Variations in Competencies Across Models.} 
Our skill framework exposes model-specific strengths and failures that a single aggregate score hides.
For instance, 23 of the 30 models evaluated (Appendix~\ref{sec:full-model-results}) score lower on implement QAs than on design QAs, underscoring the need to target all of these higher-order skills to holistically assess architecture competency.
\looseness=-1
\vspace{-0.1in}

\begin{table}
\centering
\caption{Per-generation accuracy (\%) by \ours{} Skill. Best performing models highlighted \first{first}, \second{second}, and \third{third}.}
\label{tab:skill-accuracy}
\small
\resizebox{\columnwidth}{!}{%
\begin{tabular}{lrrrr}
    \toprule
    Model & & \multicolumn{3}{c}{\ours{}-\textsc{Reasoning}} \\
    \cmidrule(lr){3-5}
    & Recall & Analyze & Design & Implement \\
    \midrule
    \textit{Multimodal Models} \\
    \modelname{GPT-5.2} & \third{88.4} & \first{73.0} & \first{79.4} & \first{72.4} \\
    \modelname{GPT-5} (Non-Reasoning) & 86.4 & 48.9 & 57.1 & 39.5 \\
    \modelname{Gemini-3-Pro} & \first{90.9} & \third{71.4} & \second{73.8} & \third{65.6} \\
    \modelname{Gemini-3-Flash} & \second{90.1} & \second{72.8} & \third{66.7} & \second{66.3} \\
    \modelname{Claude-Sonnet-4} & 85.3 & 48.8 & 50.0 & 41.2 \\
    \modelname{Claude-3.7-Sonnet-Thinking} & 85.7 & 52.8 & 44.4 & 45.2 \\
    \modelname{Llama-4-Maverick} & 85.2 & 34.6 & 26.2 & 28.0 \\
    \modelname{Mistral-Medium-3.1} & 84.2 & 34.6 & 38.1 & 24.5 \\
    \midrule
    \textit{Text-only Models} \\
    \modelname{GPT-OSS-120B} & 84.5 & 63.9 & 66.7 & 58.1 \\
    \modelname{DeepSeek-R1} & 86.9 & 56.6 & 47.8 & 44.4 \\
    \bottomrule

\end{tabular}}
\end{table}

\subsection{Key Insights \& Observations} 
\label{sec:key_insights}

Results on \ours{} illustrate a clear gap in model capabilities. Based on extensive grading performed by experts (Section~\ref{sec:llm-judge-analysis}), we synthesize key insights into specific failure modes observed.

\textbf{(1) Struggles with architecture-semantics of code execution.} 
Architectural semantics of code execution refers to the deep understanding of how high-level code interacts with the underlying hardware architecture (e.g., memory access patterns, instruction scheduling, etc.)~\citep{tschand2025swizzleperf}.
Our analysis reveals that LLMs struggle with these nuanced aspects of code execution, failing to accurately predict or analyze the architectural implications of code snippets (Appendix~\ref{sec:failure_mode_1}).
\textbf{(2) Assuming unconventional architectural properties.}
In computer architecture, decades of practice have cemented certain system designs, such as byte-addressable memory, as de facto defaults unless otherwise specified.
However, our analysis exposes a misalignment: when prompts fail to state conventions explicitly, we observe LLMs defaulting to unconventional choices, such as word-level addressing (Appendix~\ref{sec:failure_mode_2}).
Models are able to succeed when provided with the default conventions explicitly, highlighting that practitioners leveraging LLMs in this domain must identify their implicit assumptions to guide the model effectively.
\textbf{(3) Modeling and tracking system state.} 
Building an intuition and mental model of how system components interact and the implications of their interactions is central to computer architecture.
In general-domain QA, this corresponds to situational world modeling~\citep{rogers2023qa}: instantiating entities, tracking their locations and states, and inferring temporal and causal relations to answer queries about an evolving scenario.
We find that frontier models often fail to maintain consistent system state and thus misunderstand how local actions cascade into system-level effects on latency, throughput, and correctness (Appendix~\ref{sec:failure_mode_3}).

\textbf{(4) Variations in domain expertise.}
Our analysis reveals that LLMs develop specialized expertise across different domains.
For instance, within ``Implement''-Style questions, \modelname{GPT-5.2} performs well on Emerging Architectures and struggles on IP Design, while \modelname{Gemini-3-Flash} exhibits the opposite behavior. While overall, \modelname{GPT-5.2} performed noticeably better on ``Implement''-Style questions (6\% improvement), these domain expertise differences resulted in \modelname{Gemini-3-Flash} performing $\sim$50\% better on IP Design implement questions. Model capabilities are thus more nuanced than the aggregate scores of Table~\ref{tab:skill-accuracy}.
These findings provide the opportunity to create multi-model systems that combine the domain strengths of multiple LLMs rather than relying on a single ``best'' model.
Spider plots visualizing these per-topic variations for frontier models are shown in Appendix~\ref{sec:topic-spiders}.

\textbf{(5) Sensitivity to QA modality.} 
In computer architecture, visuals such as pipeline diagrams, cache hierarchies, and system interconnects convey structural relationships and spatial information that cannot be adequately captured through text descriptions alone~\citep{rtl1}. 
\camera{
Multimodal models perform on average 5\% worse on questions with images than on text-only questions (Appendix Table~\ref{tab:non-sota-results}).
}
This gap indicates that frontier models struggle with interpreting and reasoning about diagrams, schematics, and tables 
(see Appendix~\ref{sec:failure_mode_4} for failure examples).

\begin{table}[!t]
\centering
\small
\caption{Percentage of trials in which at least one proposed solution satisfied both area and energy design specifications.
}
\begin{tabular}{lcc}
\toprule
 & \textbf{Baseline} & \textbf{Fine-tuned} \\
\midrule
\textbf{\modelname{Gemma-3-27B-it}} & 6 / 20 (30\%)  & 14 / 20 (70\%) \\
\textbf{\modelname{Llama-3.3-70B-Instruct}} & 11 / 20 (55\%) & 19 / 20 (95\%) \\
\bottomrule
\end{tabular}
\label{tab:finetuning-feasibility}
\end{table}

\subsection{\icmlrevision{Knowledge Transfer Case Study: Applying \ours{}-\textsc{Reasoning} QAs to Memory Design}}
\label{sec:finetuning-case-study}
\icmlrevision{
This section investigates the following question: \textit{Can a model's ability to solve questions in QuArch translate to improved success on a concrete architectural design task?}
To answer this, we conduct a case study that asks models to design the memory hierarchy of an ML hardware accelerator for matrix multiplication, an extensively studied problem in computer architecture~\citep{sze2017efficient}.
Critically, effective memory hierarchy optimization demands reasoning grounded in core architectural principles to understand the complex interactions between design knobs that impact area–energy trade-offs.
A second case study focused on memory controller design is in Appendix~\ref{sec:memory-controller-case-study}.
}
\looseness=-1

\icmlrevision{
\textbf{Task Formulation.} Models must jointly propose hardware and software modifications (i.e., cache hierarchy organization, dataflow ordering, tiling strategies, memory layout) with the objective of reducing chip area relative to a baseline design while satisfying strict energy constraints.  
Proposed designs are evaluated using established architectural simulation and area/energy estimation tools~\citep{timeloop, accelergy}.
See Appendix~\ref{app:finetuning} for additional details.
}

\icmlrevision{
\textbf{Training Methodology.} We select two open-source models for training: \modelname{Gemma-3-27B-it} and \modelname{Llama-3.3-70B-Instruct}.
We distill answers and explanations on a subset of \ours{} (45 text-only FRQs related to memory subsystems and matrix multiplication) into these models via supervised fine-tuning (SFT)~\citep{muennighoff2025s1}. 
In total, we fine-tuned on 541 LLM responses to these 45 questions and 135 \modelname{GPT-5.2} explanations on how to solve each question, for three epochs. See Appendix~\ref{app:finetuning} for hyperparameters. 
}

\icmlrevision{
\textbf{Results.}
For both the base and fine-tuned models, we perform 20 independent trials of this task, where one trial consisted of 10 iterative turns between the model and the design simulator. 
Table~\ref{tab:finetuning-feasibility} reports the percentage of trials in which the model successfully produced a solution that met the chip area and energy budget provided for this design. 
We observe that fine-tuning models on \ours{} leads to a significant increase in the percentage of proposed designs that met chip area and energy budgets compared to the two baseline models. 
With an additional epoch of fine-tuning on the same dataset, the fine-tuned Gemma model is able to discover a design requiring much less chip area at $3608.89\mu m^2$, a $1.86\times$ improvement over its base model counterpart (Appendix Table~\ref{tab:appendix-cache-case-study-results}). Similarly, the Llama model finds a solution requiring only $3637.89\mu m^2$ ($1.99\times$ improvement over its base model), albeit at the trade-off of lower success rate across trials. 
\textit{This case study exemplifies how the knowledge required to answer QuArch QAs can be applied to improve on tasks asked of computer architects.}
}

\subsection{LLM-as-a-Judge Analysis}
\label{sec:llm-judge-analysis}

\begin{table}[!t]
\centering
\small
\caption{Confusion matrix comparing LLM-as-a-Judge with domain-expert human grading on FRQ. %
}
\resizebox{\columnwidth}{!}{%
\begin{tabular}{lcc}
\toprule
 & \textbf{Human Correct} & \textbf{Human Incorrect} \\
\midrule
\textbf{LLM Judge Correct}   & 428 & 60  \\
\textbf{LLM Judge Incorrect} & 73  & 355 \\
\bottomrule
\end{tabular}
}
\label{tab:confusion-matrix}
\end{table}

\textbf{Motivation.}
Semantically equivalent and correct solutions to the same FRQ can differ in phrasing, as shown in Appendix~\ref{sec:successful_response}. Since full manual grading by domain experts is intractable, we employ LLM-as-a-judge for \ours{}.
\textbf{Human Validation.}
While LLM-as-a-judge has gained popularity for evaluating FRQ-style questions~\citep{lee2024prometheus,zhou2023lima,manas2024improving,pinckney2025comprehensive}, the approach is still relatively new.
We therefore validate the fidelity of LLM-as-a-Judge by measuring agreement rates between human expert and LLM judge verdicts on the correctness of generated FRQ responses. We randomly sampled 100 (8.9\%) freeform QAs in \ours{}, and generated one response each from 10 models\footnote{
Model list in Appendix~\ref{app:subsec:human-expert-comparison}.
}.
We tasked a cohort of 11 domain experts in computer architecture and hardware design to manually grade the resultant 908 responses\footnote{Non-multimodal models cannot generate responses for the multimodal proportion of sampled questions.} as \textsc{Correct}, \textsc{Partially-Correct}, or \textsc{Incorrect}. For analysis purposes, \textsc{Partially-Correct} is recategorized as \textsc{Incorrect}. Each question is graded independently by up to 3 experts and the majority consensus is taken.
All LLM judge evaluations in Sec.~\ref{sec:model-performance} and Appendix~\ref{sec:non-sota-models} are likewise performed 3x and the majority vote taken.

\textbf{LLM judges agree with human experts.}
We observe an agreement rate of 85.48\%
between LLM judges and humans (Table~\ref{tab:confusion-matrix}), when using \modelname{Claude-3.7-Sonnet-Thinking} as the judge.
We compare this agreement rate with the rate that expert humans disagree on verdicts. 84 of the 908 responses required a third expert to adjudicate between a correct and incorrect vote, corresponding to a human-to-human agreement rate of 90.75\%.
Since this agreement rate is comparable to the rate that LLM-as-a-Judge consensus agrees with human consensus, we argue LLM-as-a-Judge is eminently suitable for scalable and informative benchmarking of model performance on \ours{}. 
Additional experiments on human expert grading difficulty, alternative judge LLMs, and consensus rates are included in Appendix~\ref{app:subsec:human-expert-comparison}.

\section{Related Work}

\textbf{Software.}
Function-level code efficiency benchmarks~\citep{du2024mercury, huang2024effibench, shypula2024pie, waghjale2024ecco} and domain-focused performance tasks~\citep{press2025algotunelanguagemodelsspeed, ouyang2025kernelbench} evaluate correctness-preserving edits and runtime gains at the function or kernel level. Repository-scale SWE benchmarks and agentic toolchains~\citep{jimenez2024swebench,yang2024sweagent,wang2025openhands} test long-horizon code manipulation and integration. Recent QA code understanding benchmarks~\citep{gu2024cruxeval, nguyen2025codemmlu, li2024infibench,dinella2024crqbench} target control/data-flow semantics, behavioral equivalence, and code review comprehension. Unlike \ours{}, which primarily targets pre-implementation, system, and architectural judgment, these works focus on code artifacts, assessing code semantics rather than system-level design.
\looseness=-1

\textbf{Hardware.}
Domain-specific foundation models for chip design~\citep{chipnemo}, electronic design automation (EDA) tool interaction~\citep{eda1, eda2}, RTL generation~\citep{rtl1, rtl2, rtl3, rtl4}, design optimization~\citep{opt1, opt2, opt3}, and security-oriented tasks (bug repair and assertions)~\citep{hdlbug1,hdlbug2,hdlbug3,securitybug1,securitybug2,assertion1,assertion2} emphasize producing or improving implementation artifacts and driving tools. In contrast, \ours{} isolates the reasoning that guides implementation (e.g., architectural trade-offs)
rather than their ability to generate HDL/RTL or steer EDA flows.

\textbf{QA benchmarks.}
General-purpose and domain QA datasets~\citep{squad, trischler2016newsqa, clark2019boolq, mmlu, gpqa, ccqa, pubmedqa, gsm8k, jec-qa} have been instrumental for advancing and measuring LLMs~\citep{rogers2023qa}.
\ours{} targets advancing computer architecture specifically, with expert-verified items and skill-wise evaluation capabilities not covered by existing QA benchmarks.

\section{Conclusion}
We introduce \ours{} \camera{v1.0}, the first benchmark to directly assess computer architecture knowledge and reasoning in LLMs across four complementary skills: Recall, Analyze, Design, and Implement. 
Evaluating ten frontier models on 2,671 expert-validated QAs, we find consistently strong recall across models but reveal a pronounced gap in higher-order abilities that demand architectural reasoning. 
By providing insights into failure modes and enabling systematic tracking, \ours{} lays the groundwork for accelerating AI progress in computer architecture and, more broadly, in reasoning-centric skills for systems design.

\section*{\textcolor{black}{Acknowledgments}}
\camera{
We thank the anonymous reviewers for their thoughtful feedback and suggestions, which helped improve the quality and clarity of this manuscript. We also extend our gratitude to Derek Lockhart, Cliff Young, and James Laudon for their valuable feedback on the paper, as well as the extended team at Google DeepMind for supporting this research direction. We especially thank Kai Kleinbard for developing the \ours{} website and managing project infrastructure throughout the development of this work. 
}

\camera{
Finally, we would like to acknowledge and sincerely thank the many students and contributors who played an important role in the development of \ours{}. This project was made possible through a large collaborative community effort spanning dataset creation, question verification, infrastructure development, and evaluation.
The list that follows includes all contributors, ordered alphabetically by first name, who had participated in the \ours{} v1.0 project at the time of submission (January 28, 2026): 
Aarush Gupta,
Abhiram Ghanta,
Adarsh Sriuma,
Aditya Bhaskar,
Aditya Borse,
Aditya Mavle,
Ajay Joshi,
Akash Bommidi,
Alexander Snapp,
Andrej Vrtanoski,
Andrew Peng,
Ankith Thalanki,
Ansh Bhatti,
Anuj Bhatt,
Anurag Yadav,
Arkaprava Basu,
Arkapravo Ghosh,
Aryan Gupta,
Avani H.,
Avi Kapur Srinivasan,
Ayushi Rajpoot,
Bujji Selagamsetty,
Cheng-Jhih Shih,
Chiranjeevi Chimmili,
Daniel Terrell,
Dhruv Raj Bangad,
Divya Mahajan,
Eli Corley,
Ellen Suh,
Eugene Chu,
Euijun Chung,
Fnu Navneet,
Gaurang Upasani,
Gowsika Dharmaraj,
Han Cho,
Hanran Wu,
Haomei Liu,
Hema Chandra Kolisetty,
Himanshi Gupta,
Hongzheng Chen,
Hsueh-Yuan Chou,
Hunter Lee,
Ian Wong,
Isaac Khor,
Ishita Vohra,
Ismael Youssef,
Jackie Mac Hale,
Jae Hyung Ju,
Jagadheesvaran Tirupathi Subburayan,
James Xu,
Jarvis Jia,
Jeeho Ryoo,
Jenny Huang,
Jessica Hernandez,
Jiajie Qian,
Jiayi Qian,
Jingtian Dang,
Jinhyeok Park,
Jogesh Kumar,
Joseph Ferraro,
Joshua San Miguel,
Jun Liang Ho,
Kai Kleinbard,
Kaiyi Hu,
Kalp Vyas,
Kevin Sui,
Krishil Gandhi,
Laith Shamieh,
Lexington Whalen,
Lizy K John,
Logashree Venkatasubramanian,
Marian Verhelst,
Mayur Peshve,
Miaoyan Zhou,
Minseung Jung,
Mohamed Ghanem,
Mohnish Pai,
Muhammad Haseeb,
Nathan Duggal,
Nathan Zhong,
Nicolás Majorel Padilla,
Nishant Gadde,
Noah Bruckner,
Onur Mutlu,
Panya Bhinder,
Philip Ndikum,
Pin-Jun Chen,
Po-Han Porras Huang,
Pooria Taheri,
Prabhav Gupta,
Pramath Balisavira,
Pranaav Milaganur Mohan,
Pranay Jaggi,
Pratham Nandy,
Pravesh Sanjay Jamgade,
Priya Panda,
Puneet Bansal,
Rahul Raj,
Ramil Agliamzanov,
Renan Silva,
Ribhu Das Purkayastha,
Royce Arockiasamy,
Saketh Patel,
Samuel Xu,
Sanjay Patnala,
Santosh Pandey,
Saurabh Singh,
Seungjae Jason Lee,
Shaunak Ghatpande,
Shehab Naga,
Shengjie Lin,
Shiv Prakash,
Shreya Chivilkar,
Shreyas Grampurohit,
Siddharth Joshi,
Sidney Wright,
Soham Chausalkar,
Soham Rattan,
Sri Siddarth Chakaravarthy Prakash,
Srinath Suresh Kumar,
Srinidhi Subramaniam Pasupathy,
Steve Tang,
Suyash Dandekar,
Tianwei Jiang,
Tianyou Zhao,
Twisha Shah,
Varun Komperla,
Vimalan Krishnan Manivannan,
Wells Lu,
Yashas Ambati,
Yasmein Khalil,
Ye Chen,
Youssef Jaafar,
Yuehan Zhang,
Yuhao Zhu,
Yujie Li,
Yuming Chang,
Yunchuan Zhang,
Yunnuo Zhang,
Yuxiang Wei,
Zebin Guo,
Zekai Wang,
Zekun Li,
Zhewen Pan,
Ziyao Yin.
}

\camera{
This work was supported by NSF Grant CCF-2324862 and POSE-2346173.
This work was also supported in part by the U.S. DOE DeCoDe Project No. 84245 at PNNL and by the Columbia Center of AI Technology (CAIT). 
The authors were additionally supported by the NSF Graduate Research Fellowship Program (GRFP).
We also thank Together.AI and Amazon Research Awards (ARA) for providing research credits that supported the fine-tuning and model evaluations conducted in this work, respectively.
}

\section*{Impact Statement}
This paper presents work whose goal is to advance the field of machine learning. There are many potential societal consequences of our work, none of which we feel must be specifically highlighted here. Our statement on reproducibility is provided in Appendix~\ref{app:ethics_reproducibility_llmusage}.

\bibliography{refs}
\bibliographystyle{icml2026}

\newpage
\appendix
\onecolumn
\section*{Appendix}

\begin{center}
\Large\bfseries Table of Contents
\end{center}
\noindent\textbf{\ref{app:ethics_reproducibility_llmusage} \quad Ethics, Reproducibility, and LLM Usage} \dotfill \pageref{app:ethics_reproducibility_llmusage}\\[0.2em]
\quad \ref{app:ethics} \quad Ethics Statement \dotfill \pageref{app:ethics}\\[0.2em]
\quad \ref{app:reproducibility} \quad Reproducibility Statement \dotfill \pageref{app:reproducibility}\\[0.2em]
\quad \ref{appendix:llm} \quad LLM Usage \dotfill \pageref{appendix:llm}\\[0.2em]

\noindent\textbf{\ref{appendix:additional_results} \quad Additional Details \& Results} \dotfill 
\pageref{appendix:additional_results}\\
\quad \ref{appendix:access_usage} \quad \ours{} Benchmark Access and Usage \dotfill \pageref{appendix:access_usage}\\
\quad \ref{appendix:website} \quad \ours{} Crowdsourcing Website \dotfill \pageref{appendix:website}\\
\quad \ref{sec:topic-spiders} \quad Topic-Wise Extended Characterization \dotfill \pageref{sec:topic-spiders}\\
\quad \ref{sec:full-model-results} \quad Full Model Results by Skill \dotfill \pageref{sec:full-model-results}\\
\quad \ref{sec:non-sota-models} \quad Model Performance by Modality \dotfill \pageref{sec:non-sota-models}\\
\quad \ref{app:finetuning} \quad Case Study \#1 \dotfill \pageref{app:finetuning}\\[0.2em]
\quad \ref{sec:memory-controller-case-study} \quad Case Study \#2 \dotfill \pageref{sec:memory-controller-case-study}\\
\quad \ref{sec:partially-correct} \quad Partially Correct Judgments \dotfill \pageref{sec:partially-correct}\\
\quad \ref{app:subsec:human-expert-comparison} \quad Comparing Human Domain Experts to LLM-as-a-Judge \dotfill \pageref{app:subsec:human-expert-comparison}\\[0.2em]

\noindent\textbf{\ref{examples-appendix} \quad Example Questions} \dotfill \pageref{examples-appendix}\\
\quad \ref{sec:example_taxonomy} \quad QA Skills \dotfill \pageref{sec:example_taxonomy}\\
\quad\quad \ref{sec:example_taxonomy_recall} \quad Example 1: Recall \dotfill \pageref{sec:example_taxonomy_recall}\\
\quad\quad \ref{sec:example_taxonomy_analysis} \quad Example 2: Analysis \dotfill \pageref{sec:example_taxonomy_analysis}\\
\quad\quad \ref{sec:example_taxonomy_design} \quad Example 3: Design \dotfill \pageref{sec:example_taxonomy_design}\\
\quad\quad \ref{sec:example_taxonomy_implementation} \quad Example 4: Implementation \dotfill \pageref{sec:example_taxonomy_implementation}\\
\ref{sec:failure_mode_1} \quad Failure Mode \#1: Struggles with Architecture–Semantics of Code Execution \dotfill \pageref{sec:failure_mode_1}\\
\quad\quad \ref{sec:failure_mode_1_example_1} \quad Example 1 \dotfill \pageref{sec:failure_mode_1_example_1}\\
\quad\quad \ref{sec:failure_mode_1_example_2} \quad Example 2 \dotfill \pageref{sec:failure_mode_1_example_2}\\
\quad\quad \ref{sec:failure_mode_1_example_3} \quad Example 3 \dotfill \pageref{sec:failure_mode_1_example_3}\\
\quad\quad \ref{sec:failure_mode_1_example_4} \quad Example 4 \dotfill \pageref{sec:failure_mode_1_example_4}\\
\ref{sec:failure_mode_2} \quad Failure Mode \#2: Assuming Unconventional Architectural Properties \dotfill \pageref{sec:failure_mode_2}\\
\quad\quad \ref{sec:failure_mode_2_example_1} \quad Example 1 \dotfill \pageref{sec:failure_mode_2_example_1}\\
\quad\quad \ref{sec:failure_mode_2_example_2} \quad Example 2 \dotfill \pageref{sec:failure_mode_2_example_2}\\
\ref{sec:failure_mode_3} \quad Failure Mode \#3: Modeling and Tracking System State \dotfill \pageref{sec:failure_mode_3}\\
\quad\quad \ref{sec:failure_mode_3_example_1} \quad Example 1 \dotfill \pageref{sec:failure_mode_3_example_1}\\
\quad\quad \ref{sec:failure_mode_3_example_2} \quad Example 2 \dotfill \pageref{sec:failure_mode_3_example_2}\\
\ref{sec:failure_mode_4} \quad Failure Mode \#4: Sensitivity to QA Modality \dotfill \pageref{sec:failure_mode_4}\\
\quad\quad \ref{sec:failure_mode_4_example_1} \quad Example 1 \dotfill \pageref{sec:failure_mode_4_example_1}\\
\quad\quad \ref{sec:failure_mode_4_example_2} \quad Example 2 \dotfill \pageref{sec:failure_mode_4_example_2}\\
\ref{sec:successful_response} \quad Example Motivating LLM-as-a-Judge for Automated Evaluations \dotfill \pageref{sec:successful_response}\\
\ref{sec:case_study_questions} \quad Case Study \#2 \dotfill \pageref{sec:case_study_questions}\\
\quad\quad \ref{sec:analyzing_mem_traces_examples} \quad Examples for Analyzing Memory Traces \dotfill \pageref{sec:analyzing_mem_traces_examples}\\
\quad\quad \ref{sec:memory_controller_configuration_examples} \quad Examples for Memory Controller Configuration \dotfill \pageref{sec:memory_controller_configuration_examples}\\[0.2em]

\noindent\textbf{\ref{appendix:prompts} \quad Prompt Templates} \dotfill \pageref{appendix:prompts}\\
\quad \ref{appendix:mcq_prompt} \quad LLM Prompt for MCQs \dotfill \pageref{appendix:mcq_prompt}\\
\quad \ref{appendix:frq_prompt} \quad LLM Prompt for FRQs \dotfill \pageref{appendix:frq_prompt}\\
\quad \ref{appendix:laj_prompt} \quad LLM Prompt for LLM-as-a-Judge on FRQ Responses \dotfill \pageref{appendix:laj_prompt}\\
\quad \ref{appendix:skills_classification_prompt} \quad LLM Prompt for Skills Classification \dotfill \pageref{appendix:skills_classification_prompt}\\
\quad \ref{appendix:architecture_classification_prompt} \quad LLM Prompt for Architecture Topic Classification \dotfill \pageref{appendix:architecture_classification_prompt}\\
\quad \ref{appendix:mcq_generation_prompt} \quad LLM Prompt for Synthetic MCQ Generation \dotfill \pageref{appendix:mcq_generation_prompt}\\
\quad \ref{appendix:mcq_filtering_prompt} \quad LLM Prompts for Filtering Synthetic MCQs \dotfill \pageref{appendix:mcq_filtering_prompt}\\
\quad \ref{appendix:text_extraction} \quad LLM Prompt for Text Extraction of Exam QAs \dotfill \pageref{appendix:text_extraction}\\
\quad \ref{appendix:image_extraction} \quad LLM Prompt for Image Extraction of Exam QAs \dotfill \pageref{appendix:image_extraction}\\
\quad \ref{appendix:exam_verification} \quad LLM Prompt for Verification of Extracted Exam QAs \dotfill \pageref{appendix:exam_verification}\\
\quad \ref{appendix:finetuning_case_study_prompt} \quad LLM Prompts for Case Study \#1 \dotfill \pageref{appendix:finetuning_case_study_prompt}\\
\quad \ref{appendix:memory_controller_case_study_prompt} \quad LLM Prompt for Case Study \#2 \dotfill \pageref{appendix:memory_controller_case_study_prompt}\\

\newpage

\appendix
\section{Ethics, Reproducibility, and LLM Usage}
\label{app:ethics_reproducibility_llmusage}
\subsection{Ethics Statement}
\label{app:ethics}
\ours{} was curated from sources that permit academic use and redistribution. Synthetic items were generated from a domain corpus compiled from public materials, exam-derived items were collected from publicly accessible university course pages or contributed by instructors, and crowdsourced items were submitted through our portal with explicit contributor consent. All expert validators who participated in question review and acceptance are co-authors of this paper. We did not recruit paid crowd workers; when individuals submitted questions via our portal, they consented to inclusion under our dataset license and to public attribution (or opted to remain anonymous). We do not collect personally identifying information beyond optional contact details for acknowledgment. No student data or private repositories were used. Where third-party figures or excerpts are included, we respect the original licenses and provide attribution. We will honor takedown requests for any inadvertently mislicensed content. This project did not involve human-subject experiments or interventions and, to the best of our understanding, does not require IRB oversight.

\subsection{Reproducibility Statement}
\label{app:reproducibility}
Section~\ref{sec:quarch_construction} describes the methodology for constructing \ours{} that can be used to reproduce a dataset of similar quality and characteristics. 
Appendix~\ref{appendix:skills_classification_prompt}, 
\ref{appendix:architecture_classification_prompt},
\ref{appendix:mcq_generation_prompt},
\ref{appendix:mcq_filtering_prompt},
\ref{appendix:text_extraction}, \ref{appendix:image_extraction}, and \ref{appendix:exam_verification} each expand on the details of the methodology overview provided in Section~\ref{sec:quarch_construction}.
Additionally, exact prompts used for evaluation results are documented in Appendix~\ref{appendix:mcq_prompt}, \ref{appendix:frq_prompt}, and \ref{appendix:laj_prompt} for reproducibility.

\subsection{LLM Usage}
\label{appendix:llm}
Language models were employed to refine the prose (e.g., grammar, clarity, and style) and to check formatting compliance with venue guidelines. Apart from their explicit roles described in the paper, namely for synthetic QA generation, exam parsing assistance, and evaluation (LLM-as-a-Judge), LLMs were not used to originate substantive scholarly content. All benchmark content admitted to the final release was verified by domain experts, and all prompts used in construction and evaluation are reported in Appendix~\ref{appendix:prompts}.

\section{Additional Details \& Results}
\label{appendix:additional_results}

\subsection{\ours{} Benchmark and Leaderboard Access}
\label{appendix:access_usage}

The \ours{} benchmark and leaderboard are available at: \textcolor{blue}{\url{https://quarch.ai/}}. Benchmark access is gated to prevent inadvertent inclusion into LLM pretraining corpora and to prevent unauthorized redistribution. Users must agree to conduct evaluations with LLM-as-a-Judge models which are either locally hosted or accessed via APIs which guarantee inputs are never used for training. Our evaluation harness stores ground-truth solutions as Fernet-encrypted~\citep{cryptographyFernet2026} files on disk and only permits in-memory decryption. The \ours{} leaderboard uses Amazon AWS Bedrock APIs for LLM-as-a-Judge evaluation to guarantee solutions are not used for model training. 

To further safeguard the benchmark’s integrity, we maintain a private, representative hold-out subset of QAs that will remain unreleased. Following the established protocol of seminal benchmarks such as SQuAD~\citep{squad}, researchers must submit their model for evaluation on this hidden test set to qualify for our leaderboard at \textcolor{blue}{\url{https://quarch.ai/}}. This serves as a validation step to detect and prevent reporting of inflated results due to data leakage.

\subsection{\ours{} Crowdsourcing Website}
\label{appendix:website}
\begin{figure}
    \centering
    \includegraphics[width=\textwidth]{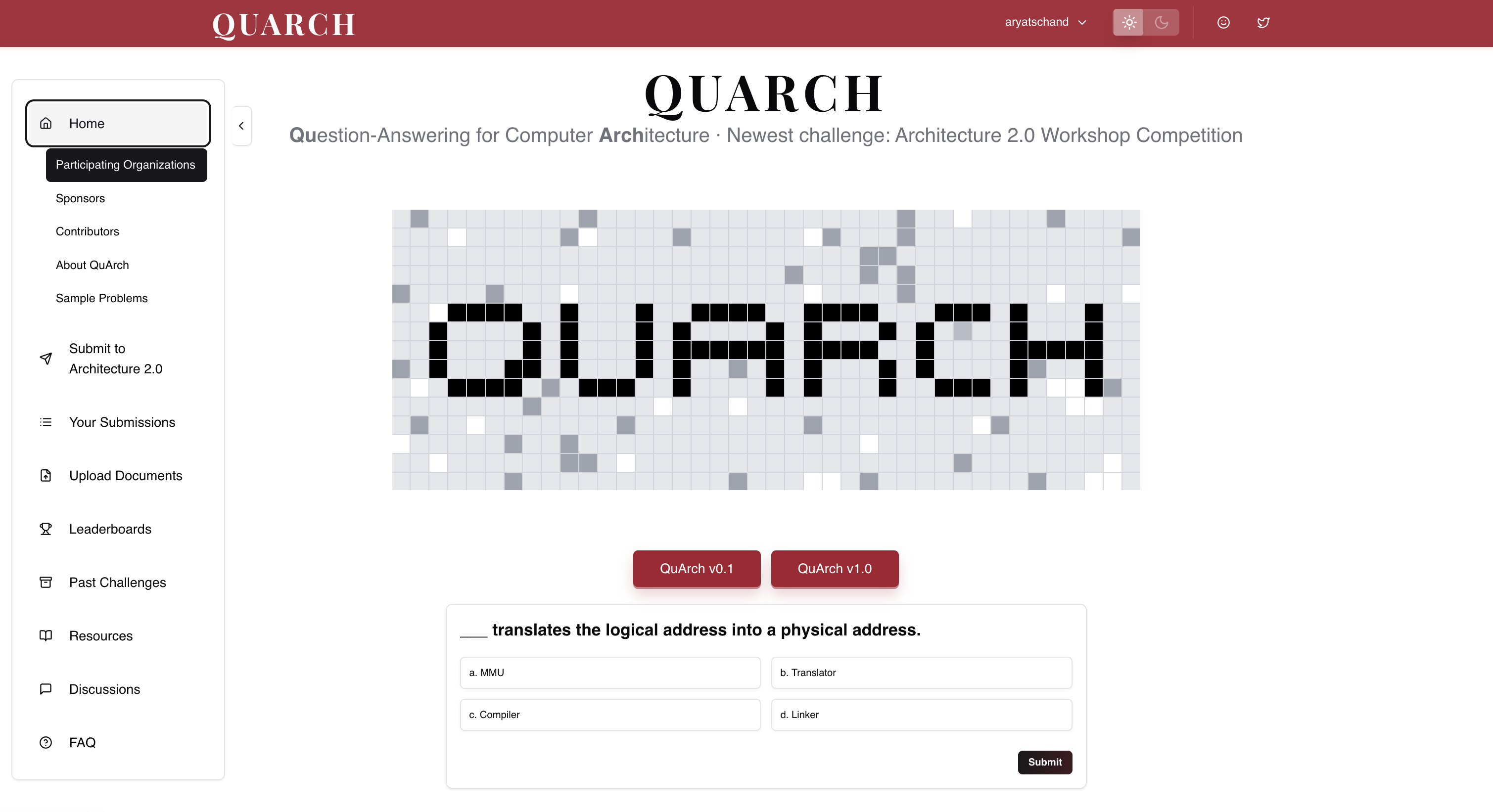}
    \caption{\ours{} Website: Home Page.}
    \label{fig:website_1}
\end{figure}
\begin{figure}
    \centering
    \includegraphics[width=\textwidth]{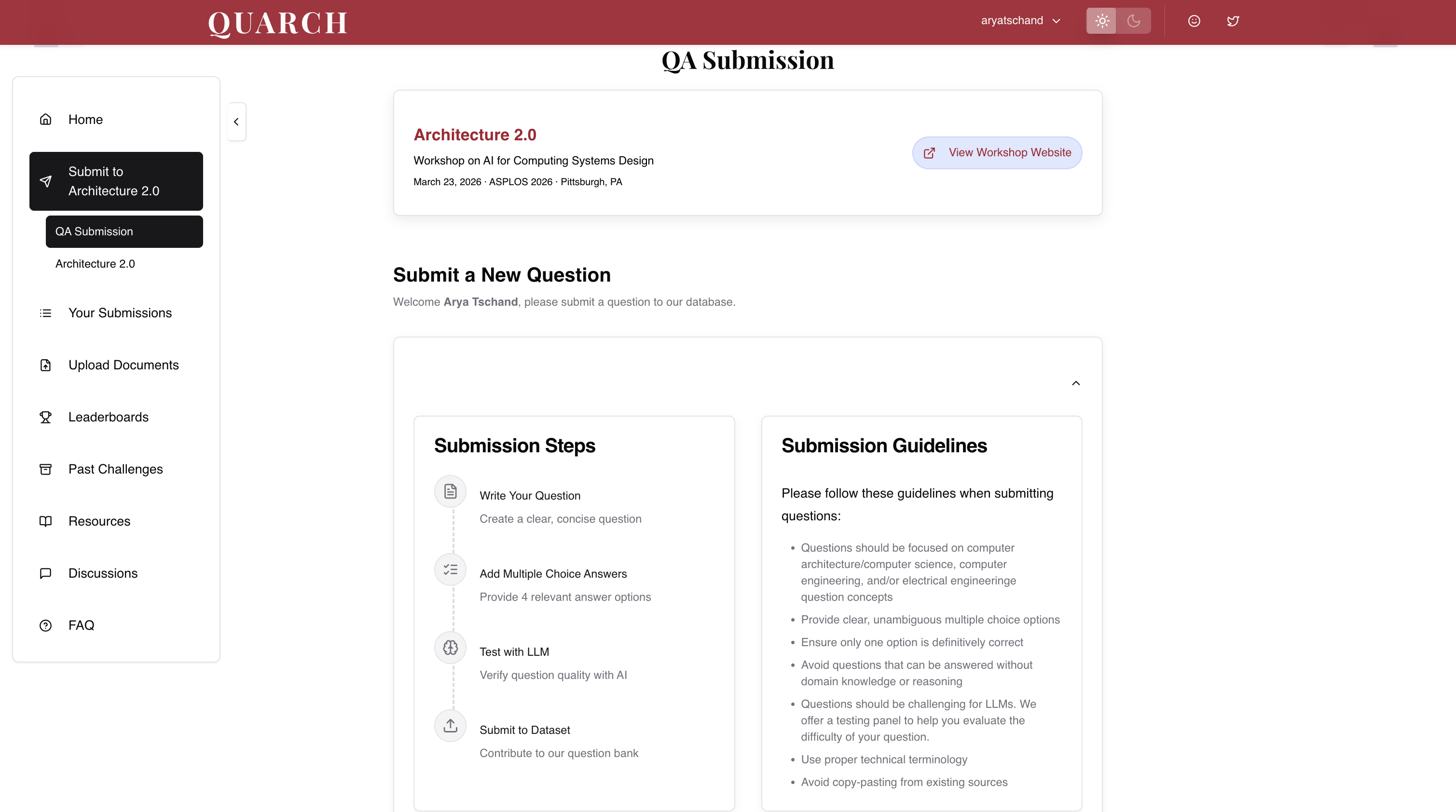}
    \caption{\ours{} Website: QA Submission Instructions.}
    \label{fig:website_2}
\end{figure}
\begin{figure}
    \centering
    \includegraphics[width=\textwidth]{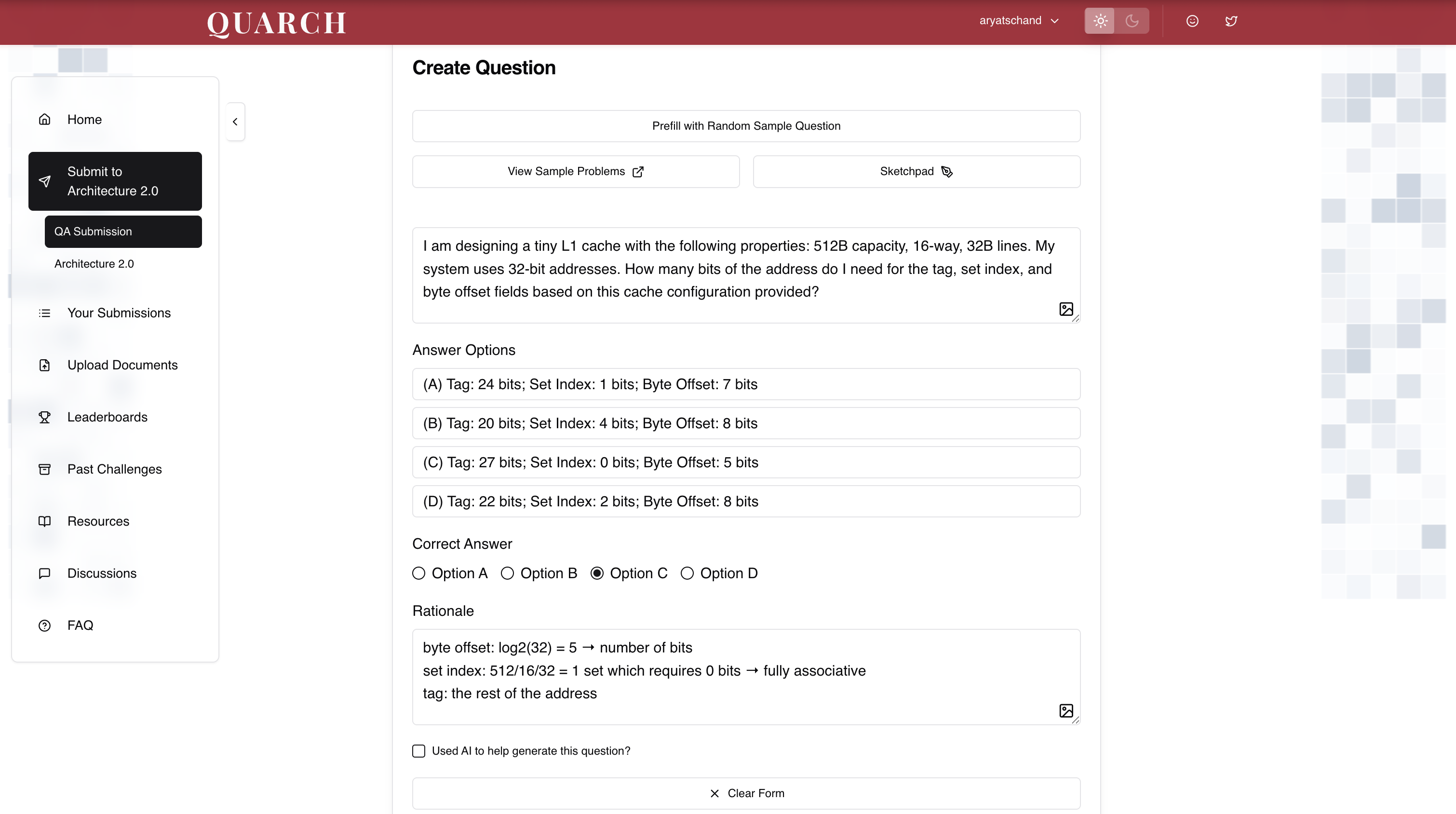}
    \caption{\ours{} Website: QA Submission Portal.}
    \label{fig:website_3}
\end{figure}
\begin{figure}
    \centering
    \includegraphics[width=\textwidth]{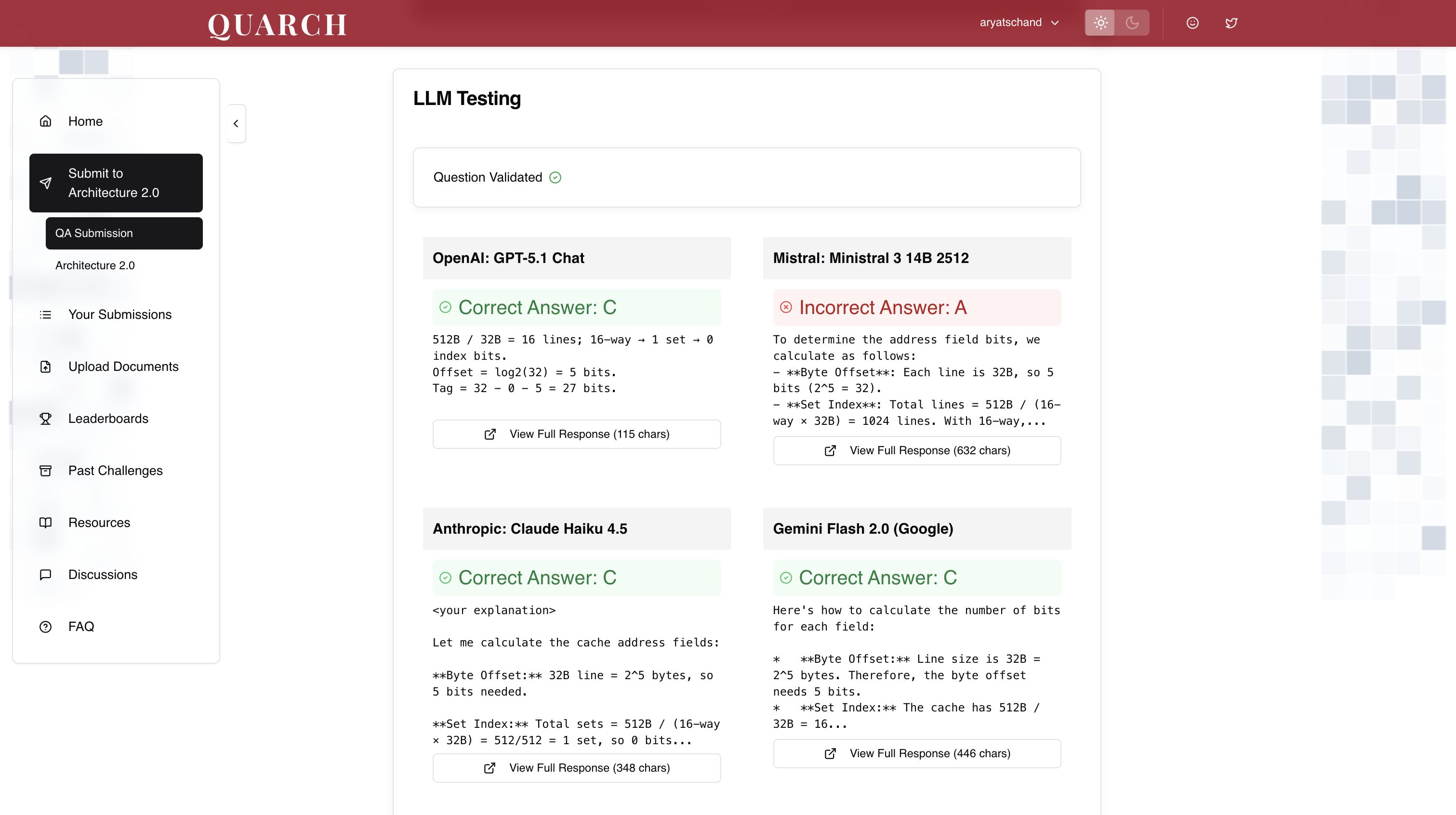}
    \caption{\ours{} Website: LLM Testing on QA Submission.}
    \label{fig:website_4}
\end{figure}
\begin{figure}
    \centering
    \includegraphics[width=\textwidth]{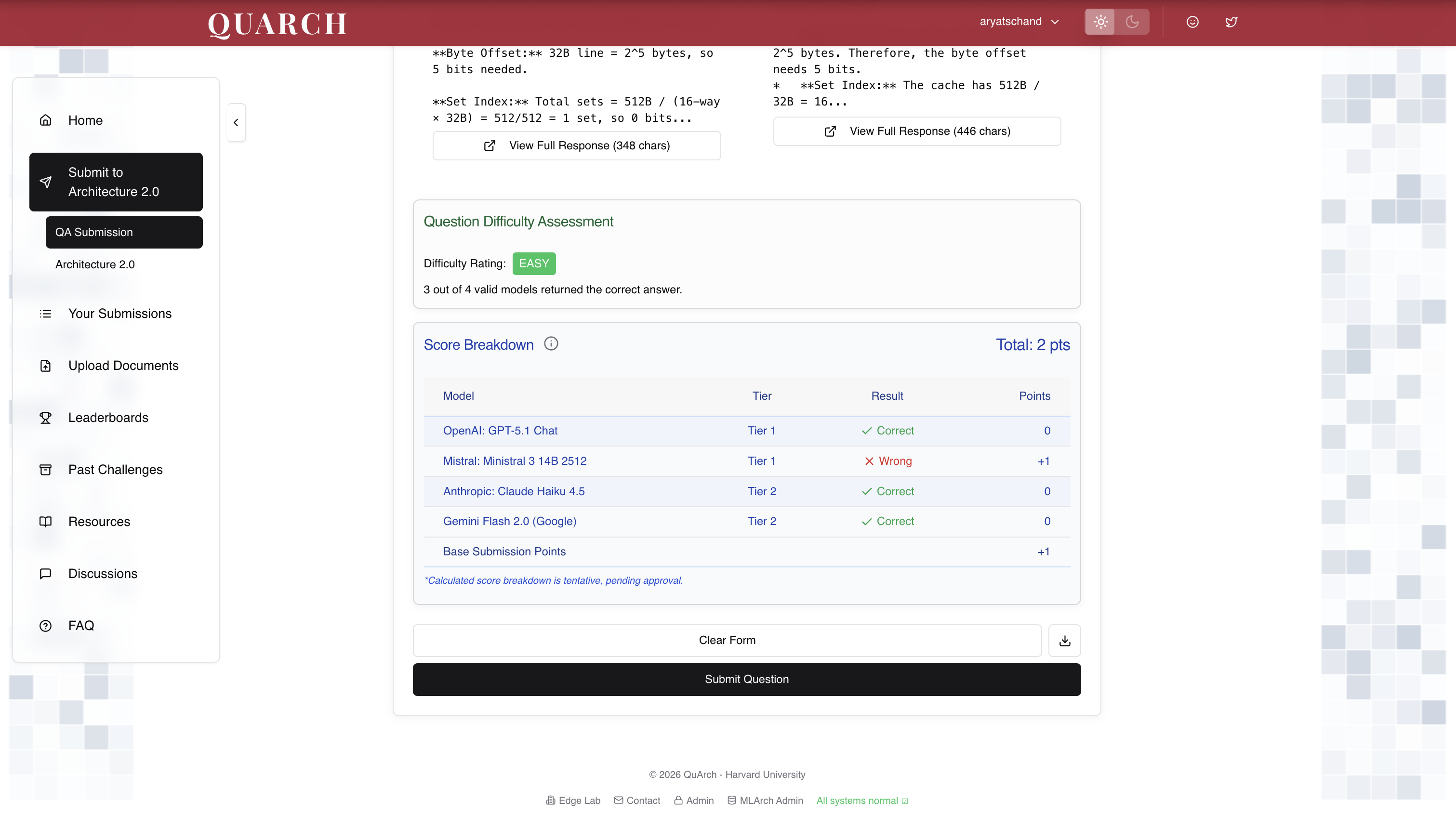}
    \caption{\ours{} Website: QA Submission Scoring.}
    \label{fig:website_5}
\end{figure}
\begin{figure}
    \centering
    \includegraphics[width=\textwidth]{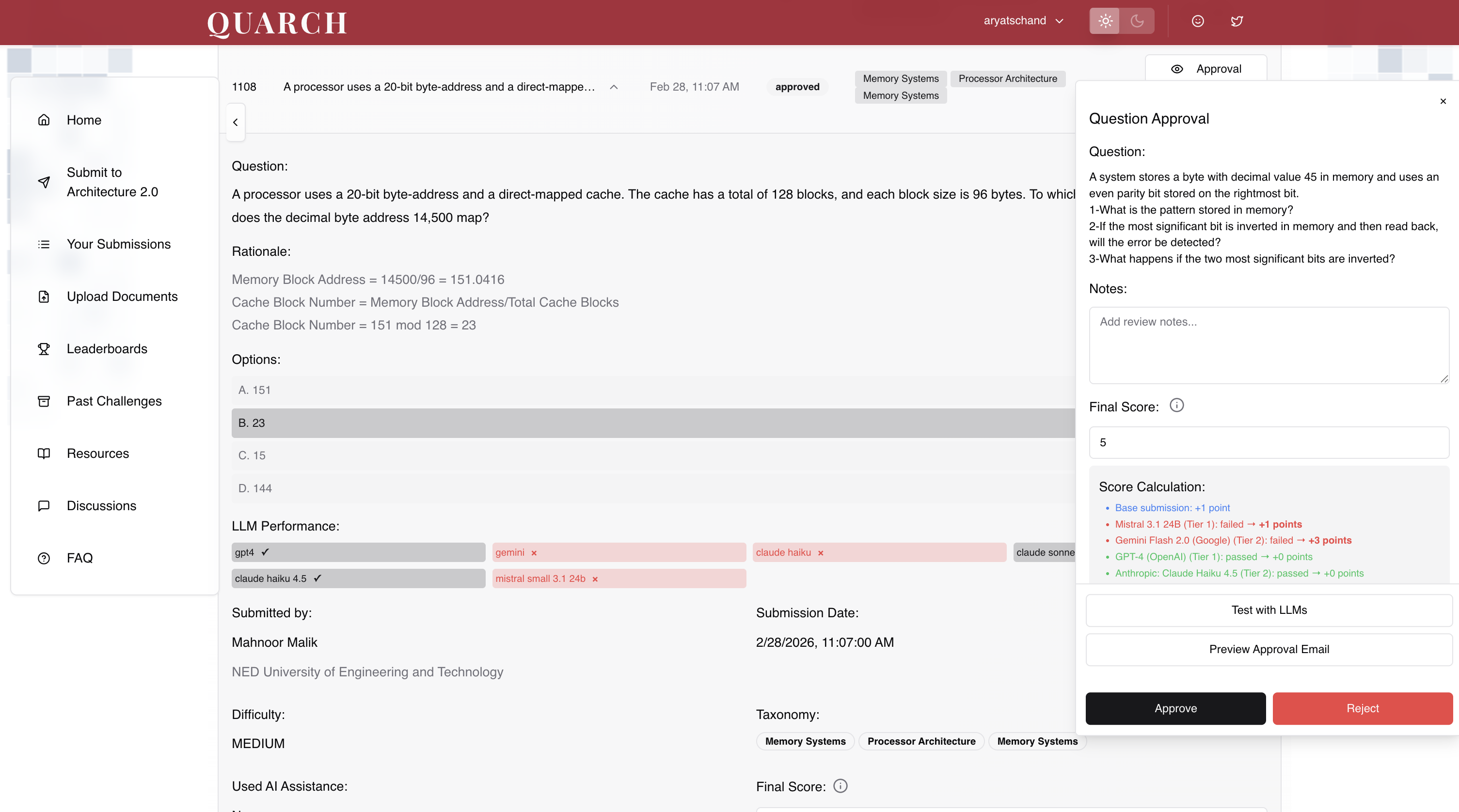}
    \caption{\ours{} Website: Administrator Approval Portal.}
    \label{fig:website_6}
\end{figure}
\begin{figure}
    \centering
    \includegraphics[width=\textwidth]{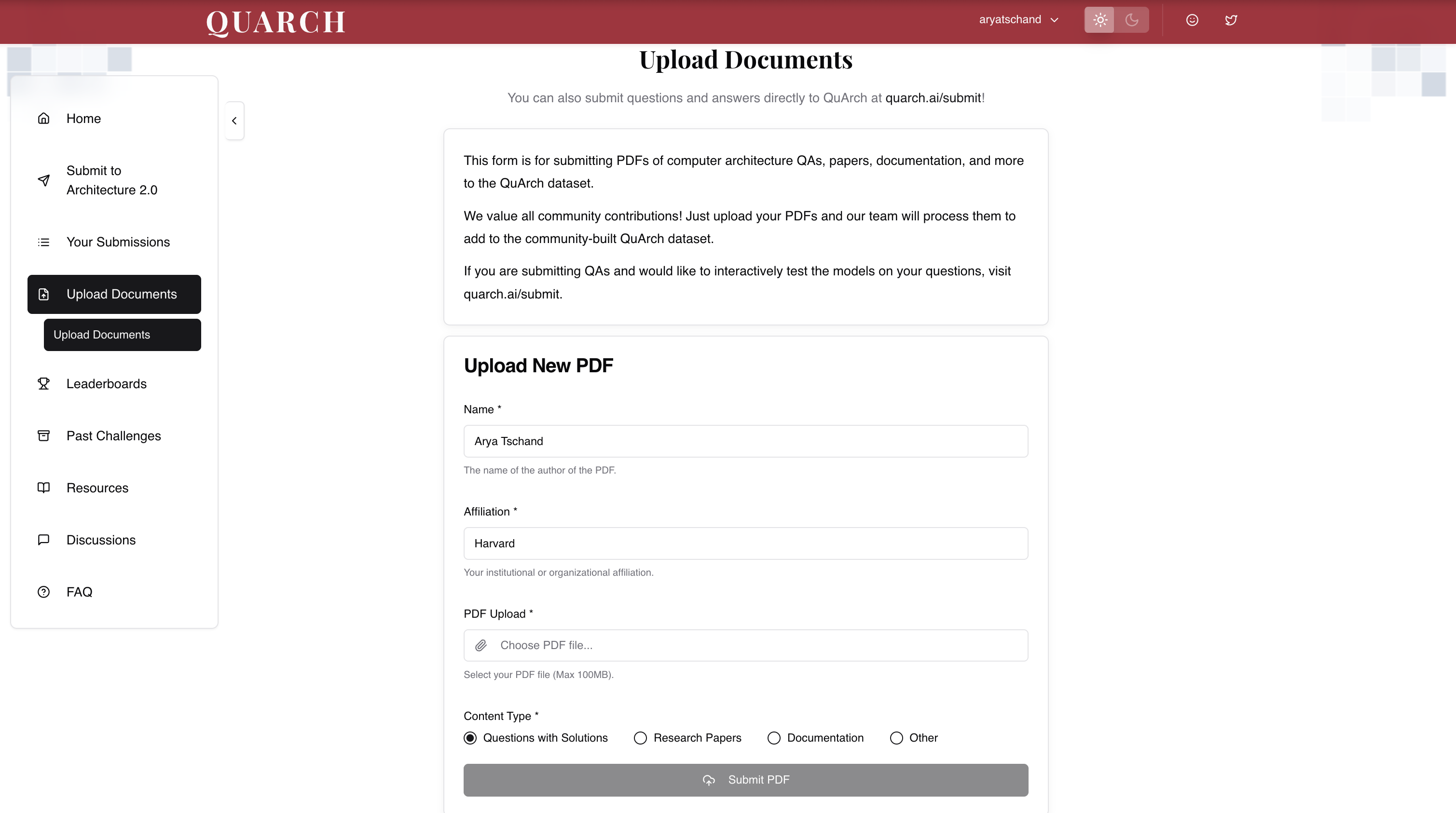}
    \caption{\ours{} Website: Exam Submission Portal.}
    \label{fig:website_7}
\end{figure}

To provide a centralized location to crowdsource questions and exams, we created the \ours{} website, shown in Figure~\ref{fig:website_1}. When a user wants to submit a question, they are presented with a set of instructions to guide accurate, relevant, and formatted questions, shown in Figure~\ref{fig:website_2}. Users submit the question, four answer options, the correct answer option, and a rationale for the correct answer, as shown in Figure~\ref{fig:website_3}. While the submitted questions are in multiple-choice format, users are instructed to not submit questions where the correct answer is ``All of the Above'' or ``None of the Above.'' This allows us to concatenate the correct answer with the rationale and format the question as a free-response questions as well. 

When a question is written, users can seamlessly test four non-frontier LLMs on correctness, shown in Figure~\ref{fig:website_4}. The question and potential answer choices are sent to the respective model using the MCQ Prompt shown in Appendix~\ref{appendix:mcq_prompt}. Users see whether the model gets the question correct and its response. By giving users early signal on whether LLMs fail, they can create difficult and correct questions that challenge a broad range of models. The LLM performance is used to assign a score for the question, where making more frontier models fail scores more points, shown in Figure~\ref{fig:website_5}. The website tracks submissions for each user and compiles a leaderboard with the most cumulative points and QA submissions.

When a question is submitted, it is queued for approval in our administrator portal, shown in Figure~\ref{fig:website_6}. The approval process involves checking the question for missing assumptions, poor relevance, or insufficient rationale. Administrators also ensure that the QA is objective and correct. Once a submission is approved, the user is notified.

In addition to individual QA submissions, users can submit exams using the website, shown in Figure~\ref{fig:website_7}. Questions, context images, and answers are then parsed out from these computer architecture course exams using the offline methodology described in Section~\ref{sec:quarch_construction}. Finally, the questions collected from the \ours{} crowdsourcing website are taxonomized by skill and funneled into the \ours{}-Reasoning benchmark when appropriate.

\subsection{Full Model Results by Skill}
\label{sec:full-model-results}

\begin{table*}[!tb]
\centering
\caption{Accuracy (\%) of all evaluated models across the four skills in \ours{}. Best performing models in each category highlighted \first{first}, \second{second}, and \third{third}.}
\label{tab:full_models_table}
\setlength{\tabcolsep}{6pt}
\renewcommand{\arraystretch}{1.1}
    \begin{tabular}{l
                    c
                    c
                    c
                    c
                    c
                    }
        \toprule
        \textbf{Model} & & \multicolumn{4}{c}{\ours{}-\textsc{Reasoning}} \\
        \cline{3-6}
        & \textbf{Recall} & \textbf{Analyze} & \textbf{Design} & \textbf{Implement} & \textbf{Overall} \\
        \midrule
        \textit{Closed-Source Multimodal Models} \\
        
        \modelname{GPT-5.2} & 88.4 & \first{73.0} & \second{79.4} & \first{72.4} & \first{73.2} \\
        \modelname{GPT-5} & \third{89.1} & \third{71.6} & \first{84.1} & \second{71.1} & \second{72.0} \\
        \modelname{GPT-5} (Non-Reasoning) & 86.4 & 48.9 & 57.1 & 39.5 & 48.4 \\
        \modelname{GPT-4o} & 83.7 & 28.1 & 21.4 & 22.8 & 27.4 \\
        \modelname{Gemini-3-Pro} & \first{90.9} & 71.4 & \third{73.8} & 65.6 & 71.0 \\
        \modelname{Gemini-3-Flash} & \second{90.1} & \second{72.8} & 66.7 & \third{66.3} & \third{71.9} \\
        \modelname{Gemini-2.5-Pro} & 87.2 & 63.1 & 56.3 & 63.6 & 62.9 \\
        \modelname{Gemini-2.5-Flash} & 83.3 & 57.0 & 58.7 & 51.4 & 56.5 \\
        \modelname{Claude-Sonnet-4} & 85.3 & 48.8 & 50.0 & 41.2 & 48.1 \\
        \modelname{Claude-3.7-Sonnet-Thinking} & 85.7 & 52.8 & 44.4 & 45.2 & 51.8 \\
        \modelname{Mistral-Medium-3.1} & 84.2 & 34.6 & 38.1 & 24.5 & 33.8 \\
        \midrule
        \textit{Open-Source Multimodal Models} \\
        \modelname{Gemma-3-27B-IT} & 75.3 & 21.8 & \second{27.0} & 15.0 & 21.4 \\
        \modelname{Gemma-3-4B-IT} & 61.6 & 8.2 & 4.0 & 3.1 & 7.5 \\
        \modelname{Qwen3-VL-32B-NonThinking-Instruct} & \second{83.7} & \first{42.3} & \first{36.5} & \first{38.8} & \first{41.8} \\
        \modelname{Llama-4-Maverick} & \first{85.2} & \second{34.6} & \third{26.2} & \second{28.0} & \second{33.7} \\
        \modelname{Llama-3.2-11B} & 68.4 & 9.5 & 6.3 & 4.7 & 9.0 \\
        \modelname{Mistral-Small-3.2-24B-Instruct} & \third{77.7} & \third{23.8} & 22.2 & \third{18.7} & \third{23.3} \\
        \midrule
        \textit{Text-Only Models} \\
        \modelname{GPT-OSS-120B} & 84.5 & \first{63.9} & \first{66.7} & \first{58.1} & \first{63.6} \\
        \modelname{DeepSeek-R1} & \first{86.9} & \third{56.6} & 47.8 & 44.4 & \third{55.4} \\
        \modelname{Llama-3.3-70B} & 79.7 & 27.0 & 4.3 & 12.8 & 25.1 \\
        \modelname{Llama-3.2-1B} & 36.1 & 2.5 & 0.0 & 0.0 & 2.2 \\
        \modelname{Mistral-Codestral-2508} & 74.6 & 30.3 & 14.5 & 23.9 & 29.2 \\
        \modelname{Mistral-Devstral-Medium} & 81.6 & 28.5 & 10.1 & 24.8 & 27.5 \\
        \modelname{Kimi-K2-0905} & 84.1 & 42.8 & 49.3 & 37.6 & 42.7 \\
        \modelname{Qwen3-Coder-480B-A35B-Instruct} & 82.6 & 41.6 & 34.8 & 25.6 & 40.2 \\
        \modelname{Qwen3-235B-A22B-Thinking} & \third{85.4} & \second{62.6} & \second{60.9} & \third{45.3} & \second{61.3} \\
        \modelname{Qwen3-235B-A22B-NonThinking-Instruct} & \second{86.3} & 56.1 & 49.3 & \third{45.3} & 55.1 \\
        \modelname{Qwen3-Next-80B-A3B-Thinking} & 84.4 & 54.5 & \third{50.7} & 39.7 & 53.3 \\
        \modelname{Qwen3-30B-A3B-Thinking} & 82.3 & 50.3 & 42.0 & 32.1 & 48.6 \\
        \modelname{Qwen3-Coder-30B-A3B-Instruct} & 77.6 & 29.5 & 15.9 & 12.0 & 27.7 \\
        \bottomrule

    \end{tabular}
\end{table*}

Table~\ref{tab:full_models_table} provides the complete set of results for all evaluated models across the four skill categories in \ours{}. The table illustrates how models perform differently on factual recall compared to higher-order reasoning, design, and implementation. This comprehensive view allows for comparison across both small and large language models, highlighting overall trends and providing a foundation for tracking progress over time. The table captures the broader landscape of model performance and makes clear the varying degrees of capability across skills that are critical for reasoning in computer architecture.  

\textbf{Performance of Small Language Models}

Small language models (SLMs) keep pace with LLMs on recall-style questions, but their performance drops on \ours{}-Reasoning QAs, especially when multimodal reasoning is required. This gap suggests that parameter scale (and capacity for long-horizon reasoning and state tracking) matters more for higher-order architectural problem solving (analyze, design, implement) than for factual retrieval. In practice, SLMs are well-suited for low-latency, cost-efficient assistants that handle definitions, quick checks, and targeted lookups, while agentic systems design, trade-off analysis, and figure/table interpretation still benefit from larger models or strong tool scaffolding. A pragmatic path is a cascaded workflow: route recall to SLMs, escalate complex reasoning to LLMs, and bolster SLMs with retrieval and simulators rather than relying on scale alone.

\subsection{Topic-Wise Extended Characterization}
\label{sec:topic-spiders}

\begin{figure*}
    \centering
    \includegraphics[width=\textwidth]{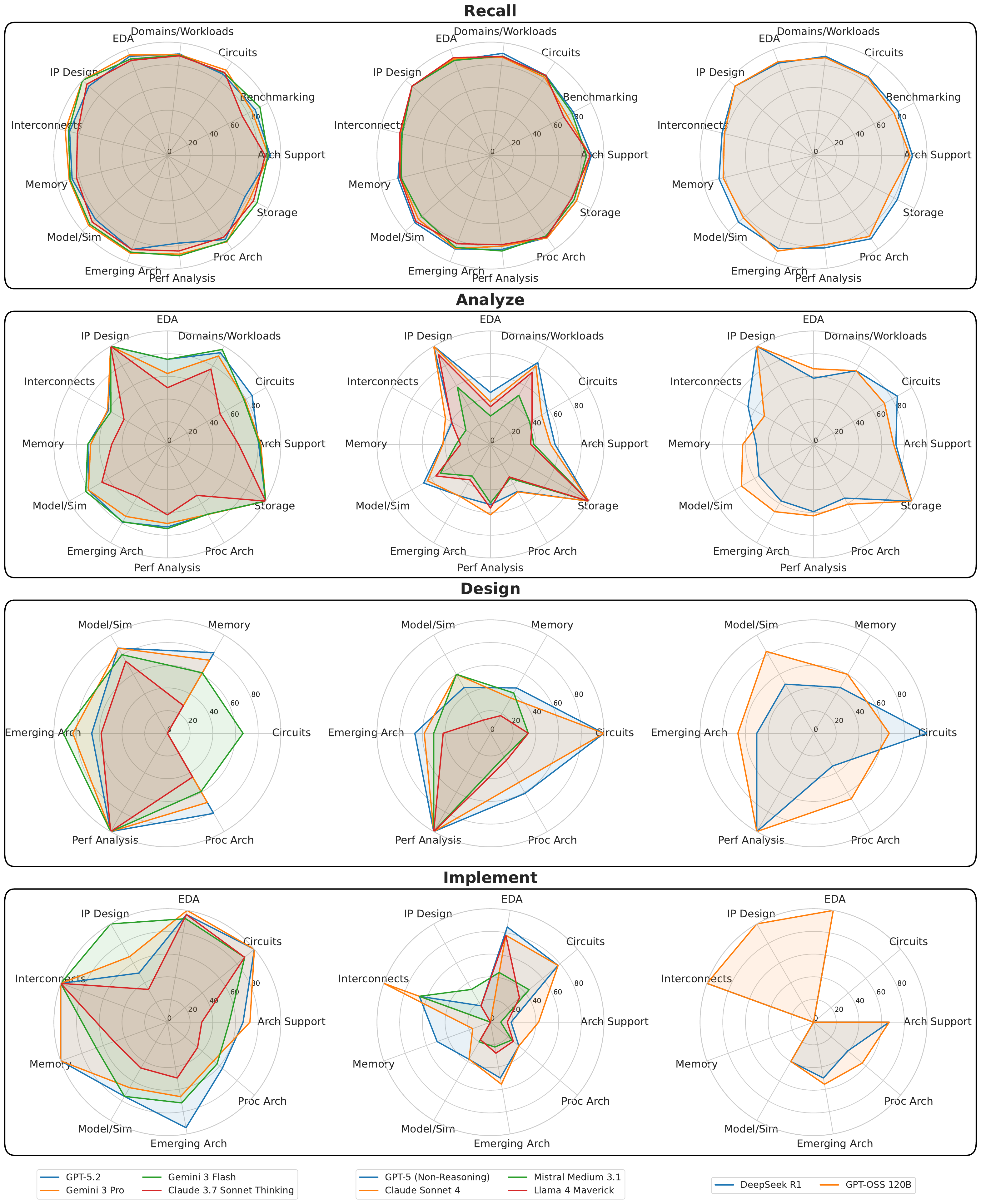}

    \caption{Topic-wise characterization of frontier models, categorized by skill. In order, the groupings of 3 radar plots correspond to ``Recall'', ``Analyze'', ``Design'', and ``Implement'' questions. Within each grouping, leftmost plot contains the best performing multimodal models, the middle plot contains the worst performing multimodal models, and the rightmost plot contains the two text-only models.}
    \label{fig:big-spiders}
\end{figure*}

Figure~\ref{fig:big-spiders} visualizes the topic-wise performance of the frontier models on \ours{}. As performance across all models on \ours{}-Recall is very high, topic-wise performance variability is not noticeable. However, on the higher-order skills of \ours{}-Reasoning (Analyze, Design, and Implement), models exhibit surprising heterogeneity in per-topic performance.

\subsection{Model Performance by Modality}
\label{sec:non-sota-models}

\begin{table*}[!tb]
\caption{Per-generation accuracy (\%) by evaluation type and modality. Best performing models in each category highlighted \first{first}, \second{second}, and \third{third}.}
\label{tab:non-sota-results}
\centering
\renewcommand{\arraystretch}{1.1}

\begin{tabular}{lcccc}
\toprule
Model & \multicolumn{2}{c}{Text-only} & Multimodal & All \\
\cline{2-3}
    & MCQ & FRQ & FRQ & FRQ \\
\midrule
\textit{Closed-Source Multimodal Models} \\

\modelname{GPT-5.2} & 88.5 & \second{73.6} & \first{73.5} & \first{73.5} \\
\modelname{GPT-5} & \third{89.0} & \first{74.7} & 70.5 & \second{72.7} \\
\modelname{GPT-5} (Non-Reasoning) & 86.4 & 53.8 & 44.8 & 49.4 \\
\modelname{GPT-4o} & 84.3 & 31.3 & 24.6 & 28.1 \\
\modelname{Gemini-3-Pro} & \first{90.9} & 71.2 & \third{71.7} & 71.4 \\
\modelname{Gemini-3-Flash} & \second{90.2} & \third{71.9} & \second{72.7} & \third{72.3} \\
\modelname{Gemini-2.5-Pro} & 87.5 & 63.9 & 62.1 & 63.0 \\
\modelname{Gemini-2.5-Flash} & 83.2 & 59.3 & 55.4 & 57.4 \\
\modelname{Claude-Sonnet-4} & 85.6 & 52.1 & 45.4 & 48.8 \\
\modelname{Claude-3.7-Sonnet-Thinking} & 85.9 & 53.8 & 51.3 & 52.6 \\
\modelname{Mistral-Medium-3.1} & 84.7 & 41.2 & 27.7 & 34.6 \\
\midrule
\textit{Open-Source Multimodal Models} \\
\modelname{Gemma-3-27B-IT} & 75.7 & 26.8 & \third{17.7} & 22.4 \\
\modelname{Gemma-3-4B-IT} & 63.2 & 9.4 & 4.1 & 6.9 \\
\modelname{Qwen3-VL-32B-NonThinking-Instruct} & \second{84.2} & \first{44.0} & \first{38.5} & \first{41.3} \\
\modelname{Llama-4-Maverick} & \first{85.9} & \second{36.1} & \second{32.1} & \second{34.2} \\
\modelname{Llama-3.2-11B} & 70.2 & 9.5 & 5.6 & 8.0 \\
\modelname{Mistral-Small-3.2-24B-Instruct} & \third{78.6} & \third{29.1} & \third{17.7} & \third{23.5} \\
\midrule
\textit{Text-Only Models} \\
\modelname{GPT-OSS-120B} & 84.1 & \first{65.3} & - & - \\
\modelname{DeepSeek-R1} & \first{87.1} & \third{56.2} & - & - \\
\modelname{Llama-3.3-70B} & 81.0 & 23.6 & - & - \\
\modelname{Llama-3.2-1B} & 37.1 & 0.8 & - & - \\
\modelname{Mistral-Codestral-2508} & 75.7 & 28.0 & - & - \\
\modelname{Mistral-Devstral-Medium} & 82.5 & 27.3 & - & - \\
\modelname{Kimi-K2-0905} & 84.2 & 44.2 & - & - \\
\modelname{Qwen3-Coder-480B-A35B-Instruct} & 83.3 & 40.1 & - & - \\
\modelname{Qwen3-235B-A22B-Thinking} & \third{85.6} & \second{61.7} & - & - \\
\modelname{Qwen3-235B-A22B-NonThinking-Instruct} & \second{86.6} & 55.7 & - & - \\
\modelname{Qwen3-Next-80B-A3B-Thinking} & 84.5 & 54.1 & - & - \\
\modelname{Qwen3-30B-A3B-Thinking} & 82.7 & 48.8 & - & - \\
\modelname{Qwen3-Coder-30B-A3B-Instruct} & 78.6 & 26.9 & - & - \\
\bottomrule
\end{tabular}
\end{table*}

Table~\ref{tab:non-sota-results} performs detailed comparisons between text-only QA performance, image-only performance, and image-text performance across all evaluated models. See Section~\ref{sec:key_insights} for interpretation and analysis of sensitivity to input modalities. We observe consistent trends across model scales, including significant gaps in higher-order reasoning and multimodal tasks between SLMs and LLMs. By including this wider range of models, we provide a more complete picture of the landscape and enable future work to track progress not only at the frontier but also in more lightweight, cost-efficient models.

\subsection{Case Study \#1: Additional Details and Results}
\label{app:finetuning}
\textbf{Task Interface and Evaluation Protocol.}
We first provide further details on the task interface and evaluation protocol used in the case study described in Section~\ref{sec:finetuning-case-study}. 
Each trial consists of an iterative interaction between the language model and simulator, in which the model proposes a memory hierarchy design for a workload. 
Designs are specified via a structured JSON configuration that encodes the hardware and software decisions listed in Section~\ref{sec:finetuning-case-study}.
Each proposed configuration is evaluated using the Accelergy tool~\citep{accelergy} to estimate chip area and energy consumption.
After each proposal, the model receives this quantitative feedback on the resulting area and energy, which it may use to refine subsequent designs. 
Each trial consists of 10 turns to allow the model to iterate on its proposals.
A trial is considered successful if the model proposes at least one design within the 10 iterations that consumes at most one millijoule of energy and at most 15{,}000\,$\mu$m$^2$ of chip area; the model is instructed to optimize as much as possible for area while remaining within the energy budget. 
Importantly, the baseline design provided to the model at the start of each trial does not meet these specifications, requiring models to actively reason about architectural trade-offs in order to meet the design constraints.

\textbf{Fine-Tuning Details and Additional Results.}
To construct the fine-tuning dataset, we filtered FRQs from \ours{} related to memory design and matrix multiplication workloads with \modelname{Claude-3.7-Sonnet-Thinking} using the filtering prompt listed in~\ref{appendix:finetuning_case_study_prompt}. This resulted in 83 questions, 47 of which are text-only. Out of the 47 text-only questions, 45 had at least one correctly-judged response among the benchmarked LLMs in Table~\ref{tab:full_models_table} and we selected 676 of these responses for our SFT dataset. In particular, 135 of the 676 samples included \modelname{GPT-5.2} generated explanations of how to arrive at the correct answer, using the three didactic prompts in Section~\ref{appendix:finetuning_case_study_prompt}. Our approach to distillation was inspired by~\citet{lampinen2022can, muennighoff2025s1, zhang2024distillation}. Fine-tuning on all unfiltered FRQs did not provide a similar lift in case study task performance. We hypothesize that for smaller language models, SFT datasets specialized to a single architectural task are simpler to distill from, compared to a diverse mix of closely and distantly related architecture topics, without performing a correspondingly large hyperparameter search.

Table~\ref{tab:finetuning_hparams} lists the two hyperparameter configurations we used for finetuning \modelname{Gemma-3-27B-it} and \modelname{Llama-3.3-70B-Instruct}. We used the same 676 SFT samples for both models, and finetuned both models under each set of hyperparameters using Together AI's fine-tuning API. 

The first configuration trains each model for 3 epochs, and significantly increases the success rate on the case study task as reported in Table~\ref{tab:appendix-cache-case-study-results}. With 4 epochs and a higher learning rate, the second configuration is a more aggressive hyperparameter regime and results in models that are able to discover solutions with much smaller chip area, but at the cost of not significantly improving the success rate (Table~\ref{tab:appendix-cache-case-study-results}). 
The success rate reflects how consistently a model is able to navigate the architectural design space to satisfy strict design constraints, while the best-area metric captures the most area-efficient design discovered across all trials. 
Together, these results illustrate how fine-tuning on \ours{} questions improves both the reliability and quality of architecture design proposals.

\begin{table}[t]
\centering
\begin{tabular}{lcc}
\toprule
\textbf{Hyperparameters} & \textbf{Config 1} & \textbf{Config 2} \\
\midrule
SFT Epochs            & 3      & 4      \\
Learning Rate              & $1\times10^{-6}$ & $2\times10^{-6}$ \\
LR Scheduler               & Cosine & Cosine \\
Warmup Ratio               & 0.30   & 0.03   \\
Min.\ LR Ratio             & 0.00     & 0.40   \\
Weight Decay               & 0.00   & 0.01   \\
\bottomrule
\end{tabular}
\vspace{0.5em}
\caption{Hyperparameters used for fine-tuning experiments in Case Study \#1.}
\label{tab:finetuning_hparams}
\end{table}

\begin{table}[t]
\centering
\small
\begin{tabular}{lcc}
\toprule
\textbf{Model} & \textbf{Success Rate} & \textbf{Best Area Across Trials ($\mu m^2$)} \\
\midrule
\multicolumn{3}{l}{\textit{\modelname{Llama-3.3-70B-Instruct}}} \\
Base Model & 55\% (11/20) & 7246.78 \\
Fine-tuned Model (Config 1)   & \textbf{95\% (19/20)} & 7275.78 \\
Fine-tuned Model (Config 2)   & 35\% (7/20)  & \textbf{3637.89} \\
\midrule
\multicolumn{3}{l}{\textit{\modelname{Gemma-3-27B-it}}} \\
Base Model        & 30\% (6/20)  & 6728.49 \\
Fine-tuned Model (Config 1) & \textbf{70\% (14/20)} & 6695.84 \\
Fine-tuned Model (Config 2)   & 50\% (10/20) & \textbf{3608.89} \\
\bottomrule
\end{tabular}
\vspace{0.5em}
\caption{Results from the Case Study \#1 (Section~\ref{sec:finetuning-case-study}), grouped by model family. We report the fraction of trials in which each model proposes designs that satisfy both area and energy design constraints (i.e., success rate), along with the best (minimum) chip area discovered across all runs.}
\label{tab:appendix-cache-case-study-results}
\end{table}

\subsection{Case Study \#2: Designing a Memory Controller}
\label{sec:memory-controller-case-study}
\begin{figure}%
    \centering
    \includegraphics[width=0.6\linewidth]{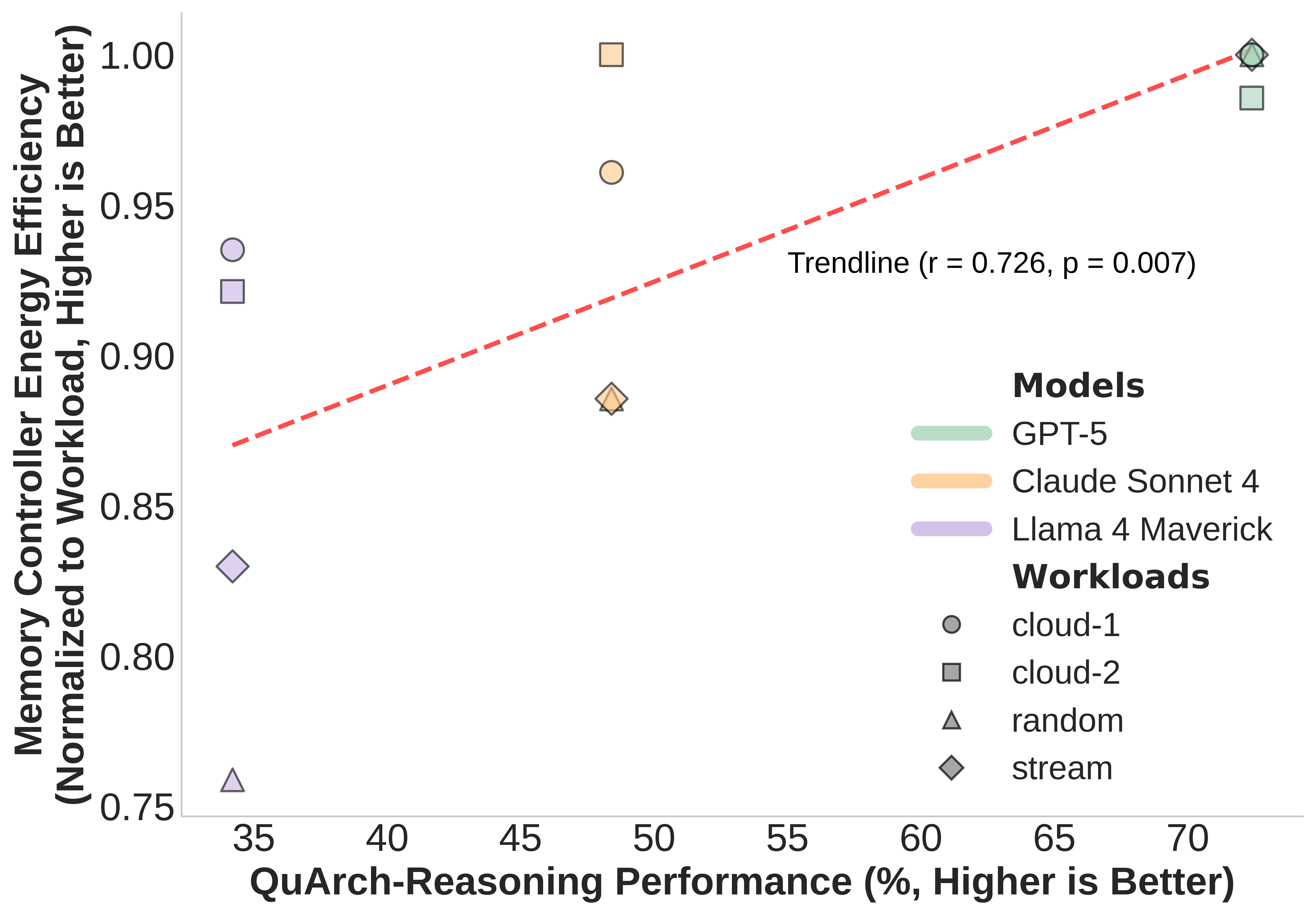}
    \caption{
    Correlation between \ours{}-\textsc{Reasoning} and DRAM controller design quality. Each point is a model-workload pair, with the x-axis showing \ours{}-\textsc{Reasoning} accuracy and the y-axis showing normalized memory-controller energy efficiency (power$^{-1}$).
    }
    \label{fig:quarch_dram_correlation}
\vspace{-0.2in}
\end{figure}
Case Study \#1 (Section~\ref{sec:finetuning-case-study}, Appendix~\ref{app:finetuning}) demonstrated knowledge transfer between \ours{} QAs and a downstream cache design task.
To futher examine how \ours{}-\textsc{Reasoning} performance translates to real architectural decision-making, we conduct an additional case study in this section on Dynamic Random Access Memory (DRAM) controller design, a canonical target of architecture design space exploration tasks~\citep{archgym}.
In this task, the model must propose DRAM controller configurations (e.g. scheduling policies, refresh behavior, etc.) that would lead to energy efficient designs.
The prompt (Appendix~\ref{appendix:memory_controller_case_study_prompt}) lists all design parameters and explicitly asks the model to analyze the workload's memory trace and reason about how the controller design choices affect power consumption before proposing a concrete configuration.
Critically, (1) effectively answering these questions requires architectural reasoning and influences optimal design choices, and (2) similar QAs about memory trace analysis and memory controller configurations are assessed in \ours{} (Appendix~\ref{sec:case_study_questions}).
Designs are evaluated using a DRAM simulator~\citep{jung2015dramsys}.
We run 10 independent rounds per model across 4 different workloads under a fixed evaluation budget (40 proposed designs per round). We evaluate three models with varying \ours{}-\textsc{Reasoning} performance: \modelname{GPT-5}, \modelname{Claude-Sonnet-4}, and \modelname{Llama-4-Maverick}.
Figure~\ref{fig:quarch_dram_correlation} reports the mean energy efficiency (power$^{-1}$, normalized within each workload) of each model's best controller designs across the 10 runs versus their \ours{}-\textsc{Reasoning} performance.
We observe a strong and statistically significant positive correlation between \ours{}-\textsc{Reasoning} accuracy and normalized energy efficiency (\emph{r} = 0.726, \emph{p} = 0.007), indicating that models which reason better on \ours{} also discover more energy-efficient configurations. This correlation is important, as general QA benchmarks (e.g. MMLU) do not always correlate with downstream task-focused agentic benchmarks (e.g. SWE-Bench)~\citep{qinAPTBenchBenchmarkingAgentic2025}.
This case study finds that improvement on \ours{} directly reflects improvement on computer architecture tasks.

\subsection{Partially Correct Judgments}
\label{sec:partially-correct}

\begin{table*}[t]
\centering
\caption{Addition of ``Partially Correct" Judgments in LLM-as-a-judge rubric. Results are pass@1 (each model was given one attempt at each QA) and using a single LLM-as-a-judge response rather than from consensus.}
\label{tab:partially-correct-results}

    \begin{tabular}{l c c c}
    \toprule
    \textbf{Model} & \textbf{Correct (\%)} & \textbf{Partially Correct (\%)} & \textbf{Incorrect (\%)} \\
    \midrule
    \textit{Closed-Source Multimodal Models} \\
    \modelname{GPT-5.2} & 74.1 & 17.6 & 8.3 \\
    \modelname{GPT-5} & 70.4 & 19.2 & 10.2 \\
    \modelname{GPT-5} (Non-Reasoning) & 48.5 & 31.6 & 20.0 \\
    \modelname{GPT-4o} & 28.5 & 40.2 & 31.3 \\
    \modelname{Gemini-2.5-Pro} & 61.9 & 24.1 & 13.8 \\
    \modelname{Gemini-2.5-Flash} & 56.7 & 26.7 & 16.5 \\
    \modelname{Claude-Sonnet-4} & 49.6 & 31.8 & 18.6 \\
    \modelname{Claude-3.7-Sonnet-Thinking} & 52.1 & 30.7 & 16.9 \\
    \modelname{Mistral-Medium-3.1} & 33.5 & 34.5 & 31.8 \\
    \midrule
    \textit{Open-Source Multimodal Models} \\
    \modelname{Gemma-3-27B-Instruct} & 24.0 & 35.2 & 40.9 \\
    \modelname{Gemma-3-4B-Instruct} & 6.6 & 30.9 & 62.5 \\
    \modelname{Qwen-VL-32B-NonThinking-Instruct} & 43.4 & 27.0 & 29.5 \\
    \modelname{Llama-4-Maverick} & 33.7 & 37.9 & 28.5 \\
    \modelname{Llama-3.2-11B} & 8.6 & 27.9 & 63.5 \\
    \modelname{Mistral-Small-3.2-24B-Instruct} & 24.5 & 36.4 & 39.0 \\
    \midrule
    \textit{Text-Only Models} \\
    \modelname{GPT-OSS-120B} & 66.7 & 17.5 & 15.8 \\
    \modelname{DeepSeek-R1} & 55.9 & 25.2 & 18.8 \\
    \modelname{Llama-3.3-70B} & 24.0 & 39.5 & 36.2 \\
    \modelname{Llama-3.2-1B} & 0.3 & 13.4 & 86.3 \\
    \modelname{Mistral-Codestral-2508} & 27.9 & 37.7 & 33.9 \\
    \modelname{Mistral-Devstral-Medium} & 27.0 & 37.6 & 35.2 \\
    \modelname{Kimi-K2-0905} & 45.4 & 25.8 & 28.8 \\
    \modelname{Qwen-3-Coder-480B-Instruct} & 40.7 & 34.8 & 24.5 \\
    \modelname{Qwen-3-235B-A22B-Thinking} & 57.7 & 16.6 & 25.6 \\
    \modelname{Qwen-3-235B-A22B-NonThinking-Instruct} & 54.4 & 26.5 & 18.5 \\
    \modelname{Qwen-3-Next-80B-Thinking} & 56.1 & 20.6 & 23.2 \\
    \modelname{Qwen-3-30B-A3B-Thinking} & 48.0 & 21.7 & 30.1 \\
    \modelname{Qwen-3-Coder-30B-A3B-Instruct} & 28.9 & 34.7 & 36.2 \\
    \bottomrule
    \end{tabular}
\end{table*}

Table~\ref{tab:partially-correct-results} reports results when we extend the LLM-as-a-judge rubric to include a “Partially Correct” category. We observe that many models, particularly weaker or smaller ones, produce answers that are not fully correct but demonstrate partial understanding (e.g., identifying the right concept while failing to complete all reasoning steps). Incorporating this intermediate category reveals a richer distribution of model behavior: some models that appear very weak under a strict correct/incorrect rubric (e.g., sub-30\% accuracy) show substantially higher rates of partially correct answers, suggesting they are closer to reaching full correctness than raw accuracy alone fully captures. At the same time, the strongest models still cluster most of their output into “Correct,” with only modest use of the partially correct band. This analysis highlights that while partial correctness is less useful in practice for computer architecture tasks that often require precise answers, capturing it can provide a more diagnostic view of model progress and failure modes.

\subsection{Comparing Human Domain Experts to LLM-as-a-Judge}
\label{app:subsec:human-expert-comparison}

In this section, we provide additional details to complement Section~\ref{sec:llm-judge-analysis} on validating the fidelity of LLM judges for \ours{} QA evaluation, by comparing LLM-as-a-Judge against human expert evaluations of student responses from 10 models considered frontier in September 2025 (\modelname{GPT-5}, \modelname{GPT-5} (Non-Reasoning), \modelname{Gemini-2.5-Pro}, \modelname{Gemini-2.5-Flash}, \modelname{Claude-Sonnet-4}, \modelname{Claude-3.7-Sonnet-Thinking}, \modelname{Llama-4-Maverick}, \modelname{Mistral-Medium-3.1}, \modelname{GPT-OSS-120B}, and \modelname{DeepSeek-R1}).
We instruct LLM-as-a-Judge to reason about the accuracy of each FRQ response with respect to the ground truth answer as though the response is from a student completing an academic exam. See Appendix~\ref{appendix:laj_prompt} for prompts. The judge is instructed to grade each response as \textsc{Correct}, \textsc{Partially-Correct}, or \textsc{Incorrect}. For our reported evaluations (Section~\ref{sec:model-performance}), we recategorize each LLM-as-a-Judge assessment into a binary \textsc{Correct} or \textsc{Incorrect} by rounding down \textsc{Partially-Correct} judge assessments to \textsc{Incorrect}. The \textsc{Partially-Correct} category serves two purposes: it disincentivize the judge from rounding up a nearly-correct answer to correct, and it enables analysis of fine-grained knowledge (Appendix~\ref{sec:partially-correct}).

We generate multiple samples per question and multiple judgments per sample to control for model stochasticity, and report estimated \textit{pass@k}=1 across 3 samples ($n=3$) as defined in~\citet{pinckney2025revisiting}. For each question in \ours{}, each model under evaluation (student $s$) generates 3 responses using the model's default generation parameters. For each individual student response, judge model $j$ generates up to 3 assessments until a majority vote consensus is reached. For example, if on a given problem, two student model samples are each judged by majority vote to be \textsc{Correct}, and the third student sample is majority vote \textsc{Incorrect}, \textit{pass@k}=1 on that problem is $\tfrac{2}{3}$.  

\textbf{Alternative LLM Judges:} In this section we justify our selection of \modelname{Claude-3.7-Sonnet-Thinking} as the LLM judge used for all benchmark evaluations by comparing Claude against two alternative judge models. We compare Claude against \modelname{Gemini-2.5-Pro} and \modelname{Qwen-3-VL-235B-A22B-Instruct} as alternate candidates for our LLM judge (using a consensus size of 3 for all three judge models). Across approximately 80,000 responses from 38 student models,  we find cross-model agreement rates to closely match (Claude-Gemini: 90.4\%, Claude-Qwen: 91.2\%, Gemini-Qwen: 89.8\%, Table~\ref{tab:judge-confusion-matrices}). Critically, these rates align with our measured human-to-human agreement rates of 90.75\% (Section~\ref{sec:llm-judge-analysis}), suggesting that LLMs exhibit similar consistency as human experts when judging QAs.

We also compare our three candidate LLM-as-a-Judge models directly against human expert evaluators. Claude had an agreement rate with human judgments of 85.48\% (Table~\ref{tab:cm-claude-human}), while Gemini and Qwen had slightly lower agreement rates at 84.64\% (Table~\ref{tab:cm-gemini-human}) and 84.04\% (Table~\ref{tab:cm-qwen-human}), respectively. Claude's slightly higher human agreement rate motivates our choice as the primary LLM judge. For completeness, we include an alternative version of Table~\ref{tab:skill-accuracy} using Gemini and Qwen as the judge LLM, shown in Appendix Tables~\ref{tab:gemini-llm-judge-results} and~\ref{tab:qwen-llm-judge-results}.

\begin{table}
    \caption{Per-generation accuracy (\%) of SoTA models across the four skills of \ours{} using \modelname{Gemini-2.5-Pro} as the LLM judge. Best performing models in each category highlighted \first{first}, \second{second}, and \third{third}.}
    \label{tab:gemini-llm-judge-results}
    \centering
    \begin{tabular}{l
                    c
                    c
                    c
                    c
                    c
                    }
    \toprule
    \textbf{Model} & & \multicolumn{4}{c}{\ours{}-\textsc{Reasoning}} \\
    \cline{3-6}
    & \textbf{Recall} & \textbf{Analyze} & \textbf{Design} & \textbf{Implement} & \textbf{Overall} \\
    \midrule
    \textit{Multimodal Models} \\
    \modelname{GPT-5.2} & \third{88.8} & \third{79.2} & \second{81.7} & \second{80.3} & \third{79.4} \\
    \modelname{GPT-5} (Non-Reasoning) & 86.8 & 49.5 & 56.3 & 42.5 & 49.1 \\
    \modelname{Gemini-3-Pro} & \first{91.5} & \third{79.2} & \first{82.5} & \first{81.0} & \second{79.5} \\
    \modelname{Gemini-3-Flash} & \second{90.7} & \first{80.0} & \third{77.0} & \third{78.9} & \first{79.8} \\
    \modelname{Claude-Sonnet-4} & 86.0 & 51.3 & 51.6 & 50.0 & 51.2 \\
    \modelname{Claude-3.7-Sonnet-Thinking} & 86.1 & 55.3 & 53.2 & 50.7 & 54.8 \\
    \modelname{Mistral-Medium-3.1} & 84.5 & 31.1 & 31.0 & 27.6 & 30.8 \\
    \modelname{Llama-4-Maverick} & 85.8 & 33.2 & 31.0 & 30.9 & 32.9 \\
    \midrule
    \textit{Text-Only Models} \\
    \modelname{GPT-OSS-120B} & 84.7 & 60.9 & 56.5 & 56.4 & 60.4 \\
    \modelname{DeepSeek-R1} & 87.5 & 58.2 & 39.1 & 52.1 & 57.0 \\
    \bottomrule
    \end{tabular}
\end{table}

\begin{table}
    \caption{Per-generation accuracy (\%) of SoTA models across the four skills of \ours{} using \modelname{Qwen-3-VL-235B-A22B-Thinking} as the LLM judge. Best performing models in each category highlighted \first{first}, \second{second}, and \third{third}.}
    \label{tab:qwen-llm-judge-results}
    \centering
    \begin{tabular}{l
                    c
                    c
                    c
                    c
                    c
                    }
    \toprule
    \textbf{Model} & & \multicolumn{4}{c}{\ours{}-\textsc{Reasoning}} \\
    \cline{3-6}
    & \textbf{Recall} & \textbf{Analyze} & \textbf{Design} & \textbf{Implement} & \textbf{Overall} \\
    \midrule
    \textit{Multimodal Models} \\
    \modelname{GPT-5.2} & \third{88.7} & \first{77.0} & \first{80.2} & \first{82.7} & \first{77.7} \\
    \modelname{GPT-5} (Non-Reasoning) & 86.8 & 53.2 & 59.5 & 57.1 & 53.8 \\
    \modelname{Gemini-3-Pro} & \first{91.1} & \third{74.5} & \second{78.6} & \second{73.8} & \second{74.6} \\
    \modelname{Gemini-3-Flash} & \second{90.7} & \second{75.0} & \third{71.4} & \third{70.7} & \third{74.5} \\
    \modelname{Claude-Sonnet-4} & 85.8 & 51.6 & 51.6 & 53.4 & 51.7 \\
    \modelname{Claude-3.7-Sonnet-Thinking} & 86.1 & 53.6 & 46.0 & 51.0 & 53.1 \\
    \modelname{Mistral-Medium-3.1} & 84.7 & 35.9 & 45.2 & 35.4 & 36.2 \\
    \modelname{Llama-4-Maverick} & 85.5 & 34.1 & 28.6 & 30.9 & 33.6 \\
    \midrule
    \textit{Text-Only Models} \\
    \modelname{GPT-OSS-120B} & 84.8 & 67.4 & 75.4 & 71.8 & 68.1 \\
    \modelname{DeepSeek-R1} & 87.4 & 57.8 & 53.6 & 53.0 & 57.3 \\
    \bottomrule
    \end{tabular}
\end{table}

\textbf{Majority Adjudication Frequency:} We also inspect the frequency of necessitated tie-breaking under our consensus size of 3 (where a tie consists of one \textsc{Correct} and one \textsc{Incorrect} verdict) across \ours{}'s entire QA benchmark and all models assessed in the main text and appendix. We observe across 65,659 responses, the first two judgments from \modelname{Claude-3.7-Sonnet-Thinking} matched 89.0\% of the time (and hence did not require a third judgment to adjudicate). 

\textbf{Fine-Grained Analysis on Self-Judgment Bias:} We also seek to investigate whether LLM-as-a-Judge favors responses produced by itself (i.e., if a bias is present when the same model is both the student and the judge). We analyze self-judgment behavior on the human-verified subset of student model responses. In \Cref{tab:self-judgment-confusion-matrices}, we compare \modelname{Claude-3.7-Sonnet-Thinking} and \modelname{Gemini-2.5-Pro} when each model judges its own generations against human expert judgments, grouping partially-correct responses with incorrect responses to match our binary FRQ accuracy evaluation. To study potential bias in LLMs, we focus on false positive rate which captures cases where human experts mark a response as incorrect, but LLM-as-a-Judge marks its own response as correct. Claude exhibits a low self-judgment false positive rate of $2.6\%$ ($1/39$), suggesting that our primary judge does not exhibit substantial self-preference on this subset. In contrast, Gemini exhibits a higher self-judgment false positive rate of $29.2\%$ ($7/24$), indicating that self-judgment bias can vary across judge models. Together with the cross-judge consistency results in \Cref{tab:judge-confusion-matrices}, this analysis suggests that \modelname{Claude-3.7-Sonnet-Thinking} does not exhibit pronounced self-preferential bias.

\begin{table*}[!tb]
\captionsetup{singlelinecheck=false}
\caption{Pairwise confusion matrices comparing LLM-as-a-Judge agreement counts for our three candidate judge models across approximately 80,000 student model responses. Here, agreement denotes that the same judgment (``correct'' or ``incorrect'') is reached by both judge models under majority consensus. Agreement rates between models are broadly similar and mirror human-to-human agreement rates.
}
\label{tab:judge-confusion-matrices}
\centering
\renewcommand{\arraystretch}{1.12}
\setlength{\tabcolsep}{4.5pt}

\subcaptionbox{
\modelname{Claude-3.7-Sonnet-Thinking} vs.\\
\modelname{Gemini-2.5-Pro}
\label{tab:cm-claude-gemini}
}[0.31\linewidth]{
\centering
\small
\begin{tabular}{lrr}
\toprule
 & \multicolumn{2}{c}{Gemini} \\
\cmidrule(lr){2-3}
Claude & Correct & Incorrect \\
\midrule
Correct   & 31{,}915 & 3{,}039  \\
Incorrect & 4{,}549  & 39{,}610 \\
\bottomrule
\end{tabular}
}
\hfill
\subcaptionbox{
\modelname{Claude-3.7-Sonnet-Thinking} vs.\\
\modelname{Qwen-3-VL-235B-A22B-Instruct}
\label{tab:cm-claude-qwen}
}[0.31\linewidth]{
\centering
\small
\begin{tabular}{lrr}
\toprule
 & \multicolumn{2}{c}{Qwen} \\
\cmidrule(lr){2-3}
Claude & Correct & Incorrect \\
\midrule
Correct   & 34{,}541 & 2{,}073  \\
Incorrect & 5{,}131  & 40{,}233 \\
\bottomrule
\end{tabular}
}
\hfill
\subcaptionbox{
\modelname{Gemini-2.5-Pro} vs.\\
\modelname{Qwen-3-VL-235B-A22B-Instruct}
\label{tab:cm-gemini-qwen}
}[0.31\linewidth]{
\centering
\small
\begin{tabular}{lrr}
\toprule
 & \multicolumn{2}{c}{Qwen} \\
\cmidrule(lr){2-3}
Gemini & Correct & Incorrect \\
\midrule
Correct   & 33{,}101 & 3{,}339  \\
Incorrect & 4{,}737  & 37{,}841 \\
\bottomrule
\end{tabular}
}
\end{table*}

\begin{table*}[!tb]
\captionsetup{singlelinecheck=false}
\caption{Confusion matrices comparing three LLM-as-a-Judge candidate models against human experts on the 908 human-verified student model responses, under majority consensus. \modelname{Claude-3.7-Sonnet-Thinking} exhibits a slightly higher agreement rate of 85.48\% with human experts compared to the other candidate judge models.}
\label{tab:judge-human-confusion-matrices}
\centering
\renewcommand{\arraystretch}{1.12}
\setlength{\tabcolsep}{4.5pt}

\subcaptionbox{
\modelname{Claude-3.7-Sonnet-Thinking} vs.\\
\hspace*{1.4em}Human Experts
\label{tab:cm-claude-human}
}[0.31\linewidth]{
\centering
\small
\begin{tabular}{lrr}
\toprule
 & \multicolumn{2}{c}{Human Experts} \\
\cmidrule(lr){2-3}
Claude & Correct & Incorrect \\
\midrule
Correct   & 428 & 60  \\
Incorrect & 73  & 355 \\
\bottomrule
\end{tabular}
}
\hfill
\subcaptionbox{
\modelname{Gemini-2.5-Pro} vs. Human Experts
\label{tab:cm-gemini-human}
}[0.31\linewidth]{
\centering
\small
\begin{tabular}{lrr}
\toprule
 & \multicolumn{2}{c}{Human Experts} \\
\cmidrule(lr){2-3}
Gemini & Correct & Incorrect \\
\midrule
Correct   & 443 & 82  \\
Incorrect & 59  & 334 \\
\bottomrule
\end{tabular}
}
\hfill
\subcaptionbox{
\modelname{Qwen3-VL-235B-A22B-Instruct}\\
\hspace*{1.4em}vs. Human Experts
\label{tab:cm-qwen-human}
}[0.31\linewidth]{
\centering
\small
\begin{tabular}{lrr}
\toprule
 & \multicolumn{2}{c}{Human Experts} \\
\cmidrule(lr){2-3}
Qwen & Correct & Incorrect \\
\midrule
Correct   & 444 & 90  \\
Incorrect & 56  & 325 \\
\bottomrule
\end{tabular}
}
\end{table*}

\begin{table*}[!tb]
\captionsetup{singlelinecheck=false}
\caption{
Confusion matrices comparing self-judgments by LLM-as-a-Judge against human expert judgments on the human-verified subset.
}
\label{tab:self-judgment-confusion-matrices}
\centering
\renewcommand{\arraystretch}{1.12}
\setlength{\tabcolsep}{7pt}

\subcaptionbox{
\modelname{Claude-3.7-Sonnet-Thinking} Self-Judgments
\label{tab:cm-human-claude-self}
}[0.45\linewidth]{
\centering
\small
\begin{tabular}{lrr}
\toprule
 & \multicolumn{2}{c}{Human Experts} \\
\cmidrule(lr){2-3}
\modelname{Claude-3.7-Sonnet-Thinking} & Correct & Incorrect \\
\midrule
Correct   & 45 & 2 \\
Incorrect & 9  & 43 \\
\bottomrule
\end{tabular}
}
\hfill
\subcaptionbox{
\modelname{Gemini-2.5-Pro} Self-Judgments
\label{tab:cm-human-gemini-self}
}[0.45\linewidth]{
\centering
\small
\begin{tabular}{lrr}
\toprule
 & \multicolumn{2}{c}{Human Experts} \\
\cmidrule(lr){2-3}
\modelname{Gemini-2.5-Pro} & Correct & Incorrect \\
\midrule
Correct   & 66 & 11  \\
Incorrect & 5  & 17 \\
\bottomrule
\end{tabular}
}
\end{table*}

\textbf{Grading Difficulty:} In order to understand if LLM judge disagreements with human experts increased in frequency for questions that were more challenging to grade, we additionally asked each human expert to assign each QA a score between 1-5 on the difficulty of grading the question (not the difficulty of the question itself). We did not observe any such trend in our data; the domain expertise and familiarity with academic content in the human cohort led to 1 and 2 being the most frequently assigned scores for grading difficulty. We believe this preliminary result leads to three potential directions for future work in alignment between the performance of LLM-as-a-Judge and domain experts: (1) judge prompt optimization by both domain experts and automated methods~\citep{opsahl2024optimizing}, such as by informing the judge that students may try to earn extra points on a question they can't answer by including relevant-sounding jargon to mimic understanding as we observe this behavior exhibited by some SLMs, (2) characterizing question difficulty in \ours{} and exploring whether harder questions are also harder to accurately grade by both LLMs and humans, and (3) investigating the tradeoff across LLM judge generation parameters between verdict determinism and verdict accuracy under consensus when comparing against human expert verdicts as ground truth.

\section{Example Questions}
\label{examples-appendix}
\subsection{QA Skills}
\label{sec:example_taxonomy}

\subsubsection{Example 1: Recall}
\label{sec:example_taxonomy_recall}

\begin{tcolorbox}[breakable,
  colback=gray!5,
  colframe=gray!40!black,
  enhanced jigsaw,
  sharp corners,
  title=Storage Systems]
\textbf{Question:} Moving compute closer to the \_\_\_ in solid state drives (SSDs) offers higher bandwidth but introduces challenges in managing frequent errors. \\

\textbf{Options:} \\(a) controller \\
(b) NAND dies \\
(c) cache \\
(d) DRAM \\

\textbf{Correct Solution:} \\(b) NAND dies \\
\end{tcolorbox}

\textbf{Rationale:} This question falls under the recall category as it requires retrieval of factual knowledge about SSD architecture, specifically the trade-offs between bandwidth optimization and error management when positioning compute resources relative to different storage components.

\subsubsection{Example 2: Analysis}
\label{sec:example_taxonomy_analysis}

\begin{tcolorbox}[breakable,
  colback=gray!5,
  colframe=gray!40!black,
  enhanced jigsaw,
  sharp corners,
  title=Systolic Array]

\includegraphics[width=0.9\linewidth]{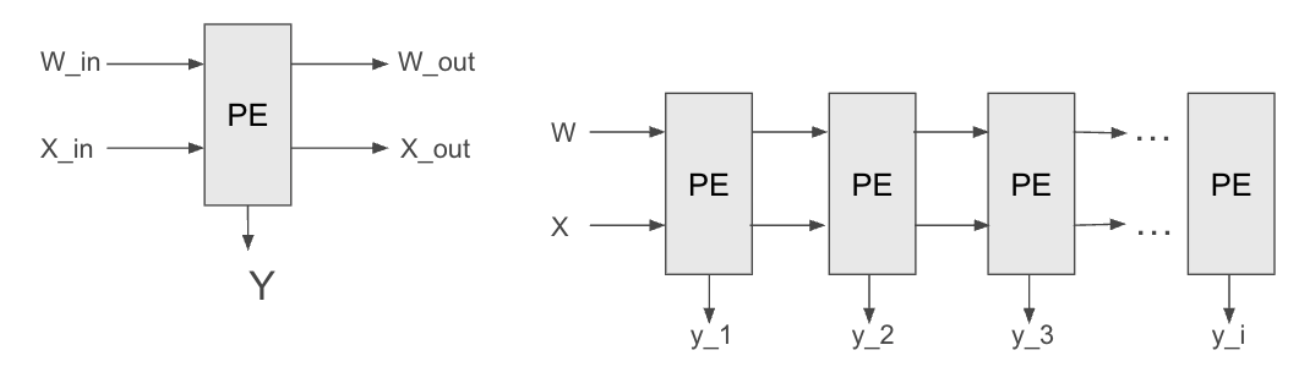}

\textbf{Context:} Given arrays $X = [x_1, x_2, x_3, \dots, x_n]$ and $W = [w_1, w_2, w_3, \dots, w_k]$, you want to compute \\
$Y = [y_1, y_2, y_3, \dots, y_{(n+1-k)}]$ using the formula: \\
$y_i = w_1x_i + w_2x_{(i+1)} + w_3x_{(i+2)} + \dots + w_kx_{(i+k-1)}$ \\[0.5em]

The figure shows a systolic array of processing elements (PEs) and their input-output behavior. \\

\textbf{Question:} What should be the relative speeds at which $X$ and $W$ values flow through the array to end up with the correct result for all $Y$ values? \\[0.5em]

\textbf{Correct Solution:} 2:1\\

This question is formulated using discussion on Fig 7 in the paper ``Why Systolic Architectures?" by HT Kung.\\

Citation: Kung, ``Why systolic architectures?," in Computer, vol. 15, no. 1, pp. 37-46, Jan. 1982, doi: 10.1109/MC.1982.1653825.\\

Please refer to paper for full discussion. \\

Intuition below:\\

The formula for y\_i is y\_i = w1*x\_i + w2*x\_(i+1) + w3*x\_(i+2) + … + w\_k*x\_(i+k-1). The number of PEs is equal to the number of Y values. Each PE computes one Y value. A single PE performs the computation Y ← Y + W\_in * X\_in.\\

Since w\_1 and x\_1 are fed as input on the first cycle, the first PE computes y\_1. The next PE needs to compute y\_2 = w\_1*x\_2 + w\_2*x\_3 + ... w\_k * x\_(k+1). So, after the first cycle, x\_2 and w\_1 should enter the second PE. This means that x should flow at twice the speed of w to align correctly for the calculations of each y\_i. Therefore, the relative speed is 2:1.
\end{tcolorbox}

\textbf{Rationale:} This is an analysis question because it requires breaking down the systolic array's computational flow and examining how data dependencies between X and W arrays must be synchronized across multiple processing elements.

\subsubsection{Example 3: Design}
\label{sec:example_taxonomy_design}
\begin{tcolorbox}[breakable,
  colback=gray!5,
  colframe=gray!40!black,
  enhanced jigsaw,
  sharp corners,
  title=Cache Partitioning and Associativity]
  
\textbf{Context:} Suppose we have a system with 32 cores that share a physical second-level cache. Assume each core is running a single single-threaded application, and all 32 cores are concurrently running applications. Assume that the page size of the architecture is 8KB, the block size of the cache is 128 bytes, and the cache uses LRU replacement. We would like to ensure each application gets a dedicated space in this shared cache without any interference from other cores. We would like to enforce this using the utility based cache partitioning (UCP) to partition the cache. Assume we would like to design a 4MB cache with a 128-byte block size. Recall that UCP aims to minimize the cache miss rate by allocating more cache ways to applications that obtain the most benefit from more ways, as we discussed in lecture.\\

\textbf{Question:} Consider the maximum associativity of the cache such that each application is guaranteed a minimum amount of space without interference. Is it desirable to implement UCP on a cache with this maximum associativity? Why, why not? Explain.
\\

\textbf{Correct Solution:}  No, it is not desirable to implement UCP with this maximum associativity because the overhead of UCP for 32 applications on this cache will likely outweigh its benefits. UCP will only work with LRU replacement policy. But implementing LRU on top of a 32 k-way cache is impractical. Also the number of counters needed by UCP and the partitioning solution space for UCP are very large for such a cache.
\end{tcolorbox}

\textbf{Rationale:} This question qualifies as a design question because it requires the analysis and formulation of architectural strategies for cache partitioning in multicore systems. Rather than executing a specific algorithm or implementation, the focus is on evaluating system-level trade-offs, exploring alternative approaches, and proposing optimal solutions under varying associativity constraints.

\subsubsection{Example 4: Implementation}
\label{sec:example_taxonomy_implementation}
\begin{tcolorbox}[breakable,
  colback=gray!5,
  colframe=gray!40!black,
  enhanced jigsaw,
  sharp corners,
  title=Linked-List Manipulation via Self-Modifying Code]

\textbf{Context:} In this question, you will implement linked-list operations using self-modifying code on an EDSACjr machine. The memory layout is shown in the figure on the right. You have access to the named memory locations as indicated. Linked-list nodes consist of two words: the first is an integer value, the second is an address pointing to the next node. \_HEAD contains the address of the first node of the list (or \_INVALID if it is empty). The next field of the last node is \_INVALID. All valid addresses are positive. You may create new local and global labels as explained in the EDSACjr handout. Table A-1 shows the EDSACjr instruction set.

\noindent
\begin{minipage}{0.49\linewidth}
    \centering
    \includegraphics[width=\linewidth]{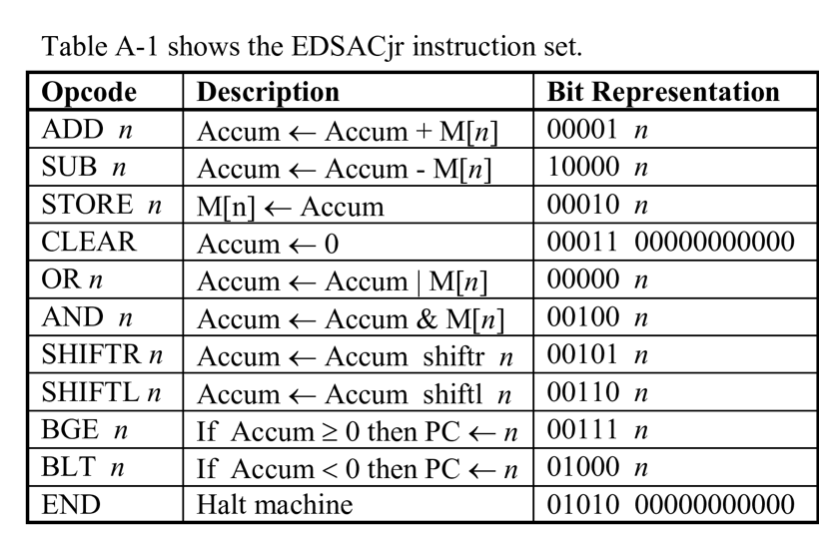}
\end{minipage}
\hfill
\begin{minipage}{0.49\linewidth}
    \centering
    \includegraphics[width=\linewidth]{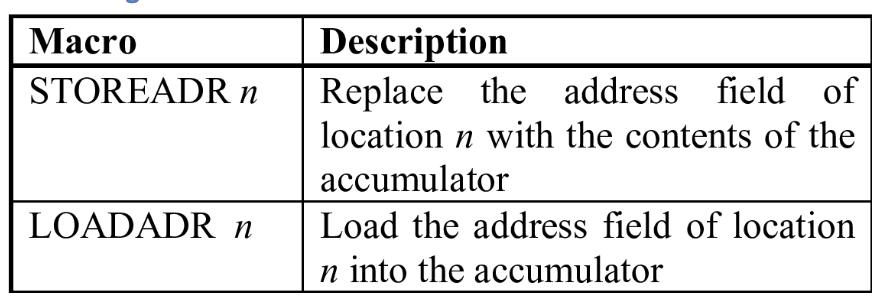}
\end{minipage}

You may also use the following macros if required:
\begin{itemize}
    \item \textbf{STOREADR \textit{n}}: Replace the address field of location \textit{n} with the contents of the accumulator
    \item \textbf{LOADADR \textit{n}}: Load the address field of location \textit{n} into the accumulator
\end{itemize}

\includegraphics[width=0.6\linewidth]{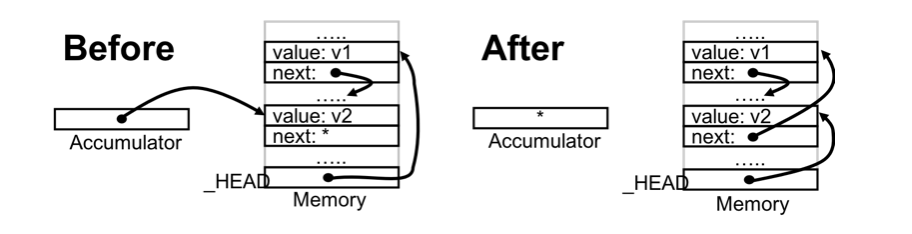}

Write a macro for \textbf{LISTPUSH}, which pushes the node pointed to by the accumulator to the head of the list. LISTPUSH takes one argument, the memory address of the new node, which is available in the accumulator. As shown in the figure below, LISTPUSH stores the current \_HEAD pointer in the new node's next field, and updates the \_HEAD pointer to point to the new node. Implement the macro using the EDSACjr instruction set and macros provided above. Do not refer to ``value" or ``next"; they are for illustration only. You need not worry about memory allocation; the new node's address is provided in the accumulator.

\includegraphics[width=0.4\linewidth, angle=90]{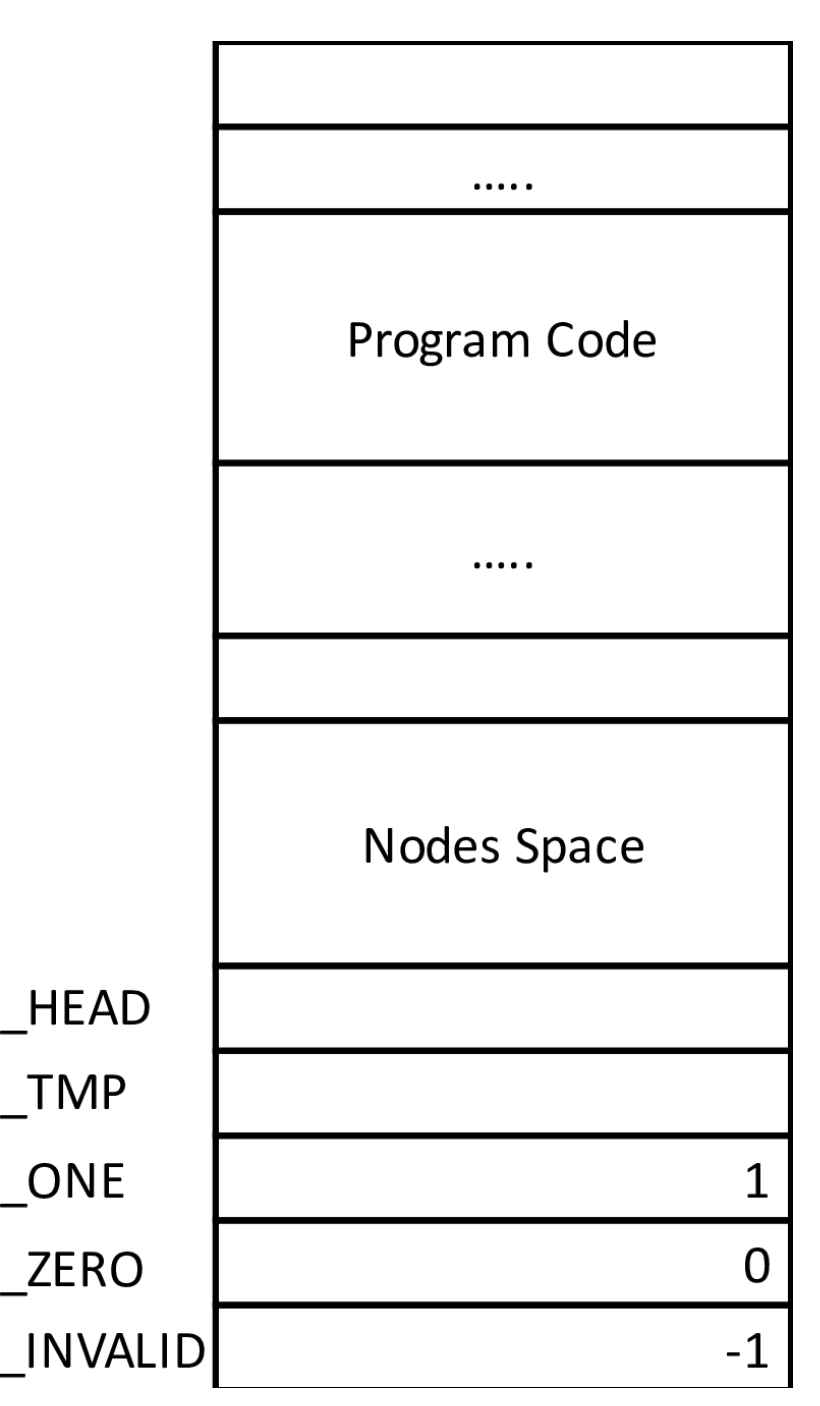}

\textbf{Question:} Write a macro for LISTPUSH.

\begin{lstlisting}[language={},basicstyle=\ttfamily]
.macro LISTPUSH
    STORE _TMP   ;; store accumulator (address of the new node)
.end
\end{lstlisting}

\textbf{Correct Solution:}

\begin{lstlisting}[language={},basicstyle=\ttfamily]
.macro LISTPUSH
    STORE _TMP    ;; store accumulator (address of the new node)
    ADD _ONE      ;; accum <- address of the new node's next field
    STOREADR _STN ;; address field of location _STN has the address
                  ;; of the new node's next field
    CLEAR
    ADD _HEAD     ;; accum <- M[_HEAD], current head pointer

_STN: STORE 0     ;; 0 will be replaced with the node's next field 
                  ;; address. M[_TMP + 1] <- accum
    CLEAR
    ADD _TMP      ;; retrieve address of new node in accumulator
    STORE _HEAD   ;; M[_HEAD] <- accum; Update the head pointer
                  ;; to the new node
.end
\end{lstlisting}
\end{tcolorbox}

\textbf{Rationale:} This is an implementation question because it requires constructing actual, working code or macros using the EDSACjr instruction set to perform specific memory operations. Rather than theorizing or analyzing abstract concepts, it requires to translate the algorithm into concrete steps that the hardware can execute.

\subsection{Failure Mode \#1: Struggles with Architecture-Semantics of Code Execution}

\label{sec:failure_mode_1}
\subsubsection{Example 1}
\label{sec:failure_mode_1_example_1}
\begin{tcolorbox}[breakable,
  colback=gray!5,
  colframe=gray!40!black,
  enhanced jigsaw,
  sharp corners,
  title=Reuse Intervals in Matrix Multiplication]

\textbf{Context:}\\
For a given compute kernel, we define a tensor’s reuse interval (RI) as the number of different elements of that tensor that have been referenced between each re-reference of the same element. For example, consider the following:

\begin{verbatim}
for m in [0, M)
  for n in [0, N)
    Z[m, n] = A[m] * B[n]
\end{verbatim}

Since A’s element is used at every iteration of the inner loop, its RI is 1. Each element of B is re-referenced after N references, so its RI is N. Z has “infinite” reuse interval (i.e., no data reuse) since no element is re-referenced throughout the computation:\\
RI of A = 1\\
RI of B = N\\
RI of Z = infinite / no reuse

Consider the following Matrix-Matrix multiply pseudocode, which multiplies two dense matrices A and B to produce Z:
\begin{itemize}
\item Multiply two matrices A and B to produce Z
\item First matrix A is MxK
\item Second matrix B is KxN
\item Thus, resulting matrix Z is MxN
\end{itemize}

\begin{verbatim}
for m in [0, M)
  for n in [0, N)
    for k in [0, K)
      Z[m, n] += A[m, k] * B[k, n]
\end{verbatim}

\textbf{Question:}\\
What are the reuse intervals for the three matrices? Provide your answers in terms of M, N, and K.
\\

\textbf{Correct Solution:}\\
RI of A = K\\
RI of B = N*K\\
RI of Z = 1
\\

\textbf{Incorrect Frontier Model Response:}\\
To determine the reuse intervals (RIs) for the matrices A, B, and Z in the given Matrix–
Matrix multiply pseudocode, let's analyze the loop structure and how each matrix is
accessed.

\medskip
\noindent The pseudocode is as follows:
\begin{lstlisting}
for m in [0, M)
  for n in [0, N)
    for k in [0, K)
      Z[m, n] += A[m, k] * B[k, n]
\end{lstlisting}

\textit{1. Matrix A (MxK).}
\begin{itemize}
  \item The element \texttt{A[m, k]} is accessed within the innermost loop (\texttt{for k in [0, K)}).
  \item For a fixed \texttt{m}, the row \texttt{A[m, k]} is accessed for each \texttt{k} from \texttt{0} to \texttt{K-1}.  
  Thus, for each iteration of the outer loop (\texttt{for m in [0, M)}), the same row is not 
  re-referenced until \texttt{M} iterations have passed.
  \item Within the inner loop, each element of a row of \texttt{A} is accessed once.  
  So for a given \texttt{m}, elements are accessed once as \texttt{k} varies from \texttt{0} to \texttt{K-1}.  
  For the next \texttt{m}, the same element is re-referenced after \texttt{K} iterations of the inner loop.
  \item More precisely: the re-reference happens after the inner loop (\texttt{for k}) has completed 
  for all \texttt{N} iterations of \texttt{n}, because \texttt{m} is fixed during those \texttt{N} iterations.  
  Hence the reuse interval for \texttt{A} is \texttt{N*K}.
  \item A finer analysis shows that each element \texttt{A[m, k]} is accessed for every \texttt{n} in \texttt{[0, N)}.  
  Thus, it is reused \texttt{N} times before the next \texttt{k} is accessed.
\end{itemize}
Hence, the correct RI for \texttt{A} is \texttt{N}.

\textit{2. Matrix B (KxN).}
\begin{itemize}
  \item \texttt{B[k, n]} is accessed within the innermost loop (\texttt{for k in [0, K)}).
  \item For a fixed \texttt{n}, \texttt{B[k, n]} is accessed for each \texttt{k}.  
  Each element of a column is accessed once as \texttt{k} varies, and re-referenced 
  after \texttt{K} iterations for a different \texttt{k}.
  \item Considering the loop structure: for a given \texttt{k} and \texttt{n}, \texttt{B[k, n]} is 
  accessed for every \texttt{m} in \texttt{[0, M)}.  
  Thus, it is reused \texttt{M} times.
\end{itemize}
Hence, the RI for \texttt{B} is \texttt{M}.

\textit{3. Matrix Z (MxN).}
\begin{itemize}
  \item \texttt{Z[m, n]} is accessed within the innermost loop (\texttt{for k in [0, K)}).
  \item For a fixed \texttt{m, n}, \texttt{Z[m, n]} is updated \texttt{K} times.  
  It is not re-referenced until the computation for that pair is completed, 
  which happens after \texttt{K} iterations of the innermost loop.
\end{itemize}
Hence, the RI for \texttt{Z} is \texttt{K}.

\medskip
\noindent \textit{Therefore, the reuse intervals are:}
\begin{itemize}
  \item RI of A = \texttt{N}
  \item RI of B = \texttt{M}
  \item RI of Z = \texttt{K}
\end{itemize}

These values represent how many times each element of the respective matrices is reused
before being re-referenced.

\end{tcolorbox}

\textbf{Question Summary and Relevance:} This question examines understanding of reuse distance analysis in nested loop kernels, a fundamental concept in computer architecture for predicting cache behavior and memory bandwidth requirements. The reuse interval (RI) counts the number of distinct memory references between successive accesses to the same data element, which affects locality and cache hit rates. 

For matrix multiplication, the dominant kernel in scientific computing and deep learning, accurate reuse analysis guides critical optimization decisions such as cache blocking factors, prefetch distances, and loop transformation strategies. The question specifically tests whether models can trace memory access sequences through triple-nested loops and correctly apply the reuse interval definition to determine that A has $\text{RI} = K$ (reused after each inner loop completion), B has $\text{RI} = N \times K$ (reused only after all $N \times K$ iterations), and Z has $\text{RI} = 1$ (immediately reused for accumulation).

\textbf{Analysis of Model Response:} The model's errors reveal a fundamental conceptual confusion between \textit{reuse count} (how many times an element is accessed) and \textit{reuse interval} (distinct accesses between reuses). For example, when analyzing matrix $A$, the model incorrectly reasons that because ``$A[m,k]$ is accessed for every $n$ in $[0,N)$. Thus, it is reused $N$ times...,'' concluding $\text{RI}=N$. This conflates usage frequency with reuse distance; the correct analysis requires tracing the actual reference stream to see that $A[m,k]$ is accessed once, then $K$ different elements are accessed before returning to $A[m,k]$ in the next $n$ iteration, giving $\text{RI}=K$. Similarly, for matrix $B$, the model focuses on $B$ being ``reused $M$ times'' across different $m$ values, missing that the same $B[k,n]$ element isn't accessed again until after $N \times K$ intervening references.

\subsubsection{Example 2}
\label{sec:failure_mode_1_example_2}
\begin{tcolorbox}[breakable,
  colback=gray!5,
  colframe=gray!40!black,
  enhanced jigsaw,
  sharp corners,
  title=Timing Side Channel from Short-Circuit Evaluation]

\textbf{Question:}\\
Consider the code given below:

\begin{verbatim}
bool detect_duplicate(int* elements, int length){
    bool cond, dup = false;
    for(int i=0; i<length; i++) {
        for(int j=0; j<length; j++) {
            cond = (elements[i] == elements[j]) && (i!=j);
            dup  = CMOV(cond, cond, dup);
        }}
    return !dup;
}
\end{verbatim}

Assume that CMOV is implemented as the \texttt{cmov} x86 instruction, a secure constant-time instruction. Assume the length of the array is already known publicly and is not sensitive information. Does this code have timing side channels that can reveal insights into the contents of the \texttt{elements} array?
\\

\textbf{Correct Solution:}\\
Yes, there is a timing side channel.
\\

\textbf{Incorrect Frontier Model Response:}\\

Analysis of Timing Side Channels in the Code:

The provided code aims to detect duplicates in an array without leaking information through timing side channels. Below, I analyze whether the code has timing side channels that could reveal insights into the contents of the \texttt{elements} array.\\\\
\textit{Key Features of the Code.}
\begin{itemize}
  \item The function \texttt{detect\_duplicate} uses two nested loops that iterate exactly \texttt{length}~$\times$~\texttt{length} times, regardless of the array values. This fixed iteration count ensures that the loop structure itself does not introduce timing variations based on the data.
  \item The critical operation inside the inner loop is a comparison (\texttt{elements[i] == elements[j]}) and a logical AND with (\texttt{i != j}), followed by a conditional move (CMOV) instruction. The CMOV instruction is specified as a secure constant-time instruction (like the x86 \texttt{cmov}), meaning it executes in constant time regardless of the condition outcome.
  \item The memory access pattern is determined solely by the indices \texttt{i} and \texttt{j}, which are fixed for a given \texttt{length}. Since \texttt{length} is public and non-sensitive, the sequence of memory accesses (to \texttt{elements[i]} and \texttt{elements[j]}) is identical for every invocation with the same \texttt{length}. Thus, cache behavior (e.g., hits or misses) is predictable and does not depend on the array values.
  \item There are no conditional branches that depend on the array values. All instructions (comparisons, logical operations, and CMOV) are executed unconditionally, minimizing the risk of variable execution time due to branch prediction or speculation.
\end{itemize}

\textit{Reasoning.}
\begin{itemize}
  \item Constant-Time Operations: The comparison \texttt{elements[i] == elements[j]} is typically implemented as a constant-time integer comparison on modern processors. Similarly, the logical AND and CMOV operations are constant-time under the given assumptions. Therefore, the time taken per inner iteration is invariant to the array values.
  \item Fixed Workload: The total number of iterations is \texttt{length\textsuperscript{2}}, which is constant for a given \texttt{length}. Since the work per iteration is constant-time, the overall execution time is determined solely by \texttt{length}, not by the array contents.
  \item Cache Considerations: While cache timing could theoretically vary based on access patterns, here the access pattern is fixed for a given \texttt{length}. The values in the array do not influence the addresses accessed, so cache behavior cannot reveal information about the values themselves.
  \item Return Value: The function returns \texttt{!dup}, but this computation occurs after the loops and is trivial (a single negation). It does not affect the timing of the main computation.
\end{itemize}

\textit{Conclusion.}
Under the assumptions that:
\begin{itemize}
  \item CMOV is a secure constant-time instruction,
  \item integer comparisons are constant-time,
  \item and \texttt{length} is public and non-sensitive,
\end{itemize}
the code does not have timing side channels that can reveal insights into the contents of the \texttt{elements} array. The execution time is predictable and depends only on \texttt{length}, not on the presence or absence of duplicates or the specific values in the array.\\\\
\textit{Additional Note.}
If the processor exhibits timing variations for CMOV or comparisons due to microarchitectural factors (e.g., value-dependent aliasing), these are considered negligible or non-exploitable for side channels in this context, given the problem constraints. However, in absolute terms, constant-time code should be verified on specific hardware, but the code here is designed to be constant-time.

\end{tcolorbox}

\textbf{Question Summary and Relevance}: This question tests understanding of \textit{timing side-channel vulnerabilities} in security-critical code, a paramount concern in computer security and cryptographic implementations. The problem presents a duplicate detection function that attempts to be constant-time using the \texttt{CMOV} (conditional move) instruction, a hardware primitive specifically designed to avoid branch-based timing attacks.

The question is particularly clever because it combines a secure primitive (CMOV) with an insecure pattern (short-circuit evaluation), testing whether one can reason about the interaction between hardware features and high-level programming language semantics. Such vulnerabilities have been exploited in the real-world with attacks against RSA, AES, and other cryptographic implementations.

\textbf{Analysis of Model Response:} The model concluded that the code was secure, citing the constant-time nature of CMOV and the public array length, but it failed to account for the short-circuit evaluation of the logical AND operator in C. This oversight demonstrates that the model failed to effectively reason about the architecture-semantics interplay in code execution. Specifically, the model did not incorporate how code semantics (like conditional evaluation) introduce data-dependent timing variations, leading to an incorrect assessment of side channels. 

\subsubsection{Example 3}
\label{sec:failure_mode_1_example_3}
\begin{tcolorbox}[breakable,
  colback=gray!5,
  colframe=gray!40!black,
  enhanced jigsaw,
  sharp corners,
  title=SIMD Utilization and Warp Divergence]

\textbf{Context:}\\
We define the SIMD utilization of a program that runs on a GPU as the fraction of SIMD lanes that are kept busy with active threads during the run of a program. As we saw in lecture and practice exercises, the SIMD utilization of a program is computed across the complete run of the program. The following code segment is run on a GPU. A warp in the GPU consists of 64 threads, and there are 64 SIMD lanes in the GPU. Each thread executes a single iteration of the shown loop. Assume that the data values of the arrays A and B are already in vector registers so there\\
are no loads and stores in this program. Both A and B are arrays of integers. (Hint: notice that there are 6 instructions in each thread.)

\begin{verbatim}
for (i = 0; i < 4096; i++) {
    if (B[i] < 8888) {      // Instruction 1
        A[i] = A[i] * C[i]; // Instruction 2
        A[i] = A[i] + B[i]  // Instruction 3
        C[i] = B[i] + 1;    // Instruction 4
    }

    if (B[i] > 8888) {      // Instruction 5
        A[i] = A[i] * B[i]; // Instruction 6
    }
}
\end{verbatim}

\textbf{Question:}\\
What needs to be true about array B to achieve the minimum possible SIMD utilization?\\
Show your work. (Please cover all cases in your answer.)
\\

\textbf{Correct Solution:}\\
1 in every 64 of B’s elements are greater than 8888, and 1 in every 64 of B’s elements are less than 8888, and the rest of the elements are 8888. \\

\textbf{Incorrect Frontier Model Response: } \\
Looking at this problem, I need to analyze how SIMD execution works with conditional branches and determine what values in array $B$ would minimize the number of active threads across all instructions.\\

\textit{Understanding the Code Structure}

The code has 6 instructions total:
\begin{enumerate}
    \item if ($B[i] < 8888$) (condition check)
    \item $A[i] = A[i] * C[i]$ (executed if $B[i] < 8888$)
    \item $A[i] = A[i] + B[i]$ (executed if $B[i] < 8888$)
    \item $C[i] = B[i] + 1$ (executed if $B[i] < 8888$)
    \item if ($B[i] > 8888$) (condition check)
    \item $A[i] = A[i] * B[i]$ (executed if $B[i] > 8888$)
\end{enumerate}

\textit{SIMD Execution with Branches}

With 4096 iterations and 64 threads per warp, we have 64 warps total.

In SIMD execution:
\begin{itemize}
    \item All threads in a warp execute the same instruction simultaneously
    \item When branches occur, threads not taking a path become inactive
    \item The warp must execute all paths that any thread takes
\end{itemize}

\textit{Analysis by Cases}

For each thread, based on $B[i]$ value:

\textit{Case 1: $B[i] < 8888$}
\begin{itemize}
    \item Thread executes: Instructions 1, 2, 3, 4, 5 (instruction 6 skipped)
    \item Active for 5 out of 6 instructions
\end{itemize}

\textit{Case 2: $B[i] = 8888$}
\begin{itemize}
    \item Thread executes: Instructions 1, 5 only (both conditional blocks skipped)
    \item Active for 2 out of 6 instructions
\end{itemize}

\textit{Case 3: $B[i] > 8888$}
\begin{itemize}
    \item Thread executes: Instructions 1, 5, 6 (instructions 2, 3, 4 skipped)
    \item Active for 3 out of 6 instructions
\end{itemize}

\textit{Calculating Active Thread-Instructions per Warp}

For a warp with:
\begin{itemize}
    \item $n_{1}$ threads where $B[i] < 8888$
    \item $n_{2}$ threads where $B[i] = 8888$
    \item $n_{3}$ threads where $B[i] > 8888$
    \item $n_{1} + n_{2} + n_{3} = 64$
\end{itemize}

Active thread-instructions:
\begin{itemize}
    \item Instruction 1: 64 threads (all execute condition)
    \item Instructions 2,3,4: $n_{1}$ threads each $= 3n_{1}$ total
    \item Instruction 5: 64 threads (all execute condition)
    \item Instruction 6: $n_{3}$ threads
\end{itemize}

Total per warp $= 64 + 3n_{1} + 64 + n_{3} = 128 + 3n_{1} + n_{3}$\\

\textit{Minimizing SIMD Utilization}

To minimize active thread-instructions, minimize: $3n_{1} + n_{3}$

Since $n_{1} + n_{2} + n_{3} = 64$:
\begin{itemize}
    \item Set $n_{1} = 0$ (no threads with $B[i] < 8888$)
    \item Set $n_{3} = 0$ (no threads with $B[i] > 8888$)
    \item Set $n_{2} = 64$ (all threads with $B[i] = 8888$)
\end{itemize}

\textit{Answer}

For minimum SIMD utilization: All elements of array $B$ must equal 8888

\textit{Verification}
\begin{itemize}
    \item Total possible thread-instruction slots $= 64 \times 6 \times 64 = 24{,}576$
    \item With all $B[i] = 8888$: Active thread-instructions $= 64 \times 128 = 8{,}192$
    \item Minimum SIMD utilization $= \dfrac{8{,}192}{24{,}576} = \dfrac{1}{3} \approx 0.333$
\end{itemize}

This occurs because when $B[i] = 8888$, threads only execute the two condition checks (instructions 1 and 5) and skip both conditional blocks entirely, minimizing the number of active thread-instructions across the program execution.

\end{tcolorbox}

\textbf{Question Summary and Relevance:} This problem assess understanding of SIMD utilization and warp divergence in GPUs, a central topic in computer architecture. SIMD (Single Instruction, Multiple Data) execution keeps all lanes busy when threads follow the same path, but utilization drops when conditional branches cause divergence. The question asks what values in array $B$ minimize utilization across all warps, emphasizing the cost of divergent execution paths. The relevance is twofold: (1) it tests understanding of control flow divergence and its impact on performance and (2) it highlights how data distribution (values in $B$) directly affects instruction activity.
Such reasoning connects hardware execution models to software efficiency, which is a core skill in computer systems design.

\textbf{Analysis of Model Response:} The model response misinterpreted the optimization goal. It assumed that 
minimizing the \emph{number of active instructions per thread} would directly minimize SIMD utilization, 
and therefore concluded that setting all $B[i] = 8888$ yields the minimum utilization. 
However, this reasoning overlooks warp-level execution semantics: 
if all threads skip the same conditional blocks, the warp does not diverge, and utilization is actually 
higher (per executed instruction) than in the case where threads diverge. 

The core mistake was conflating \emph{per-thread activity minimization} with 
\emph{warp-level utilization minimization}. The correct answer requires mixing values of $B$ such that 
different threads in a warp diverge (some with $B[i] < 8888$, some with $B[i] > 8888$, and the rest with 
$B[i] = 8888$). This forces the warp to execute all branches, but with minimal active participation in 
each, thereby reducing the overall SIMD utilization.

\subsubsection{Example 4}
\label{sec:failure_mode_1_example_4}
\begin{tcolorbox}[breakable,
  colback=gray!5,
  colframe=gray!40!black,
  enhanced jigsaw,
  sharp corners,
  title=Combinational vs. Sequential from Verilog]

\textbf{Question:}\\
\includegraphics[width=0.85\linewidth]{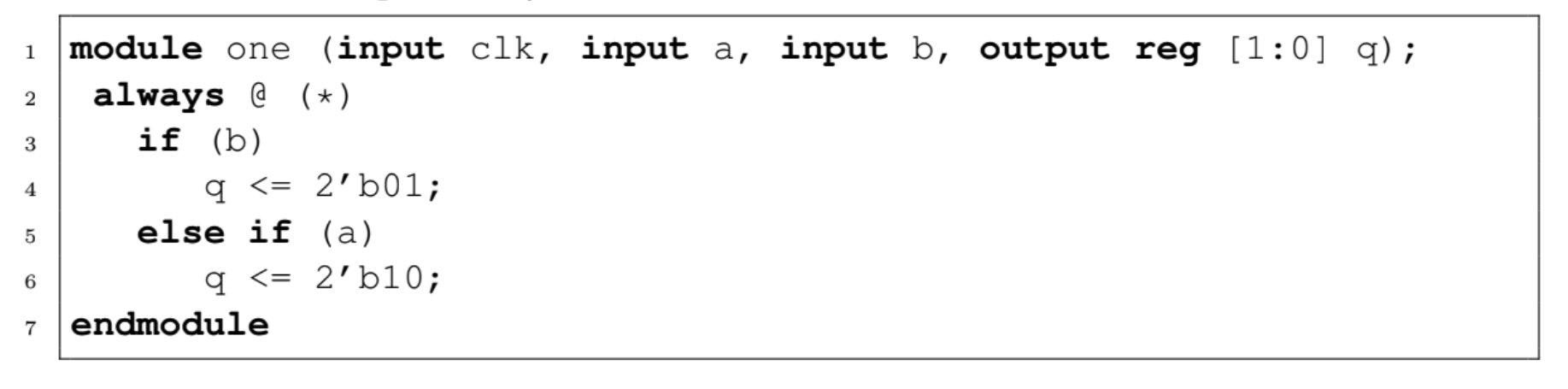}

Does the following code result in a sequential circuit or a combinational circuit? Please explain why.
\\

\textbf{Correct Solution:}\\
This code results in a \textbf{sequential} circuit because a latch is required to store the old value of \texttt{q} if both conditions are not satisfied. \\

\textbf{Incorrect Frontier Model Response:}\\
\textit{Answer and Concise Explanation:} \\
This code results in a combinational circuit, but with an important caveat - it creates an unintended latch. \\

\textit{Reasoning:}

\begin{enumerate}
    \item \textit{Sensitivity List Analysis}: The \verb|always @ (*)| statement creates a combinational sensitivity list that triggers whenever any input used within the always block changes. This is the key indicator of combinational logic.
    
    \item \textit{No Clock Edge}: There is no clock edge (like \verb|posedge clk| or \verb|negedge clk|) in the sensitivity list. Although \verb|clk| is declared as an input, it is never used in the always block, so it doesn't affect the circuit behavior.
    
    \item \textit{Data Type Clarification}: The \verb|reg| data type for output \verb|q| does not imply sequential behavior - it's simply a data type that can hold values in procedural blocks.
\end{enumerate}

\textit{Critical Issue - Incomplete Case Coverage:} \\

The code has a significant problem:
\begin{itemize}
    \item When \verb|b = 1|: \verb|q = 2'b01| 
    \item When \verb|b = 0| and \verb|a = 1|: \verb|q = 2'b10| 
    \item When \verb|b = 0| and \verb|a = 0|: No assignment to q 
\end{itemize}

This incomplete case coverage creates an unintended latch because when both inputs are 0, the output \verb|q| retains its previous value, introducing memory behavior. \\

\textit{Conclusion:} \\

While this synthesizes as combinational logic (due to the \verb|always @ (*)| construct), it creates a problematic latch that should be avoided in good design practice. To make it purely combinational, an \verb|else| clause should be added to handle the case when both \verb|a| and \verb|b| are 0.

\end{tcolorbox}

\textbf{Question Summary and Relevance:} The given Verilog code asks whether the design results in a sequential or combinational circuit. At first glance, the use of \texttt{always @(*)} suggests combinational intent since the block triggers on input changes rather than a clock edge. However, because not all input conditions assign a value to the output \texttt{q}, the synthesis tool infers a latch to hold the previous value when both \texttt{a} and \texttt{b} are \texttt{0}. This subtlety makes the design sequential, even without an explicit clock.  

In the context of computer architecture, understanding such distinctions is critical. Hardware description languages (HDLs) like Verilog are used to describe datapaths, control units, and memory systems. Misclassifying sequential vs. combinational logic can lead to unintended latches, incorrect timing behavior, or hardware inefficiencies. This impacts pipeline stages, critical path delays, and ultimately system performance and reliability, which are important considerations in architecture design.  

\textbf{Analysis of Model Response:} The model’s response incorrectly concluded that the code produces a purely combinational circuit with an unintended latch. While it correctly identified the incomplete case coverage, it failed to reconcile the key fact that \emph{incomplete assignments within an \texttt{always @(*)} block result in latch inference}. This is not ``still combinational with a caveat,'' but a defining feature of sequential logic, since the latch introduces memory behavior.   

\subsection{Failure Mode \#2: Assuming Unconventional Architectural Properties}
\label{sec:failure_mode_2}

\subsubsection{Example 1}
\label{sec:failure_mode_2_example_1}

\begin{tcolorbox}[breakable,
  colback=gray!5,
  colframe=gray!40!black,
  enhanced jigsaw,
  sharp corners,
  title=Page Table Entry Sizing]

\textbf{Context:} \\
Assume a system which has the following characteristics: 
\begin{enumerate}
    \item Virtual address space is 64 KB and is byte addressable 
    \item Physical RAM is 16 KB and is byte addressable 
    \item Page size is 256 Byte 
    \item One level page table, where each page table entry contains a valid bit, a dirty bit, and the physical page number 
    \item Integer is 32 bits 
    \item PTBR is 0x2F5C 
    \item There is a data TLB that stores two page table entries
\end{enumerate}

\textbf{Question:}\\
What would be the size of the page table? \\

\textbf{Correct Solution:}\\
256 * 8 bits = 256 bytes \\

\textbf{Incorrect Frontier Model Response: }\\
Looking at this step-by-step: \\

\textit{Step 1: Calculate the number of virtual pages} \\

Number of virtual pages = Virtual address space ÷ Page size \\
Number of virtual pages = 64 KB ÷ 256 bytes = 65,536 ÷ 256 = **256 pages** \\

\textit{Step 2: Determine page table entries needed} \\

Since we have a **one-level page table**, we need one page table entry for each virtual page.\\
Therefore: **256 page table entries** are required.\\

\textit{Step 3: Calculate the size of each page table entry} \\

Each page table entry contains:
\begin{itemize}
    \item Valid bit: 1 bit
    \item Dirty bit: 1 bit  
    \item Physical page number: Need to calculate required bits
\end{itemize}

For the physical page number:
\begin{itemize}
    \item Physical RAM = 16 KB = 16,384 bytes
    \item Number of physical pages = 16,384 ÷ 256 = 64 pages
    \item Bits needed for physical page number = $log_2(64)$ = **6 bits**
\end{itemize}
Minimum bits per entry = 1 + 1 + 6 = 8 bits \\

However, the problem states that "Integer is 32 bits," which indicates that each page table entry is stored as a 32-bit integer for alignment and efficiency purposes. \\

Therefore: **Each page table entry = 32 bits = 4 bytes** \\

\textit{Step 4: Calculate total page table size} \\

Total page table size = Number of entries × Size per entry \\
Total page table size = 256 entries × 4 bytes = **1,024 bytes = 1 KB** \\

\textit{Answer} \\

The size of the page table is **1,024 bytes (1 KB)**.
\end{tcolorbox}

\textbf{Question Summary and Relevance:}
This question asks to calculate the amount of memory required to store a 1-level page table for a system with known virtual and physical memory sizes. This requires calculating the necessary number of bits per page table entry, along with the total number of entries required to represent the entire virtual address space. Additional unnecessary information about the underlying system is provided in the question context, so this question also tests the student's ability to recognize what information is relevant to the question at hand.

\textbf{Analysis of Model Response:}
While the frontier model correctly identifies minimum number of bits required per page table entry (6 physical frame bits + 1 valid bit + 1 dirty bit), it makes an assumption that all page table entries must be aligned to the size of an integer (32 bits). This assumption is unconventional: there is no benefit for aligning page table entries to the boundary of an unrelated data type. In fact, doing so would leave 24 bits of each page table entry unused and wasted. This assumption causes the model to assume that page table entries must be 32 bits---rather than the required 8 bits---leading the frontier model to incorrectly calculate a required size of 1 KB.

\subsubsection{Example 2}
\label{sec:failure_mode_2_example_2}

\begin{tcolorbox}[breakable,
  colback=gray!5,
  colframe=gray!40!black,
  enhanced jigsaw,
  sharp corners,
  title=Branch Misprediction Penalty Cycles]

\textbf{Context:} \\
Ben Bitdiddle is designing a processor with the complex pipeline illustrated below: \\

\includegraphics[width=0.5\linewidth]{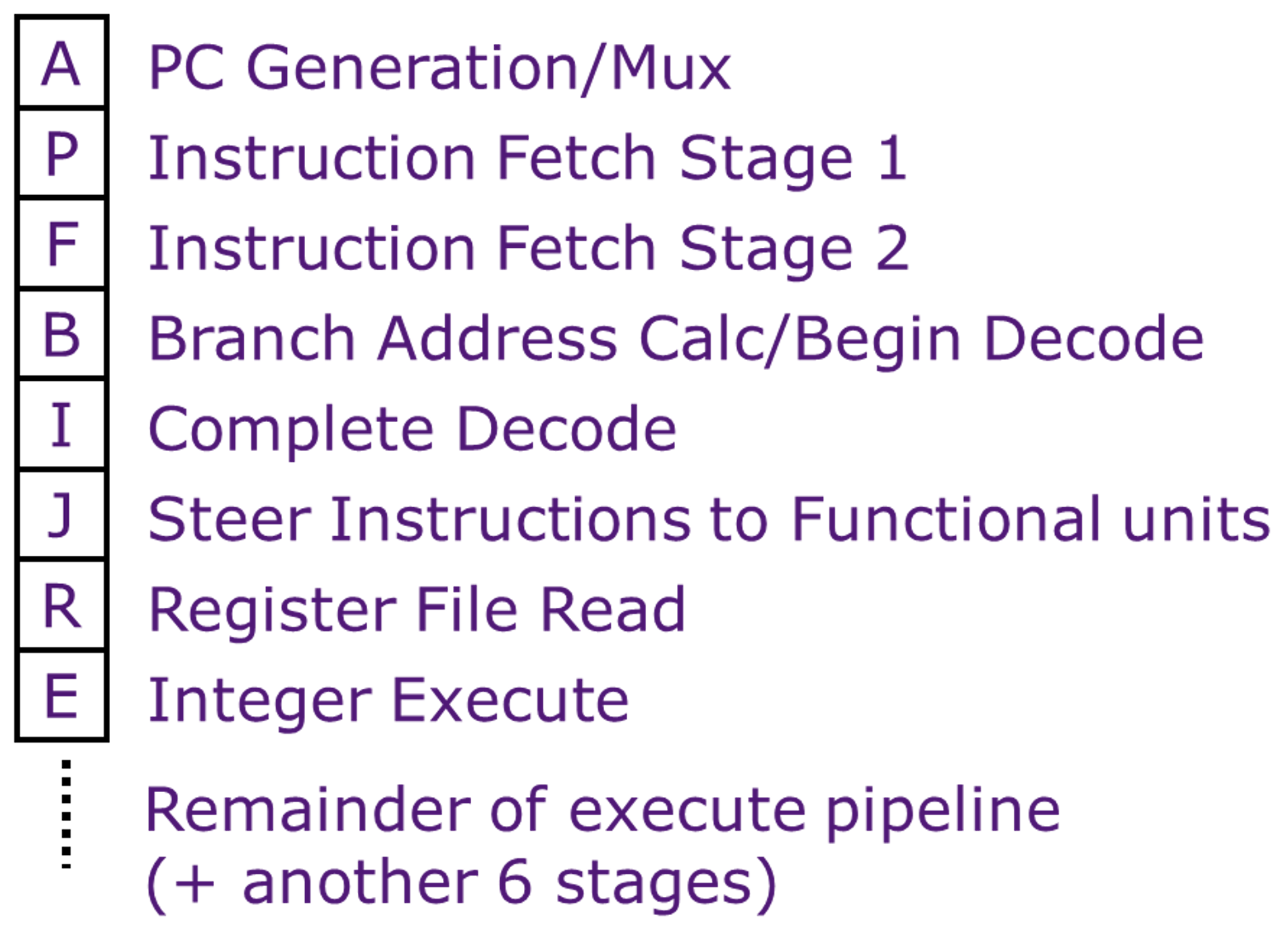}

The processor has the following characteristics:
\begin{itemize}
    \item Issues at most one instruction per cycle.
    \item Branch addresses are known at the end of the B stage (Branch Address Calc/Begin Decode).
    \item Branch conditions (taken/not taken) are known at the end of the R stage (Register File Read).
    \item Branches always go through the pipeline without any stalls or queuing delays.
\end{itemize}

Ben's target program is shown below: \\
\begin{verbatim}
for(int i = 0; i <= 1000000; i++)
{
    if(i % 2 == 0) //Branch B1
    { //Not taken
       (Do something A)
    }
    if(i % 4 == 0) //Branch B2
    { //Not taken
       (Do something B)
    }
} //Branch LP


\end{verbatim} \\

\begin{verbatim}
      ANDi R1 0
LOOP: MODi R2 R1 2
      BNE  R2 M4      // B1
      (Do something A)
      ... ...
M4:   MODi R3 R1 4
      BNE  R3 END     // B2
      (Do something B)
      ... ... 
END:  SUBi R4 R1 1000000
      BNE  R4 LOOP    // LP
      ... ...
      
\end{verbatim} \\

The MODi (modulo-immediate) instruction is defined as follows:
\begin{verbatim}
MODi Rd Rs imm: Rd <- Rs Mod imm

\end{verbatim} \\

\textbf{Question:}\\
In steady state, how many cycles per iteration are lost on average if the processor always speculates that every branch is not taken (i.e., next PC is PC+4)? \\

\textbf{Correct Solution:}\\
Penalty for miss prediction = 6 cycles \\
6 * 0.5 + 6 * 0.75 + 6 * 1 = 13.5 \\

\textbf{Incorrect Frontier Model Response: }\\

Answer: 6.75 cycles per iteration (on average) \\

Reasoning:
\begin{enumerate}
    \item Misprediction penalty with “always not taken”
    \begin{itemize}
        \item Branch target known at end of B.
        \item Branch direction known at end of R.
        \item If the branch is actually taken but we speculated not taken, redirection waits from B → R, i.e., 3 stages (I, J, R).
        \item Therefore, penalty per taken branch = 3 cycles.
    \end{itemize}
    
    \item Branch outcomes per iteration
    \begin{itemize}
        \item B1: i\%2==0 $\rightarrow$ BNE is taken when i is odd. Taken 1/2 of iterations.
        \item B2: i\%4==0 $\rightarrow$ BNE is taken when i\%4$\neq$0. Taken 3/4 of iterations.
        \item LP (loop back): taken every iteration in steady state.
    \end{itemize}

    \item Average cycles lost per iteration \\  
    = 3 × (P[taken B1] + P[taken B2] + P[taken LP]) \\
    = 3 × (0.5 + 0.75 + 1) \\
    = 3 × 2.25 \\
    = 6.75 cycles/iteration. \\
\end{enumerate}

Note: “Steady state” ignores the initial and final iterations, where LP is not taken once; including that would change the average negligibly.
\end{tcolorbox}

\textbf{Question Summary and Relevance:}
This question asks the model to calculate the average number of wasted cycles a pipelined processor would incur on a target program when using a simple branch prediction policy. Solving this question requires understanding of the penalty of a branch misprediction on the provided custom processor pipeline, along with reasoning about the average number of mispredictions the target program would incur. Notably, the provided processor pipeline is non-standard, so this question also tests a student's ability to generalize their understanding of the consequences of branch mispredictions.

\textbf{Analysis of Model Response:}
The frontier model correctly understands the average number of branch mispredictions that will occur in the target program (50\% of the time for the first \texttt{if} statement, 75\% of the time for the second \texttt{if} statement, and 100\% of the time for the \texttt{for} loop conditional). However, the model fails to correctly calculate the misprediction penalty, claiming the penalty is 3 cycles instead of 6 cycles. Looking at the reasoning trace, the model incorrectly assumes that during a misprediction, only instructions between stages B and R need to be considered. While this assumption can be valid in simpler pipelines, it doesn't apply here, since instructions in earlier stages (A, P, F) also need to be flushed during a misprediction.

\subsection{Failure Mode \#3: Modeling and Tracking System State}
\label{sec:failure_mode_3}

\subsubsection{Example 1}
\label{sec:failure_mode_3_example_1}

\begin{tcolorbox}[breakable,
  colback=gray!5,
  colframe=gray!40!black,
  enhanced jigsaw,
  sharp corners,
  title=Test-and-Set States ]
\textbf{Context:}\\
You are writing a queue to be used in a multi-producer/single-consumer application. (Producer threads write messages that are read by one consumer.) We assume here a queue with infinite space. The basic code is shown below.\\

\texttt{TST rs, Imm(rt)} is the test-and-set instruction, which atomically loads the value at Imm(rt) into rs, and if the value is zero, updates the memory location at \texttt{Imm(rt)} to 1. This atomic instruction is useful for implementing locks: a value of 1 at the memory location indicates that someone holds the lock, and a value of 0 means the lock is free. \\

Producer pushes a message onto queue: (memory operations in bold)
\begin{verbatim}
    
void push(int** tail_ptr, int* tail_write_lock, int message) {
    while (lock_try(tail_write_lock) == false);
    **tail_ptr = message;
    *tail_ptr++;
    lock_release(tail_write_lock);
}
\end{verbatim}
\begin{verbatim}

# R1 - contains address of data to enqueue
# R2 - contains the address of the tail pointer of queue
# R3 - address of tail pointer write lock
P1  SpinLock:TST   R4, 0(R3)  # try to acquire tail write lock
P2  BNEZ           R4, R4, SpinLock
P3  LD             R4, 0(R2)  # get tail pointer
P4  ST             R1, 0(R4)  # write message to tail
P5  ADD            R4, R4, 4  # update tail pointer
P6  ST             R4, 0(R2)
P7  ST             R0, 0(R3)  # release lock
\end{verbatim}

\begin{verbatim}
    
int pop(int** head_ptr, int** tail_ptr) {
    while (*head_ptr == *tail_ptr);
    int message = **head_ptr;
    *head_ptr++;
    return message;
}
\end{verbatim}
\begin{verbatim}

# R1 - will receive address contained in message
# R2 - contains the address of the head pointer of queue
# R3 - contains the address of the tail pointer of the queue  
C1 Retry: LD R4, 0(R2)  # get head pointer
C2  LD R5, 0(R3)  # get tail pointer
C3  SUB R5, R4, R5  # is there a message?
C4  BNEZ R5, Pop
C5  JMP Retry
C6 Pop: LD R1, 0(R4)  # read message from queue
C7  ADD R4, R4, 4  # update head pointer
C8  ST R4, 0(R2)

    
\end{verbatim}
\textbf{Question:}\\
We'll now focus on the tail pointer only. Assuming a MSI invalidate coherence protocol, show the state of the tail pointer in the producer and consumer cache after each operation in the sequence below. Show any data or permissions transfers, e.g. ``Memory→C" or ``C invalidates P".\\

How many state transitions occur per message in the steady state?

\includegraphics[width=0.85\linewidth]{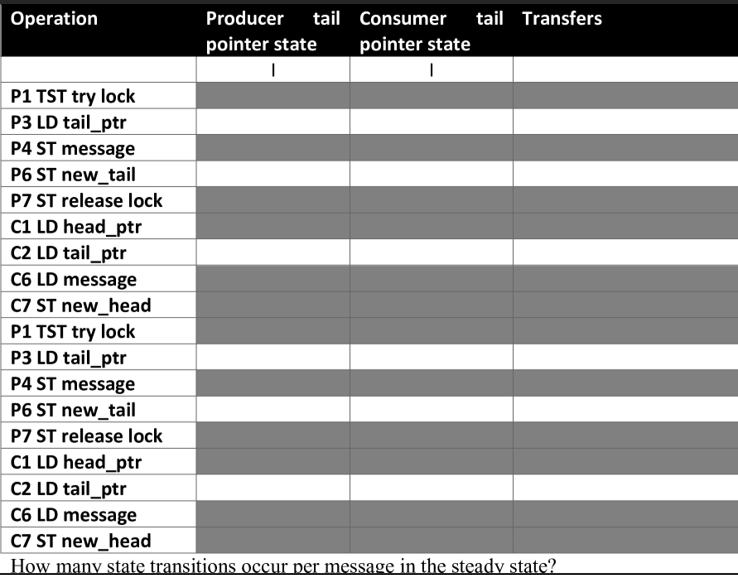}
\\

\textbf{Correct Solution:}\\
\includegraphics[width=0.85\linewidth]{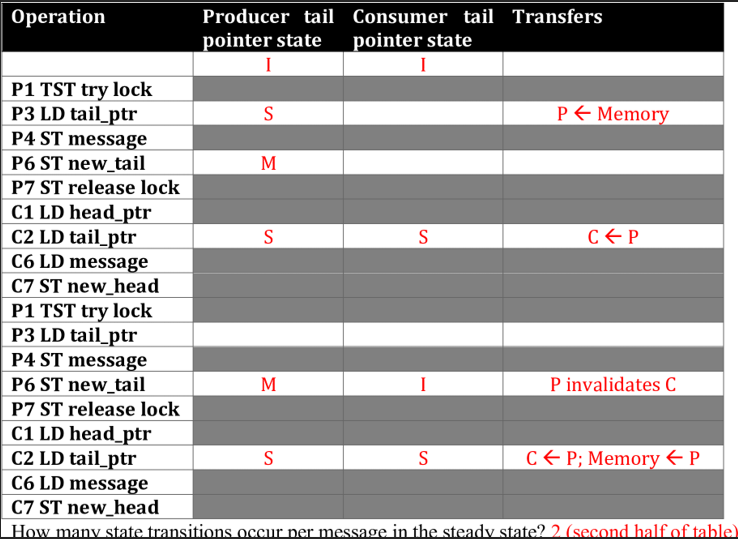}
\\

\textbf{Incorrect Frontier Model Response:}\\

To analyze the state transitions of the tail pointer in the 
producer and consumer cache, we need to carefully follow the 
sequence of operations for both the producer and the consumer, 
assuming an MSI (Modified, Shared, Invalid) 
cache coherence protocol. \\

\textit{Steady-State Analysis of Tail Pointer:} \\

\textit{Producer Side:}
\begin{enumerate}
    \item P1 TST try lock: \begin{itemize}
        \item State Transition: The producer tries to acquire the lock on the tail pointer.
        \item Cache State: If the lock is free (cache line state is 
   I or S), it will transition to M (Modified) when the 
   producer acquires the lock.
    \end{itemize}
    \item P3 LD tail\_ptr: \begin{itemize}
        \item State Transition: The producer loads the tail pointer into its cache.
        \item Cache State: If the tail pointer is not in the cache, 
   it will transition from I (Invalid) to S (Shared) 
   or E (Exclusive) if it is not shared.
    \end{itemize}
    \item P4 ST message: \begin{itemize}
        \item State Transition: 
   The producer writes the message to the tail location.
        \item Cache State: This write will transition the cache line
   to M (Modified).
    \end{itemize}
    \item P6 ST new\_tail: \begin{itemize}
        \item State Transition: 
   The producer updates the tail pointer.
        \item Cache State: The tail pointer cache line 
   will transition to M (Modified) in the producer's cache.
    \end{itemize}
    \item P7 ST release lock: \begin{itemize}
        \item State Transition: 
   The producer releases the lock.
        \item Cache State: The lock cache line will 
   transition back to I (Invalid) or S (Shared) 
   depending on other cores’ states.
   \end{itemize}
\end{enumerate}

\textit{Consumer Side:}
\begin{enumerate}
    \item C2 LD tail\_ptr: \begin{itemize}
        \item State Transition: 
   The consumer loads the tail pointer into its cache.
        \item Cache State: If the tail pointer is not 
   in the cache or is invalid, it will transition 
   from I to S (Shared) or E (Exclusive).
   \end{itemize}
\end{enumerate}

---

\textit{MSI State Transitions per Message in Steady State:}
\begin{enumerate}
\item Producer Acquires Lock (P1 TST try lock): \begin{itemize}
   \item Transition from I or S to M (1 transition).
   \end{itemize}

\item Producer Loads Tail Pointer (P3 LD tail\_ptr): \begin{itemize}
   \item Transition from I to S or E (1 transition).
   \end{itemize}

\item Producer Writes Message (P4 ST message): \begin{itemize}
   \item Transition from S or E to M (1 transition).
   \end{itemize}

\item Producer Updates Tail Pointer (P6 ST new\_tail): \begin{itemize}
   \item Transition from S or E to M (1 transition).
   \end{itemize}

\item Producer Releases Lock (P7 ST release lock): \begin{itemize}
   \item Transition from M to I (1 transition).
   \end{itemize}

\item Consumer Loads Tail Pointer (C2 LD tail\_ptr): \begin{itemize}
   \item Transition from I to S (1 transition), 
   potentially causing an invalidation in the 
   producer's cache (1 additional transition).
   \end{itemize}
\end{enumerate}
---

\textit{Data/Permission Transfers:}
- Memory to Cache (Memory $\rightarrow$ C): When the tail pointer 
is initially loaded into the cache.
- Cache Invalidation (C invalidates P): When the 
consumer loads the tail pointer, it may invalidate 
the producer's copy if it was in M state.

---

\textit{Summary of State Transitions per Message:}
\begin{enumerate}
\item Producer: \begin{itemize}
   \item Acquire lock: I/S $\rightarrow$ M (1 transition)
   \item Load tail pointer: I $\rightarrow$ S/E (1 transition)
   \item Write message: S/E $\rightarrow$ M (1 transition)
   \item Update tail pointer: S/E $\rightarrow$ M (1 transition)
   \item Release lock: M $\rightarrow$ I (1 transition)
\end{itemize}

\item Consumer: \begin{itemize}
   \item Load tail pointer: I $\rightarrow$ S (1 transition)
   \item Invalidate producer's tail pointer: M $\rightarrow$ I (1 transition)
\end{itemize}
\end{enumerate}

Total State Transitions per Message: \begin{itemize}
    \item 5 transitions on the producer side.
    \item 2 transitions on the consumer side.
\end{itemize}

Total: 7 state transitions per message in the steady state. \\

\resizebox{\linewidth}{!}{%
\begin{tabular}{|l|l|l|l|}
\hline
Operation & Producer Cache State Transition & Consumer Cache State Transition & Transfers \\
\hline
P1 TST try lock    & I/S $\rightarrow$ M & - & - \\
P3 LD tail\_ptr    & I $\rightarrow$ S/E & - & Memory $\rightarrow$ P \\
P4 ST message      & S/E $\rightarrow$ M & - & - \\
P6 ST new\_tail    & S/E $\rightarrow$ M & - & - \\
P7 ST release lock & M $\rightarrow$ I   & - & - \\
C2 LD tail\_ptr    & -                   & I $\rightarrow$ S & Memory $\rightarrow$ C, C invalidates P \\
\hline
\end{tabular}
} \\

---

\textit{Conclusion:}
In the steady state, there are 7 state transitions per 
message considering both the producer and consumer operations. 
Each message involves multiple cache line state changes 
and potential invalidations to maintain coherence.
\end{tcolorbox}

\textbf{Question Summary and Relevance:}
This question assesses whether models can concretely reason about cache coherence behavior in a multicore setting. By stepping through the producer–consumer interaction under an MSI protocol, it forces careful tracking of how load, store, and lock operations map to specific cache-line state transitions and invalidations. The question relies on architectural intuition -- seeing how simple pointer updates or lock releases induce multiple coherence transitions highlights the hidden costs of synchronization and communication, which often create a scalability bottleneck in shared-memory systems.

\textbf{Analysis of Model Response:}
The model incorrectly asserted that the producer’s test-and-set (TST) on the lock variable alters the MSI state of the tail pointer. This reflects a misunderstanding of coherence granularity: the MSI protocol tracks cache lines at the level of physical addresses, not program-level abstractions. Since the lock and the tail pointer occupy distinct memory locations, a TST on the lock can only trigger transitions on the lock’s cache line. Attributing such transitions to the tail pointer indicates a conceptual error in tracking the architectural system state, where ordering effects of synchronization are mistakenly interpreted as coherence events on unrelated data.

\subsubsection{Example 2}
\label{sec:failure_mode_3_example_2}

\begin{tcolorbox}[breakable,
  colback=gray!5,
  colframe=gray!40!black,
  enhanced jigsaw,
  sharp corners,
  title=Identifying Data Dependencies from Instruction Trace]
\textbf{Context:}
\begin{verbatim}

addi x2, x0, 17
addi x8, x0, 14
add x1, x1, x2
and x3, x4, x1
addi x1,x2, 5
add x3, x3, x3
sw x3, 4(x1)
add x7, x5, x4
sw x7, 0(x5)
add x11, x2, x3
add x9, x7, x8
add x9, x9, x9
ld x11, 12(x8)
ld x9, 0(x5) 

\end{verbatim}

\textbf{Question:}\\
Consider the above RISC-V instruction sequence. The total number of instruction pairs having Read after Write (RAW) dependencies on registers and memory is \_\_\_?
\\

\textbf{Correct Solution:}\\
There are 13 RAW dependencies on registers and 2 RAW dependencies on memory (for memory addresses 0(x5) and 26)
\\

\textbf{Incorrect Frontier Model Response:}\\

I need to analyze this RISC-V instruction sequence to identify all Read-after-Write (RAW) dependencies on both registers and memory. \\
Let me work through this step-by-step, numbering each instruction and tracking dependencies: \\

Instructions: \\
\begin{verbatim}
addi x2, x0, 17    # x2 = 0 + 17 = 17
addi x8, x0, 14    # x8 = 0 + 14 = 14
add x1, x1, x2     # x1 = x1 + x2
and x3, x4, x1     # x3 = x4 & x1
addi x1, x2, 5     # x1 = x2 + 5
add x3, x3, x3     # x3 = x3 + x3
sw x3, 4(x1)       # store x3 to memory[x1 + 4]
add x7, x5, x4     # x7 = x5 + x4
sw x7, 0(x5)       # store x7 to memory[x5 + 0]
add x11, x2, x3    # x11 = x2 + x3
add x9, x7, x8     # x9 = x7 + x8
add x9, x9, x9     # x9 = x9 + x9
ld x11, 12(x8)     # load x11 from memory[x8 + 12]
ld x9, 0(x5)       # load x9 from memory[x5 + 0]
    
\end{verbatim}
RAW Dependencies on Registers:

Instruction 3 reads x2 (written by instruction 1) → RAW: (1,3)

Instruction 4 reads x1 (written by instruction 3) → RAW: (3,4)

Instruction 5 reads x2 (written by instruction 1) → RAW: (1,5)

Instruction 6 reads x3 (written by instruction 4) → RAW: (4,6)

Instruction 7 reads x3 (written by instruction 6) → RAW: (6,7)

Instruction 7 reads x1 (written by instruction 5) → RAW: (5,7)

Instruction 9 reads x7 (written by instruction 8) → RAW: (8,9)

Instruction 10 reads x2 (written by instruction 1) → RAW: (1,10)

Instruction 10 reads x3 (written by instruction 6) → RAW: (6,10)

Instruction 11 reads x7 (written by instruction 8) → RAW: (8,11)

Instruction 11 reads x8 (written by instruction 2) → RAW: (2,11)

Instruction 12 reads x9 (written by instruction 11) → RAW: (11,12)

Instruction 13 reads x8 (written by instruction 2) → RAW: (2,13) \\

RAW Dependencies on Memory:

Instruction 14 reads from memory[x5 + 0] (written by instruction 9) → RAW: (9,14) \\

Summary:

Register RAW dependencies: 13 pairs

Memory RAW dependencies: 1 pair

Total RAW dependencies: 14

The answer is 14.
  \end{tcolorbox}
  
  \textbf{Question Summary and Relevance:} This question evaluates a student’s ability to identify and count Read-After-Write (RAW) dependencies in a given RISC-V instruction sequence, both at the register level and the memory level. All register dependencies and some memory dependencies can be found by directly statically analyzing the instructions in isolation. However, one memory dependency (address 26) requires properly tracking the values of registers \texttt{x8} and \texttt{x1}.
  
  \textbf{Analysis of Model Response:} The model successfully tracked register states but missed the more hidden memory RAW dependency on address 26. This limitation led to errors in identifying Read-after-Write (RAW) dependencies and ultimately produced an incorrect answer, illustrating the model’s weakness in handling memory-based interactions and implicit architectural information.
\subsection{Failure Mode \#4: Sensitivity to QA Modality}
\label{sec:failure_mode_4}

\subsubsection{Example 1}
\label{sec:failure_mode_4_example_1}

\begin{tcolorbox}[breakable,
  colback=gray!5,
  colframe=gray!40!black,
  enhanced jigsaw,
  sharp corners,
  title=Interpreting Bit-vectors for Page Table Entries]

\textbf{Context:} \\
A processor supports virtual memory with two regions (user and supervisor) using a VAX-like address translation approach. A breakdown of a virtual address and a page-table entry are shown below. Use these to answer each of the following questions.

\includegraphics[width=0.85\textwidth]{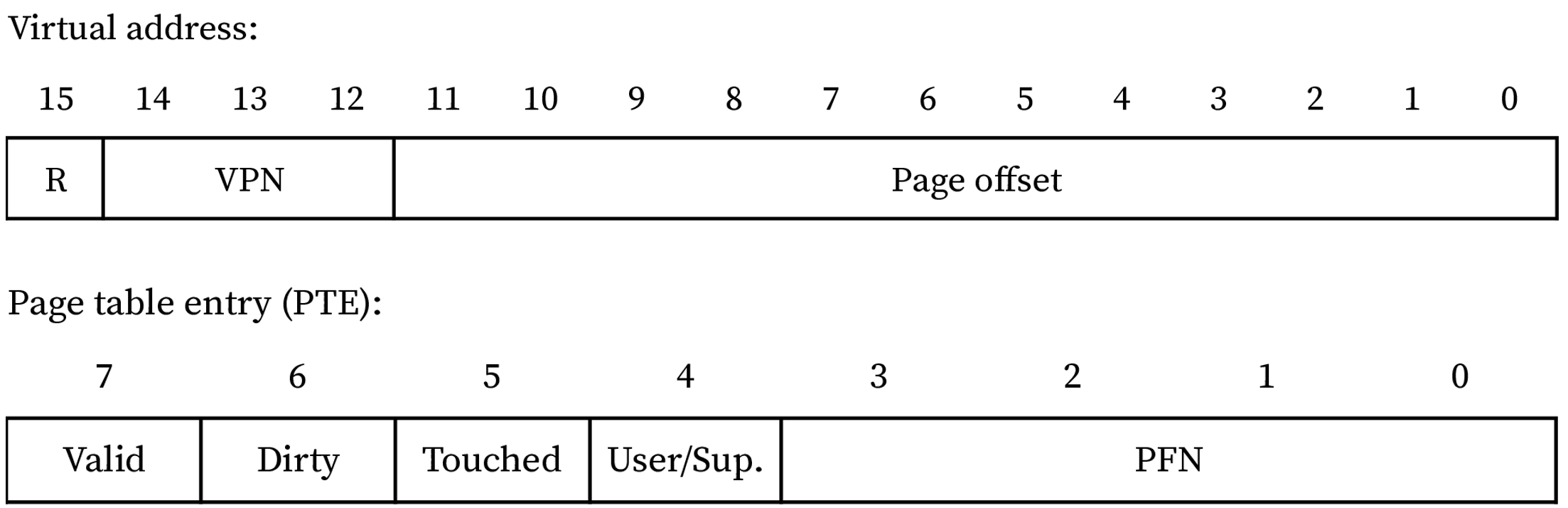} \\

\textbf{Question:}\\
What is the page size?\\

\textbf{Solution:}\\
4KiB (12 offset bits)\\

\textbf{Incorrect Frontier Model Response:}\\

\textit{Answer: Page Size Calculation}\\

To determine the page size in bytes, I need to examine the structure of the virtual address and identify how many bits are allocated to the page offset.\\

\textit{Analysis of Virtual Address Structure}\\

From the provided virtual address format:
\begin{itemize}
    \item Bit 15: R (Region bit)
    \item Bits 14-9: VPN (Virtual Page Number)
    \item Bits 8-0: Page offset
\end{itemize}

\textit{Calculation} \\

The page offset field determines the size of each page in the memory system. With a page offset of 9 bits (bits 8 through 0), the page size can be calculated as: \\

Page size = $2^{\text{number of page offset bits}}$ bytes \\
Page size = $2^9$ bytes \\
Page size = $512$ bytes \\

Therefore, the page size in this virtual memory system is 512 bytes.
\end{tcolorbox}

\textbf{Question Summary and Relevance:}
This question asks the student to interpret a bit-vector of a virtual address to calculate the appropriate page size. This requires understanding that page sizes are dependent on the size of the offset region of a virtual address. For this question, the bit-vector for a virtual address is provided as an image.

\textbf{Analysis of Model Response:}
Here, the model failed to properly interpret the provided image containing the bit-vector of a virtual address. While the image clearly shows bits 11:0 containing the page offset, the frontier model incorrectly claims the offset is in bits 8:0. Importantly, the model does execute the correct calculation for the size of a page based on the offset bit length, so this model failure was purely in the interpretation of the provided image.

\subsubsection{Example 2}
\label{sec:failure_mode_4_example_2}

\begin{tcolorbox}[breakable,
  colback=gray!5,
  colframe=gray!40!black,
  enhanced jigsaw,
  sharp corners,
  title=Deadlock-Free Channel Dependency Graphs]

\textbf{Question:}\\
The image shows a channel dependency graph for a network. What is the smallest set of edges that, when removed, eliminates deadlocks in the network?

\includegraphics[width=0.85\linewidth]{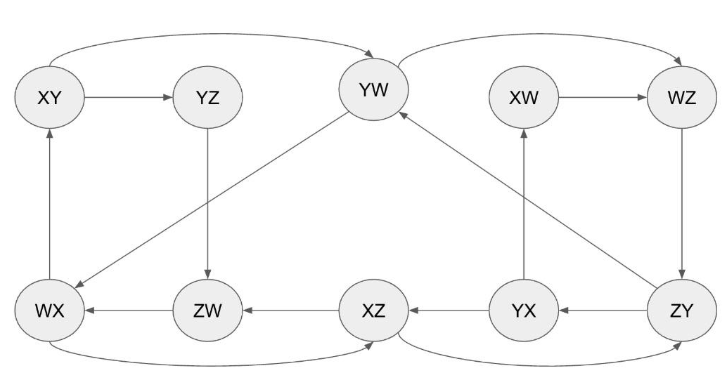}

\begin{enumerate}[A.]
    \item $ZW\rightarrow WX, ZY \rightarrow YX$
    \item $ZW \rightarrow WX, ZY \rightarrow YX, YW \rightarrow WX, ZY \rightarrow YW$
    \item $YW \rightarrow WX, ZY \rightarrow YW, XY\rightarrow YZ, XW\rightarrow WZ$
    \item $ZW\rightarrow WX$
\end{enumerate}

\textbf{Correct Solution:}\\
Eliminating edges mentioned in Option B removes all cycles from the graph and hence eliminates deadlocks. Since there are 8 cycles in the given graph, we need to eliminate at least 4 edges. 
\\

\textbf{Incorrect Frontier Model Response:}\\
Looking at this channel dependency graph, I need to identify the smallest set of edges that, when removed, eliminates all deadlocks.\\

First, let me identify the cycles in the graph, as deadlocks occur when there are circular dependencies:
\begin{itemize}
    \item XY $\rightarrow$ YZ $\rightarrow$ ZW $\rightarrow$ WX $\rightarrow$ XY (4-node cycle)
    \item YW $\rightarrow$ WX $\rightarrow$ XZ $\rightarrow$ ZY $\rightarrow$ YW (4-node cycle)
    \item XW $\rightarrow$ WZ $\rightarrow$ ZY $\rightarrow$ YX $\rightarrow$ XW (4-node cycle)
\end{itemize}

To eliminate deadlocks, I need to break all cycles. The key is finding the minimum number of edges that participate in multiple cycles, as removing such edges breaks multiple cycles simultaneously.\\

Let me analyze which edges appear in multiple cycles:
\begin{itemize}
    \item ZW$\rightarrow$WX: Appears in cycles 1 and 2
    \item ZY$\rightarrow$YX: Appears in cycles 2 and 3
    \item YW$\rightarrow$WX: Appears in cycle 2
    \item XZ$\rightarrow$ZY: Appears in cycles 2 and 3
\end{itemize}

Looking at the options:
\begin{itemize}
    \item ZW$\rightarrow$WX, ZY$\rightarrow$YX: This removes edges that break all three cycles.\\
       Removes ZW$\rightarrow$WX (breaks cycles 1 and 2).\\
       Removes ZY$\rightarrow$YX (breaks cycles 2 and 3).\\
       All cycles are broken with just 2 edges.
    \item ZW$\rightarrow$WX, ZY$\rightarrow$YX, YW$\rightarrow$WX, ZY$\rightarrow$YW: This is overkill (4 edges).
    \item YW$\rightarrow$WX, ZY$\rightarrow$YW, XY$\rightarrow$YZ, XW$\rightarrow$WZ: This is also overkill (4 edges).
    \item ZW$\rightarrow$WX: This only breaks cycles 1 and 2, leaving cycle 3 intact.
\end{itemize}

The smallest set that eliminates all deadlocks is ZW$\rightarrow$WX, ZY$\rightarrow$YX (option 1), as it breaks all cycles with just 2 edges.\\

The answer is the first option: ZW$\rightarrow$WX, ZY$\rightarrow$YX.
\end{tcolorbox}

\textbf{Question Summary and Relevance:}
The question tests students' knowledge of deadlock freedom conditions and ability to interpret directed dependency graphs. Since the graph has quite a few nodes (10 nodes), a visual representation of the dependency graph is the most natural way to represent this setting.

\textbf{Analysis of Model Response:}
The model failed to properly track the cycles in the graph and performed an oversimplified traversal, ignoring edge cases/cycles where deadlock could have occurred. This likely is due to a failure to track the cycles via the visual representation of the dependency graph.

\subsection{Example Motivating LLM-as-a-Judge for Automated Evaluations}
\label{sec:successful_response}

\begin{tcolorbox}[breakable,
  colback=gray!5,
  colframe=gray!40!black,
  enhanced jigsaw,
  sharp corners,
  title=Advantages of LdR/StC pair of instructions]

\textbf{Context:}\\  
One of the common instruction sequences used for synchronizing several processors are the
LOAD RESERVE/STORE CONDITIONAL pair (from now on referred to as LdR/StC pair). The
LdR instruction reads a value from the specified address and sets a local reservation for the address. The StC attempts to write to the specified address provided the local reservation for the address is still held. If the reservation has been cleared the StC fails and informs the CPU.
\\

\textbf{Question:}\\
Give two reasons why the LdR/StC pair of instructions is preferable over atomic read-test-
modify instructions such as the TEST \& SET instruction.
\\

\textbf{Correct Solution:}
\begin{enumerate}
    \item Bus doesn't need to be aware of them.
    \item Everything is local.
    \item No ping-pong.
    \item No extra hardware (tied to 1)
\end{enumerate}

\textbf{Correct Frontier Model Response:}\\
The LOAD RESERVE/STORE CONDITIONAL (LdR/StC) pair is preferable over atomic read-test-modify instructions like TEST\&SET for the following reasons:
\\

\begin{enumerate}
    \item Reduced Bus Contention and Cache Invalidation\\
    TEST\&SET performs an atomic read-modify-write operation, which always writes to memory even if the lock acquisition fails. This results in frequent bus traffic and cache invalidations for contending processors, degrading performance.
    
    LdR/StC avoids unnecessary writes: \begin{itemize}
        \item The StC instruction only writes to memory if the reservation is still valid.  
        \item Failed StC attempts do not generate writes, reducing bus traffic and cache coherence overhead.
    \end{itemize}

    \item No Bus Locking During Critical Section\\
    TEST\&SET requires locking the memory bus for the entire atomic operation, blocking all other memory accesses during execution. This creates a scalability bottleneck in multi-processor systems.

    LdR/StC operates without bus locking: \begin{itemize}
        \item LdR is a standard load instruction, and StC is a conditional store.  
        \item The hardware tracks reservations passively (e.g., via cache coherence protocols), allowing concurrent memory accesses by other processors between the LdR and StC.  
        \item This enables better parallelism and scalability for contended synchronization.
    \end{itemize} 
\end{enumerate}
\end{tcolorbox}

The example above illustrates a scenario where a model generates the correct answer to a question in the dataset, and where simple pattern-matching would not suffice for evaluation. The frontier model's response aligns conceptually with the answer key by addressing the same advantages using more verbose phrasing. The model's point about ``Reduced Bus Contention and Cache Invalidation" corresponds to ``No ping-pong" in the answer key - both describe how LdR/StC prevents unnecessary cache line bouncing between processors by avoiding writes on failed attempts. Similarly, the emphasis on ``No Bus Locking During Critical Section" matches ``Everything is local" - both highlight that LdR/StC operations can be managed through local reservation tracking without requiring global bus coordination or centralized locking mechanisms. 

This example highlights how strict string matching or n-gram-based methods~\cite{papineni2002bleu,lin2004rouge} may fail to capture semantically correct answers. We therefore elect to use LLM-as-a-judge (Section~\ref{sec:llm-judge-analysis}) to evaluate FRQ-style questions.

\subsection{Case Study \#2}
\label{sec:case_study_questions}

This section presents example questions from \ours{} that are similar to the types of questions asked in the memory controller design case study (Section~\ref{sec:memory-controller-case-study}). Please refer to Appendix~\ref{appendix:memory_controller_case_study_prompt} for the prompt used for this case study. The prompt asks for the LLM to analyze the memory trace and design parameters to understand their architectural implications on power. The questions in this section require similar reasoning skills.

\subsubsection{Examples for Analyzing Memory Traces}
\label{sec:analyzing_mem_traces_examples}

\begin{tcolorbox}[breakable,
  colback=gray!5,
  colframe=gray!40!black,
  enhanced jigsaw,
  sharp corners,
  title=DRAM Row Access Trace Optimization ]
  
\textbf{Context:}

Recall from your required reading on Tiered-Latency DRAM that there is a near and far segment, each containing some number of rows. Assume a very simplified memory model where there is just one bank and there are two rows in the near segment and four rows in the far segment. The time to activate and precharge a row is 25ns in the near segment and 50ns in the far segment. The time from start of activation to reading data is 10ns in the near segment and 15ns in the far segment. All other timings are negligible for this problem. Given the following memory request stream, determine the optimal assignment (minimize average latency of requests) of rows in the near and far segment (assume a fixed mapping where rows cannot migrate, a closed-row policy, and the far segment is inclusive).\\

time 0ns: row 0 read

time 10ns: row 1 read

time 100ns: row 2 read

time 105ns: row 1 read

time 200ns: row 3 read

time 300ns: row 1 read\\

\textbf{Question:}

What rows would you place in near segment? Hint: draw a timeline.\\

\textbf{Correct Solution:}

rows 0 and 2. see above

\end{tcolorbox}

\begin{tcolorbox}[breakable,
  colback=gray!5,
  colframe=gray!40!black,
  enhanced jigsaw,
  sharp corners,
  title=Tiered-Latency DRAM Design Based on Access Pattern ]

\textbf{Context:}

You would like to understand the configuration of the DRAM subsystem of a computer using reverse engineering techniques. Your current knowledge of the particular DRAM subsystem is limited to the following information:

* The physical memory address is 16 bits.

* The DRAM subsystem consists of a single channel and 4 banks.

* The DRAM is byte-addressable.

* The most-significant 2 bits of the physical memory address determine the bank.

* The DRAM command bus operates at 500 MHz frequency.

* The memory controller issues commands to the DRAM in such a way that no command for servicing a later request is issued before issuing a READ command for the current request, which is the oldest request in the request buffer. For example, if there are requests A and B in the request buffer, where A is the older request and the two requests are to different banks, the memory controller does not issue an ACTIVATE command to the bank that B is going to access before issuing a READ command to the bank that A is accessing.\\

You realize that you can observe the memory requests that are waiting to be serviced in the request buffer. At a particular point of time, you take the snapshot of the request buffer and you observe the following requests in the request buffer.\\

Requests in the request buffer (in descending order of request age, where the oldest request is on the top):

Read 0x4C80

Read 0x0140

Read 0x4EC0

Read 0x8000

Read 0xF000

Read 0x803F

Read 0x4E80\\

At the same time you take the snapshot of the request buffer, you start probing the DRAM command bus. You observe the DRAM command type and the cycle (relative to the first command) at which the command is seen on the DRAM command bus. The following are the DRAM commands you observe on the DRAM bus while the requests above are serviced.\\

Cycle 0 --- PRECHARGE

Cycle 6 --- ACTIVATE

Cycle 10 --- READ

Cycle 11 --- READ

Cycle 21 --- PRECHARGE

Cycle 27 --- ACTIVATE

Cycle 31 --- READ

Cycle 32 --- ACTIVATE

Cycle 36 --- READ

Cycle 37 --- READ

Cycle 38 --- READ

Cycle 42 --- PRECHARGE

Cycle 48 --- ACTIVATE

Cycle 52 --- READ\\

To improve performance, you decide to implement the idea of Tiered-Latency DRAM (TL-DRAM) in the DRAM chip. Assume that a bank consists of a single subarray. With TL-DRAM, an entire bank is divided into a near-segment and far-segment. When accessing a row in the near-segment, the ACTIVATE-to-READ latency reduces by 2 cycles and the ACTIVATE-to-PRECHARGE latency reduces by 5 cycles. When accessing a row in the far-segment, the ACTIVATE-to-READ latency increases by 1 cycle and the ACTIVATE-to-PRECHARGE latency increases by 2 cycles.\\

Assume that the rows in the near-segment have smaller row ids compared to the rows in the far-segment. In other words, physical memory row addresses 0 through N - 1 are the near-segment rows, and physical memory row addresses N through M - 1 are the far-segment rows.\\

\textbf{Question:}

If the above DRAM commands are issued 5 cycles faster with TL-DRAM compared to the baseline (the last command is issued in cycle 47), how many rows are in the near-segment? Show your work.\\

\textbf{Correct Solution:}

59 rows have to be in the near segment.\\

Explanation. There should 59 rows in the near-segment (rows 0 to 58) since rows until row id 58 need to be accessed with low latency to get 5 cycle reduction. Rows 59 and 192 are in the far-segment, thus latency for accessing them increases slightly.\\

Here is the new command trace:

Cycle 0 -- PRECHARGE - Bank 1

Cycle 6 -- ACTIVATE - Bank 1, Row 50, near segment

Cycle 8 -- READ - Bank 1

Cycle 9 -- READ - Bank 0

Cycle 16 -- PRECHARGE - Bank 1

Cycle 22 -- ACTIVATE - Bank 1, Row 59, far segment

Cycle 27 -- READ - Bank 1

Cycle 28 -- ACTIVATE - Bank 2, Row 0

Cycle 30 -- READ - Bank 2

Cycle 31 -- READ - Bank 3

Cycle 32 -- READ - Bank 2

Cycle 39 -- PRECHARGE - Bank 1

Cycle 45 -- ACTIVATE - Bank 1, Row 58, near segment

Cycle 47 -- READ - Bank 1

\end{tcolorbox}

\subsubsection{Examples for Memory Controller Configuration}
\label{sec:memory_controller_configuration_examples}

\begin{tcolorbox}[breakable,
  colback=gray!5,
  colframe=gray!40!black,
  enhanced jigsaw,
  sharp corners,
  title=DRAM Command Design ]

\textbf{Context:}

You are given a memory system that has four channels, and each channel has two ranks of DRAM chips. A separate memory controller controls each memory channel. Each rank of DRAM contains eight banks. A bank contains R rows. Each row in one bank is 8KB. The minimum retention time among all DRAM rows in the system is 64 ms. In order to ensure that no data is lost, every DRAM row is refreshed once per 64 ms. Refresh of each row is initiated by a command from the memory controller. The command refreshes only the specified row. The command occupies the command bus on the associated memory channel for 5 ns and the associated bank for 40 ns.\\

We define refresh utilization of a resource (such as a bus or a memory bank) as the fraction of total time for which a resource is occupied by a refresh command.\\

\textbf{Question:}

How can you reduce the command bus utilization due to refreshes? You cannot change the refresh rate when answering this question.\\

\textbf{Correct Solution:}

Have each command be responsible for multiple row refreshes.

\end{tcolorbox}

\begin{tcolorbox}[breakable,
  colback=gray!5,
  colframe=gray!40!black,
  enhanced jigsaw,
  sharp corners,
  title=DRAM Hierarchy Configuration ]

\textbf{Context:}

Recall from your required reading on Tiered-Latency DRAM that there is a near and far segment, each containing some number of rows. Assume a very simplified memory model where there is just one bank and there are two rows in the near segment and four rows in the far segment. The time to activate and precharge a row is 25ns in the near segment and 50ns in the far segment. The time from start of activation to reading data is 10ns in the near segment and 15ns in the far segment. All other timings are negligible for this problem. Given the following memory request stream, determine the optimal assignment (minimize average latency of requests) of rows in the near and far segment (assume a fixed mapping where rows cannot migrate, a closed-row policy, and the far segment is inclusive).\\

time 0ns: row 0 read

time 10ns: row 1 read

time 100ns: row 2 read

time 105ns: row 1 read

time 200ns: row 3 read

time 300ns: row 1 read\\

\textbf{Question:}

Assume now that the mapping is dynamic. What are the tradeoffs of an exclusive design vs. an inclusive design? Name one advantage and one disadvantage for each.\\

\textbf{Correct Solution:}

Exclusive requires swapping, but can use nearly full capacity of DRAM. Inclusive, the opposite.

\end{tcolorbox}

\begin{tcolorbox}[breakable,
  colback=gray!5,
  colframe=gray!40!black,
  enhanced jigsaw,
  sharp corners,
  title=DRAM Sizing Optimization ]
  
\textbf{Context:}

You are given a memory system that has four channels, and each channel has two ranks of DRAM chips. A separate memory controller controls each memory channel. Each rank of DRAM contains eight banks. A bank contains R rows. Each row in one bank is 8KB. The minimum retention time among all DRAM rows in the system is 64 ms. In order to ensure that no data is lost, every DRAM row is refreshed once per 64 ms. Refresh of each row is initiated by a command from the memory controller. The command refreshes only the specified row. The command occupies the command bus on the associated memory channel for 5 ns and the associated bank for 40 ns.\\

We define refresh utilization of a resource (such as a bus or a memory bank) as the fraction of total time for which a resource is occupied by a refresh command.\\

\textbf{Question:}

Only changing the number of rows per bank, find the maximum number of rows per bank for which either the bank utilization or the command bus utilization reaches 100\%.\\

\textbf{Correct Solution:}

Because the command bus utilization will reach 100\% before the bank utilization, we will look at how changing the number of rows will affect the command bus utilization\\
R*5ns*2ranks*8banks/64ms = 1\\
R = 800,000\\
We find that with 800k rows per bank, the command bus utilization reaches 100\%.

\end{tcolorbox}

\section{Prompt Templates}
\label{appendix:prompts}

\subsection{LLM Prompt for MCQs}
\label{appendix:mcq_prompt}

\begin{tcolorbox}[breakable,
  colback=gray!5,
  colframe=gray!40!black,
  enhanced jigsaw,
  sharp corners,
  title=MCQ Prompt]

You are an expert computer architect solving multiple choice questions. Please read the following question and select the best answer from the choices provided.\\

Question:\\
\{question\}
\\

Choices:\\
\{choices\}
\\

Please conclude your response with a JSON object containing your final answer. The JSON object must match this schema: \{``final\_answer": ``$\langle$ A, B, C, or D$\rangle$"\}.
\end{tcolorbox}

This prompt is designed for multiple-choice evaluation, where correctness can be measured directly. The JSON-constrained output ensures answers are machine-readable and easy to score at scale.

\subsection{LLM Prompt for FRQs}
\label{appendix:frq_prompt}

\begin{tcolorbox}[breakable,
  colback=gray!5,
  colframe=gray!40!black,
  enhanced jigsaw,
  sharp corners,
  title=FRQ Prompt]

You are an expert computer architect taking an exam. You will be provided with a question and its context. Your task is to provide a clear, accurate, and well-reasoned answer to the question.\\

Please provide your answer in a structured format that clearly addresses the question. If the question involves calculations, show your work step-by-step. If it involves diagrams or tables, describe them clearly.\\

Remember to:
\begin{enumerate}
    \item Read the question carefully and understand what is being asked
    \item Use the provided context to inform your answer
    \item Show your reasoning and work where appropriate
    \item Be precise and accurate in your response
    \item If you're unsure about something, acknowledge the uncertainty
\end{enumerate}

Question Context: \\
\{context\} \\
\{context\_images\_placeholder\}
\\

Question: \\
\{question\}
\\

Please provide your answer:
\end{tcolorbox}

This prompt is used to simulate a ``student'' LLM, where models act as exam-takers solving architecture questions. The emphasis is on structured, step-by-step reasoning, accuracy, and clarity, mirroring how a human student would respond to technical exam questions.

\subsection{LLM Prompt for LLM-as-a-Judge on FRQ Responses}
\label{appendix:laj_prompt}

\begin{tcolorbox}[breakable,
  colback=gray!5,
  colframe=gray!40!black,
  enhanced jigsaw,
  sharp corners,
  title=LLM-as-a-Judge Prompt]

You are an expert computer architect acting as an exam grader to evaluate the quality of an answer to a computer architecture question. You will be provided with:

\begin{enumerate}
    \item The original question and context
    \item The correct solution
    \item A student's answer to the question
\end{enumerate}

Your task is to carefully evaluate whether the student's answer is correct, partially correct, or incorrect by comparing it to the provided solution.

Evaluation criteria:
\begin{itemize}
    \item CORRECT: The answer is accurate, complete, and demonstrates proper understanding
    \item PARTIALLY-CORRECT: The answer shows some understanding but has significant errors or is incomplete
    \item INCORRECT: The answer is fundamentally wrong or shows major misunderstandings
\end{itemize}

Consider:
\begin{itemize}
    \item Mathematical accuracy
    \item Conceptual understanding
    \item Completeness of the response solely in relation to the question being asked
    \item Logical reasoning
    \item Whether the answer addresses what was actually asked
\end{itemize}

Be fair but rigorous in your evaluation. If you are unsure, err on the side of being more critical.
\\

Question Context:\\
\{context\}\\
\{context\_images\_placeholder\}
\\

Question:\\
\{question\}
\\

Correct Solution:\\
\{solution\}\\
\{solution\_images\_placeholder\}
\\

Student's Answer:\\
\{student\_answer\}
\\

Please evaluate the student's answer and provide your reasoning. At the end of your response, write exactly one of the following words in all caps on a new line: CORRECT, PARTIALLY-CORRECT, or INCORRECT. If you do not end your response with a new line with exactly one of these options, you will not be paid for your work.
\end{tcolorbox}

This prompt is used to evaluate model outputs under the “LLM-as-a-Judge” paradigm. Here, the model acts as a grader, comparing student answers to reference solutions and deciding between CORRECT, PARTIALLY-CORRECT, or INCORRECT. It enables scalable evaluation of free-response questions while preserving rigor and consistency.

\subsection{LLM Prompt for Skills Classification}
\label{appendix:skills_classification_prompt}

\begin{tcolorbox}[breakable,
  colback=gray!5,
  colframe=gray!40!black,
  enhanced jigsaw,
  sharp corners,
  title=Skill Classification Prompt,
  before=\par\FloatBarrier, %
  after=\par\FloatBarrier   %
]You are a computer architecture expert who is a professional at categorizing exam questions based on the cognitive skill they are testing of a computer architect. Your task is to classify one question at a time into exactly one of the following four categories:

\textbf{Recall}: The question asks for a fact, definition, or direct retrieval of knowledge. Answering these questions should typically not require multi-step reasoning and can be directly answered in a single step. \\[0.3em]

\textbf{Analyze}: The question requires deducing, inferring, calculating, or interpreting information based on data or a specific scenario. Answering these questions typically requires some level of multi-step reasoning but does not require invention or innovating upon an existing solution. \\[0.3em]

\textbf{Design}: The question asks you to propose, invent, suggest, or improve a method, system component, or policy. It requires proposing new or improved solutions. \\[0.3em]

\textbf{Implement}: The question requires constructing, coding, or developing a full solution or system based on explicit requirements or specifications. These typically involve providing detailed instructions or actual code (i.e., programmable/executable artifacts). \\[0.8em]

Sometimes ``Analyze'' and ``Design'' can be confused if the question is open-ended. If the main effort is proposing or inventing, choose ``Design''; if it's interpreting specific data or information, choose ``Analyze''. \\[0.3em]

Additionally, sometimes ``Design'' and ``Implementation'' can also be confused. If the answer involves some sort of programmatic implementation or executable artifact then choose ``Implement''; if the answer is at a higher-level of abstraction than this, it is likely to be ``Design''. \\[0.8em]

Your output \textbf{MUST} be \textbf{ONLY} one word, chosen from the following list: \\
Recall, Analyze, Design, Implement \\[0.8em]

Find an example below: \\
\textit{Recall}: A \_\_\_\_ cache allows any block of main memory to be placed in any line, eliminating conflict misses but requiring complex associative lookup hardware. \\
\textit{Analyze}: A fully associative cache has 4 lines and uses an LRU policy. The following sequence of memory references occurs … What is the overall hit ratio? \\
\textit{Design}: Describe a cache replacement policy that would improve the performance of the hybrid memory system more than it would DRAM. \\ %
\textit{Implement}: Create a cache controller that interfaces with both a processor pipeline and a DRAM chip, following the provided Verilog port and signal specifications. \\[0.8em]

Now, classify this question:\\
\{question\}
\end{tcolorbox}

We use this prompt to consistently label each exam or benchmark question with the specific cognitive skill it targets. Our rationale in writing it was twofold: (i) provide clear, operational definitions of \emph{Recall}, \emph{Analyze}, \emph{Design}, and \emph{Implement} that are grounded in how architects approach problem solving, and (ii) minimize ambiguity by explicitly addressing common confusions between neighboring categories (e.g., Analyze vs. Design, or Design vs. Implement). This ensures that classification is reliable across reviewers and that the benchmark’s skill taxonomy aligns with real-world architectural workflows.  

\subsection{LLM Prompt for Architecture Topic Classification}
\label{appendix:architecture_classification_prompt}
To determine the architecture topic distribution of \ours{}, we employed a two-stage classification process that combines embedding-based similarity search with LLM reasoning for scalable categorization. In the first stage, the top 3 most relevant topic candidates are identified by embedding each question using OpenAI's text-embedding-3-large model and comparing against pre-computed taxonomy topic embeddings via cosine similarity. In the second stage, \modelname{GPT-4o} is used to make the final topic selections from these 3 candidates, providing both best and second-best classifications along with justifications. 

This hybrid approach effectively balances accuracy with scalability by leveraging the computational efficiency of embedding-based similarity search for initial filtering, while utilizing LLM reasoning capabilities for nuanced final classification decisions. Below is the prompt used for LLM categorization:

\begin{tcolorbox}[breakable,
  colback=gray!5,
  colframe=gray!40!black,
  enhanced jigsaw,
  sharp corners,
  title=Architecture Topic Classification Prompt,
  before=\par\FloatBarrier, %
  after=\par\FloatBarrier   %
] 
You are a computer architecture researcher and expert. You have been asked to categorize the following question into a subfield of computer architecture. Three options have been provided and you must select the top two.\\

    Question: \{question\} \\
    
    Categories: \\ 
    1: \{taxonomy\_terms[0]\} \\
    2: \{taxonomy\_terms[1]\} \\
    3: \{taxonomy\_terms[2]\} \\
    
    Please provide the exact names of the two categories that you feel best fit this question. First select the best match of the three options and then choose the second best category that matches. 
    You have been told **you have to pick no matter what from the options** and your response should be in the format without any additional text or explanation:\\
    
    [\\
       \{\\
          ``best\_selection'': ``[BEST CATEGORY HERE]", \\
           ``justification'': ``[JUSTIFICATION \#1 HERE]"\\
       \}, \\
       \{\\
          ``second\_best\_selection'': ``[SECOND BEST CATEGORY HERE]", \\
          ``justification'': ``[JUSTIFICATION \#2 HERE]"\\
       \}\\
    ]

\end{tcolorbox}

\subsection{LLM Prompt for Synthetic MCQ Generation}
\label{appendix:mcq_generation_prompt}
\begin{tcolorbox}[breakable,
  colback=gray!5,
  colframe=gray!40!black,
  enhanced jigsaw,
  sharp corners,
  title=MCQ Generation Prompt]
You are a computer architecture professor and expert researcher. You have been provided with the following excerpt about computer architecture: \\
``\{excerpt\}"
\\

You have been asked to create one difficult, paraphrased cloze-style format multiple choice question based on this excerpt to test senior computer architects.
\\

The questions will be used for creating an interview test for senior computer architects, and they will not get to read the excerpt for context, so any questions you create should not refer to anything specifically in the excerpt that would make the question unanswerable in the excerpts's absence.
\\

The questions must be precise and clear so they can be answered *definitively*, so avoid using qualifying adjectives or adverbs that would make the answer ambiguous or depend on the context; make sure there is only one correct answer to the question. 
\\

Quote *word for word* the sentence(s) of the excerpt context that are comprehensive and self-contained and could be read to justify and support each answer to each question. Your goal is to create good **conceptual** questions that test **conceptual** knowledge from the excerpt.
\\

Here is an example of a good cloze question: \\
\_\_\_\_ is the typical penalty incurred for a branch mispredict. \\
A) 100 us \\
B) 5 ms \\
C) 5 ns \\
D) 100 ms \\
Answer: C
\\

Provide your response in this *exact* format with *zero additional characters for formatting* before or after the opening and closing brackets: \\
$\text{[}$ \\
\{ \\ 
``question": ``[CLOSE QUESTION HERE]", \\
``option A": ``[OPTION A HERE]", \\
``option B": ``[OPTION B HERE]", \\
``option C": ``[OPTION C HERE]", \\
``option D": ``[OPTION D HERE]", \\
``answer":  ``\{specified\_answer\_choice[`Cloze']\}", \\
``context": ``[JUSTIFICATION HERE AS CONTEXT]", \\
``type": ``Cloze" \\
\} \\
$\text{]}$
\end{tcolorbox}

\subsection{LLM Prompts for Filtering Synthetic MCQs}
\label{appendix:mcq_filtering_prompt}

The following prompts were used to filter out poor MCQs that were synthetically generated. QAs that passed all of these filters were then finally verified by humans for quality and correctness.

\begin{tcolorbox}[breakable,
  colback=gray!5,
  colframe=gray!40!black,
  enhanced jigsaw,
  sharp corners,
  title=MCQ Filtering Prompt \#1]
You are a computer architecture expert. 
\\

You have been asked the following question: 
    
\{question\}
\\

Here are the options:\\
\{options\}
\\

As a computer architecture expert, would you need additional context to correctly answer the question? 
\\

Please answer with one word: ``YES" or ``NO".
\\

Then provide one short sentence to justify your answer reasoning.
Return your response in this exact format with zero other characters for formatting before or after: \\
\{ \\
``answer":  ``[ANSWER HERE: YES/NO]", \\
``justification": ``[JUSTIFICATION HERE]" \\
\}
\end{tcolorbox}

\begin{tcolorbox}[breakable,
  colback=gray!5,
  colframe=gray!40!black,
  enhanced jigsaw,
  sharp corners,
  title=MCQ Filtering Prompt \#2]
You are a computer architecture expert. 
\\

You have been asked the following question: \\
\{question\}
\\

Here are the options: \\
\{options\}
\\

As a computer architecture expert, of the provided options is there *only one* answer that is correct? 
\\

Please answer with one word: ``YES" or ``NO".
\\

Then provide one short sentence to justify your answer reasoning.
Return your response in this exact format with zero other characters for formatting before or after: \\
\{\\
``answer":  ``[ANSWER HERE: YES/NO]", \\
``justification": ``[JUSTIFICATION HERE]" \\
\}
\end{tcolorbox}

\begin{tcolorbox}[breakable,
  colback=gray!5,
  colframe=gray!40!black,
  enhanced jigsaw,
  sharp corners,
  title=MCQ Filtering Prompt \#3]
You are a new graduate student reading about computer architecture to learn the subject. 
\\

You have been given the following quote from a computer architecture excerpt to read for context: \\
\{context\}
\\

You have now been asked the following question: \\
\{question\}
\\

Here are the options: \\
\{options\}
\\

Does the provided context you read sufficiently help you answer the question correctly?
\\

Please answer with one word: ``YES" or ``NO".
\\

Then provide one short sentence to justify why you answered ``YES" or ``NO".
Return your response in this exact format with zero other characters for formatting before or after: \\
\{ \\
``answer":  ``[ANSWER HERE: YES/NO]", \\
``justification": ``[JUSTIFICATION HERE]" \\
\}
\end{tcolorbox}

\begin{tcolorbox}[breakable,
  colback=gray!5,
  colframe=gray!40!black,
  enhanced jigsaw,
  sharp corners,
  title=MCQ Filtering Prompt \#4]
You are a computer architecture expert. 

You have been given the following quote from a computer architecture excerpt for context: \\
\{context\}
\\

You have now been asked the following question: \\
\{question\}
\\

Here are the options: \\
\{options\}
\\

Choose the best correct option and provide your answer justification.
Return your response in this exact format with zero other characters for formatting before or after: \\
\{ \\
``answer": ``[ANSWER HERE: LETTER OF OPTION]", \\
``justification": ``[JUSTIFICATION HERE]" \\
\}
\end{tcolorbox}

\subsection{LLM Prompt for Text Extraction of Exam QAs}

This is the prompt used to extract the text for questions in our crowdsourced exams. It ensures that problem statements, sub-questions, and solutions are consistently structured into JSON, while ignoring irrelevant formatting such as point values or images. By enforcing this schema, we can standardize raw exam PDFs into machine-readable data suitable for validation, benchmarking, and downstream analysis.

\label{appendix:text_extraction}
\begin{tcolorbox}[breakable,
  colback=gray!5,
  colframe=gray!40!black,
  enhanced jigsaw,
  sharp corners,
  title=Exam Question Text Extraction Prompt]

You are a language model assisting with the digitization of academic exam content. The input is a PDF file containing one problem of a computer architecture assessment. If part of another problem is included, ignore it and only focus on \{filename\}.\\\\
The problem may include any combination of the following:\\
A context paragraph, or just a short statement (e.g., “Convert the number 42 to binary”)\\
One or more sub-questions, or be a single standalone question\\
Context for sub-questions separate from the sub-question and separate from the original problem context\\
Multiple questions within a subquestion\\
Point value associations for the problem or subproblems, including extra credit points\\
Solutions, either typed or handwritten\\
Tables, diagrams, circuit schematics, or block diagrams\\

Your task is to identify and separate each exam problem into the listed components, including context, sub-questions, and solutions. At times, a subquestion can have nested subparts. Ignore any point values for any problem, question, or sub-question. Ignore any images, charts, or figures and do not attempt to extract text from them.\\\\
If a provided image is not part of the problem in the pdf file and instead is part of another problem(s), omit it from the dictionary.
Format your response so that it can be exported into a JSON file using the template below.
If the particular exam question lacks any of the listed components, omit them from the template.\\

Template for a problem which is split up into sub-problems:\\
\begin{lstlisting}[basicstyle=\ttfamily\small, breaklines=true]
{
    "problem": "1",
    "problem_context": <Insert any introductory paragraph or description exactly as it appears. If there is no context, don't include this header>,
    "subproblems": [ 
    {
        "subproblem": "a" (Copy the part letter/number exactly as it appears on the exam),
        "subproblem_context": <Insert any introductory paragraph or description exactly as it appears. If there is no subproblem context or if the question is the only part of the subproblem, don't include this header. Replace all double quotes " here with escaped double quotes /">,
        "subproblem_question": <Insert the full question of the subproblem, exactly as it appears in the original. Replace all double quotes " here with escaped double quotes /">,
        "subproblem_solution": <Insert the full solution of the subproblem, exactly as shown in the original. Replace all double quotes " here with escaped double quotes /">
    },
    ...repeat as needed for additional subproblems within this problem
    ],
}
\end{lstlisting}

Template for a problem which is standalone and has no sub-problems:
\begin{lstlisting}[basicstyle=\ttfamily\small, breaklines=true]
{
    "problem": "1",
    "problem_context": <Insert any introductory paragraph or description exactly as it appears. If there is no context, don't include this header. Replace all double quotes " here with escaped double quotes /">,
    "problem_question": <Insert the full question of the problem, exactly as it appears in the original. Replace all double quotes " here with escaped double quotes /">,
    "problem_solution": <Insert the full solution of the problem, exactly as shown in the original. Replace all double quotes " here with escaped double quotes /">
}
\end{lstlisting}

If any double quotes ('') within strings appear within your JSON, you must replace them with escaped double quotes (/'').
Return only valid JSON.
\end{tcolorbox}

\subsection{LLM Prompt for Image Extraction of Exam QAs}
\label{appendix:image_extraction}

This is the prompt used to associate figures with their correct roles in the digitized exam problems. By structuring images into categories such as problem context, subproblem context, problem solution, and subproblem solution, we can align visual information with text-based question content. This ensures that diagrams, tables, and schematics are consistently linked to their intended question or solution, enabling precise and reproducible dataset construction.

\begin{tcolorbox}[breakable,
  colback=gray!5,
  colframe=gray!40!black,
  enhanced jigsaw,
  sharp corners,
  title=Exam Question Image Extraction Prompt]

You are a language model assisting with the digitization of academic exam content in computer architecture.\\

**Input:** \\
You are provided with: \\
1) A PDF file containing one problem of an exam. If part of another problem is included, ignore this and only focus on the current pdf file.\\
2) A .json-styled txt file containing the problem's extracted text. It may contain some or all of the following empty-list fields: ``\texttt{problem\_context\_figures}'', ``\texttt{subproblem\_context\_figures}'', ``\texttt{subproblem\_solution\_figures}'', and ``\texttt{problem\_solution\_figures}''. \\
3) PNG images containing tables, diagrams, circuit schematics, or block diagrams that may or may not pertain to this problem. The names of the images provided are as follows, in order: \{images\}.\\

The given problem may be a standalone problem or consist of multiple sub-problems. Each problem or sub-problem may contain: \\
-- Main context figures which are necessary to understanding the main problem, or ALL of the sub-problems. \\
-- Sub-problem-specific figures that are separate from both the main problem context and the other sub-problems. \\
-- Solutions to the main problem, which may be typed or handwritten. \\
-- Solutions to a certain sub-problem, which may be typed or handwritten. \\

**Your task:** \\
Match each image file name (table, diagram, circuit schematic, or block diagram) to its correct association in the exam. Each image should have one of the four possible associations: \\
-- The main problem question (``\texttt{problem\_context\_figures}'') \\
-- The sub-problem question (``\texttt{subproblem\_context\_figures}'') \\
-- The main problem solution (``\texttt{problem\_solution\_figures}'') \\
-- The sub-problem solution (``\texttt{subproblem\_solution\_figures}'') \\

At times, the context or question in the main problem/sub-problem question/solution will include phrases (such as “The table below” or “The following diagram”) that indicate a visual image falls under that category. \\

Your output should be a modified version of the given JSON, but with each of the figure fields populated with lists of the relevant image names (files ending in .png, .jpg, etc.), for example: \\
``\texttt{problem\_context\_figures}'': [``\texttt{image\_name.png}'', ``\texttt{image\_name\_2.png}'']\\  

No content, descriptions, or recreations of the image should be included in the output; only the file name should be included. If a provided image is not part of the current PDF file and instead is part of other problem(s), omit it. If the provided JSON includes image file names that are not present in the images provided, omit those too.\\

Be as precise as possible in your associations. Only include the dictionary; do not include reasoning. If none of the images given pertain to the problem, just output the original JSON given.
\end{tcolorbox}

\subsection{LLM Prompt for Verification of Extracted Exam QAs}
\label{appendix:exam_verification}

This prompt is used to audit the parser’s extracted question–answer pairs against the original exam source. It enforces strict, itemized criteria for textual fidelity, self-contained context, correct image/table extraction, and proper figure categorization, and it standardizes the verification output by toggling \texttt{passed\_llm\_verification} and adding concise reasoning where failures are detected.

\begin{tcolorbox}[breakable,
  colback=gray!5,
  colframe=gray!40!black,
  enhanced jigsaw,
  sharp corners,
  title=Exam Question Verification]

You are a reader tasked with verifying the accuracy of an automated document parser. A computer architecture exam PDF has been fed through the parser to produce standalone question–answer pairs in a certain JSON format, and you must compare the parsed questions to the original questions in the PDF.\\

**Input:** You are given (1) the original PDF of \{problem\_name\} of the exam which may be broken down into subproblems, (2) the JSON dictionary produced by the parser, and (3) images that the parser has deemed are associated with the question and/or its subproblems. The names of the images provided are as follows, in order: \{images\}. Your task is to determine if the parser has correctly extracted the text and images while staying true to the original PDF.\\

**A problem is correct only if all of the following are satisfied:**\\
a) The extracted problem text is nearly identical, word-for-word, to the original PDF's text (special characters/math symbols may appear as unicode-escaped or equivalent).\\
b) The problem's ``\texttt{question}'' and ``\texttt{solution}'' fields are both populated. Exception: if the original solution is purely an image, that image must be correctly associated in ``\texttt{solution\_figures}''.\\
c) The problem is standalone: the ``\texttt{context}'' and ``\texttt{context\_figures}'' provide all information needed, even if the original referenced prior problems.\\
d) No part of the solution is revealed in ``\texttt{context}'' or ``\texttt{context\_figures}''.\\
e) All images are extracted/cropped correctly and categorized correctly as context vs.\ solution figures.\\
f) All tables are extracted correctly. If parsed as text instead of image, the table must be recreatable from the extracted text and usable to answer the question.\\
g) For fill-in-the-blank or fill-in-the-chart, the blank version must be provided in the question/context or ``\texttt{context\_figures}''.\\

**Output format:**\\
Return a modified version of the given JSON dictionary. Set each ``\texttt{passed\_llm\_verification}'' field to \texttt{true} or \texttt{false} (unquoted). For every item marked \texttt{false}, add a ``\texttt{reasoning}'' field explaining which conditions failed. Return exactly one dictionary and nothing else.\\
**Conservatism:** Err on the side of false. Avoid false positives; false negatives are acceptable.
\end{tcolorbox}

\subsection{LLM Prompts for Case Study \#1}
\label{appendix:finetuning_case_study_prompt}

The following prompt was used to filter out \ours{} FRQs related to memory subsystems for our case study in Section~\ref{sec:finetuning-case-study}, using \modelname{Claude-3.7-Sonnet-Thinking}:

\begin{tcolorbox}[breakable,
  colback=gray!5,
  colframe=gray!40!black,
  enhanced jigsaw,
  sharp corners,
  title=LLM-assisted Filtering of Memory Subsystem FRQs]
You are an expert computer-architecture instructor. You will be given a question, its context, and the provided solution. You need to decide whether the question directly tests the necessary skills and knowledge for designing a memory subsystem with caches for a matrix multiplication workload. Full walkthroughs of selected questions will be used to teach an experienced engineer the necessary reasoning steps for optimizing memory performance for matrix multiplication. \texttt{\{question\}\{solution\}}
\end{tcolorbox}

The following three user instruction prompts were used to produce SFT samples for data distillation. For each of the 45 \ours{} FRQs filtered using the above prompt, the teacher model (\modelname{GPT-5.2}) targeted three student types via separate roles:

\begin{tcolorbox}[breakable,
  colback=gray!5,
  colframe=gray!40!black,
  enhanced jigsaw,
  sharp corners,
  title=Teacher Prompt for Student Type \#1: Junior Architect in Industry]

You are a senior computer architect that is an expert in memory subsystems. You are training a less-experienced computer architect to take over your role as a cache designer. They have brought you the following question and solution from a graduate academic computer architecture exam, and asked you to explain the underlying principles an expert needs to know about in order to determine why this solution is the correct one, including nuances that they may be unaware of even if they answered correctly and completely. In particular, you should precisely convey the architectural principles and considerations relevant specifically to this exam question, and briefly touch on any counterexamples or connections that will reinforce their expertise in cache and memory design. Reason deeply about this exam question and solution, and consider what you can teach the architect. Keep your final didactic response to under 5 paragraphs, and keep in mind they understand the fundamentals of memory subsystems and caches at a technical level already. \texttt{\{question\}\{solution\}}\\
\end{tcolorbox}

\begin{tcolorbox}[breakable,
  colback=gray!5,
  colframe=gray!40!black,
  enhanced jigsaw,
  sharp corners,
  title=Teacher Prompt for Student Type \#2: Computer Architecture Student]

You are a professor in computer architecture that is an expert in memory subsystems. A student has brought you a question and answer from a graduate academic computer architecture exam, and asked you to explain the underlying principles behind why this solution is the correct one. In particular, when teaching them how to solve this problem you should carefully and precisely bolster their ability to solve a wide array of memory subsystem problems, and reinforce their skills in cache and memory design topics. Reason deeply about this exam question and solution, and keep your final didactic response to under 5 paragraphs (do not include basic definitions and cursory explanations, as they have taken your introductory computer architecture course already). \texttt{\{question\}\{solution\}}\\
\end{tcolorbox}

\begin{tcolorbox}[breakable,
  colback=gray!5,
  colframe=gray!40!black,
  enhanced jigsaw,
  sharp corners,
  title=Teacher Prompt for Student Type \#3: Novice]

You are a professor in computer architecture that is an expert in memory subsystems. A talented university student has brought you a question and answer from a graduate academic computer architecture exam, and asked you to explain how to arrive at the solution. The student has taken an introductory computer architecture class many years ago and remembers term definitions and can provide cursory explanations of concepts, but they don’t know how to begin to solve this particular problem and require a careful walkthrough. In particular, when teaching them how to solve this problem you should bolster their ability to solve other questions of this type, topic, and difficulty, and heavily reinforce their skills in cache and memory design topics. Every word counts and you must be incredibly precise and technically dense in your response to achieve these goals. This is the culmination of your life’s work as a subject matter expert and teacher. Reason deeply about this exam question and solution, and keep your final didactic response to under 5 paragraphs. \texttt{\{question\}\{solution\}}
\end{tcolorbox}

\subsection{LLM Prompt for Case Study \#2}
\label{appendix:memory_controller_case_study_prompt}

The following prompt is used for the DRAM controller case study (Appendix Section~\ref{sec:memory-controller-case-study}). It tasks an LLM with answering 3 questions and then proposing a new memory controller design configuration. \texttt{current\_iteration} is the i-th round of the 40 evaluation budget provided to the LLM. Thirty lines of the beginning, middle, and end of the workload's memory trace are sampled to add to the prompt. Some example configurations proposed by an LLM during a trial in this case study are provided for illustration in the prompt below. Note that ellipses are used for brevity but not actually a part of the prompt.

\begin{tcolorbox}[breakable,
  colback=gray!5,
  colframe=gray!40!black,
  enhanced jigsaw,
  sharp corners,
  title=Memory Controller Design Configuration Task]

You are a computer architecture expert specializing in DRAM memory controller design.
You have been tasked with tuning the design of a custom DRAM memory controller to optimize for **POWER** (lower is better).\\

=== SEARCH BUDGET ===

You are on iteration \{\texttt{current\_iteration}\} of 40.

Consider this when deciding your strategy:\\
- Early iterations: Consider exploring diverse configurations to understand the design space.\\
- Later iterations: Consider exploiting promising regions by making smaller refinements to the best configurations.\\

=== DRAM CONTROLLER DESIGN PARAMETERS ===

You control 10 DISCRETE design parameters, each represented as an INTEGER INDEX:

0: PagePolicy, Options: [``Open'', ``OpenAdaptive'', ``Closed'', ``ClosedAdaptive'']\\
1: Scheduler, Options: [``Fifo'', ``FrFcfsGrp'', ``FrFcfs'']\\
2: SchedulerBuffer, Options: [``Bankwise'', ``ReadWrite'', ``Shared'']\\
3: RequestBufferSize, Options: [1, 2, 4, 8, 16, 32, 64, 128]\\
4: RespQueue, Options: [``Fifo'', ``Reorder'']\\
5: RefreshPolicy, Options: [``NoRefresh'', ``AllBank'']\\
6: RefreshMaxPostponed, Options: [1,2,4,8]\\
7: RefreshMaxPulledin, Options: [1,2,4,8]\\
8: Arbiter, Options: [``Simple'', ``Fifo'', ``Reorder'']\\
9: MaxActiveTransactions, Options: [1, 2, 4, 8, 16, 32, 64, 128]\\

=== CONFIGURATIONS TRIED SO FAR ===

Here are the top 10 best configurations you've found so far, ranked by Power (W) (best first):\\
\#1: [1.0, 2.0, 2.0, 1.0, 0.0, 0.0, 0.0, 0.0, 0.0, 0.0], Energy=0.000194 J, Power=1.697670 W, Latency=0.000114290 s\\
\#2: [1.0, 2.0, 2.0, 2.0, 0.0, 0.0, 0.0, 0.0, 0.0, 1.0], Energy=0.000178 J, Power=2.092590 W, Latency=0.085218750 s\\
\#3: [1.0, 2.0, 2.0, 3.0, 1.0, 0.0, 0.0, 0.0, 0.0, 2.0], Energy=0.000171 J, Power=2.408910 W, Latency=0.000070795 s\\
...\\

Here are the last 10 configurations you've tried:\\
\#1: [1.0, 2.0, 2.0, 3.0, 1.0, 0.0, 0.0, 0.0, 0.0, 2.0]\\
\#2: [1.0, 2.0, 2.0, 2.0, 0.0, 0.0, 0.0, 0.0, 0.0, 1.0]\\
\#3: [1.0, 2.0, 2.0, 1.0, 0.0, 0.0, 0.0, 0.0, 0.0, 0.0]\\
...\\

IMPORTANT: You CANNOT propose a configuration that exactly matches any of the Top 10 or Last 10 configs you've tried listed above as proposing a duplicate configuration wastes an evaluation.\\

=== MEMORY TRACE SAMPLE ===

Below is a sample of the workload's memory trace (format: ``timestamp: operation address''):\\
--- Beginning (lines 0-29 of 10000) ---\\
0:	read	0x0\\
1:	read	0x40\\
2:	read	0x80\\
3:	read	0xC0\\
4:	read	0x100\\
...\\

--- Middle (lines 4985-5014 of 10000) ---\\
4985:	read	0x190BBCC0\\
4986:	read	0x26EEB740\\
4987:	read	0x29CEE6C0\\
4988:	read	0x3771780\\
4989:	read	0x277DC1C0\\
...\\

--- End (lines 9970-9999 of 10000) ---\\
9970:	read	0x4A880\\
9971:	read	0x4A8C0\\
9972:	read	0x4A900\\
9973:	read	0x4A940\\
9974:	read	0x4A980\\
...\\

=== YOUR TASK ===

Before proposing a new configuration, use your understanding of DRAM memory controller architecture to reason through the following to make an informed choice:\\

1. **Memory Access Pattern Analysis**: Looking at the memory trace sample above, what are key characteristics of this workload that you should consider when tuning the DRAM memory controller?\\

2. **Parameter Impact Analysis**: Based on the trace sample and broader optimization target (POWER) of yours, which parameters are most critical to tune and why?\\

3. **Learning from History**: Looking at previous configurations and their metrics, what patterns do you observe? Which parameter changes led to improvements?\\

Based on your analysis from the THREE QUESTIONS above and the remaining search budget, propose one NEW configuration that you expect will MINIMIZE POWER.\\

IMPORTANT OUTPUT FORMAT:\\
- First, provide your reasoning in a \textless reasoning\textgreater block\\
- Then, output ONLY a JSON list of exactly 10 integers (the indices 0..7 or 0..3 etc) on its own line\\
- Do NOT include any extra text beyond the \textless reasoning\textgreater and JSON list of integer indices.\\

Example output format (enclosed in markdown code block):
\begin{verbatim}
```
\end{verbatim}
\textless reasoning\textgreater\\
My analysis shows that ... therefore, I propose the following configuration:\\
\textless /reasoning\textgreater\\
indices=[2, 0, 1, 3, 1, 0, 2, 1, 2, 4]
\begin{verbatim}
```
\end{verbatim}
which in this example would correspond to the following DRAM controller design configuration:\\
  - PagePolicy: ``Closed'' (index 2)\\
  - Scheduler: ``Fifo'' (index 0)\\
  - SchedulerBuffer: ``ReadWrite'' (index 1)\\
  - RequestBufferSize: 8 (index 3)\\
  - RespQueue: ``Reorder'' (index 1)\\
  - RefreshPolicy: ``NoRefresh'' (index 0)\\
  - RefreshMaxPostponed: 4 (index 2)\\
  - RefreshMaxPulledin: 2 (index 1)\\
  - Arbiter: ``Reorder'' (index 2)\\
  - MaxActiveTransactions: 16 (index 4)\\
\end{tcolorbox}

\end{document}